\documentclass[preprint2]{emulateapj}

\usepackage{amsmath}
\usepackage{natbib}
\usepackage{graphicx}
\usepackage{epsf}
\usepackage{subfigure}
\usepackage{color}
\usepackage{threeparttable}
\usepackage{comment}
\usepackage{epsfig}
\usepackage{xspace}
\usepackage{hyperref}
\usepackage[usenames,dvipsnames,svgnames]{xcolor}
\usepackage{latexsym}
\DeclareGraphicsExtensions{.jpg,.pdf,.png,.eps,.ps}

\newcommand{\omb}{\ensuremath{\Omega_{b}h^2}\xspace}
\newcommand{\omc}{\ensuremath{\Omega_{c}h^2}\xspace}
\newcommand{\neff}{\ensuremath{N_\mathrm{eff}}\xspace}
\newcommand{\yhe}{\ensuremath{Y_\mathrm{p}}\xspace}
\newcommand{\As}{\ensuremath{A_\mathrm{s}}\xspace}

\newcommand{\Alens}{\ensuremath{A_\mathrm{L}}\xspace}
\newcommand{\ns}{\ensuremath{n_\mathrm{s}}\xspace}
\newcommand{\thetas}{\ensuremath{\theta_\mathrm{s}}\xspace}
\newcommand{\thetaMC}{\ensuremath{\theta_\mathrm{MC}}\xspace}
\newcommand{\ho}{\ensuremath{H_{0}}\xspace}
\newcommand{\clamp}{\ensuremath{10^9 \As e^{-2\tau}}\xspace}
\newcommand{\sigmaeight}{\ensuremath{\sigma_{8}}\xspace}

\newcommand{\LCDM}{\mbox{$\Lambda$CDM}\xspace}
\newcommand{\LCDMAlens}{\mbox{$\Lambda$CDM+\Alens}\xspace}
\newcommand{\LCDMyhe}{\mbox{$\Lambda$CDM+\yhe}\xspace}

\newcommand{\LCDMneff}{\mbox{$\Lambda$CDM+\neff}\xspace}
\newcommand{\LCDMyheneff}{\mbox{$\Lambda$CDM+$Y_{\mathrm{p}}$+\neff}\xspace}
\newcommand{\Dps}{\ensuremath{D_{3000}^{\mathrm{PS_{EE}}}}\xspace}
 
\newcommand{\planck}{\textit{Planck}\xspace}

\newcommand{\sptpolEETE}{\textsc{SPTpol}\xspace}
\newcommand{\planckTT}{\textsc{PlanckTT}\xspace}
\newcommand{\cmb}{\textsc{PlanckTT+SPTpol}\xspace}

\def\microKsq{\mu{\mbox{K}}^2}
\def\microK{\mu{\mbox{K}}}
\newcommand{\muksq}{\ensuremath{\mu{\rm K}^2}\xspace}
\newcommand{\sqdeg}{\ensuremath{\mathrm{deg}^2}\xspace}
\newcommand{\mukarcmin}{$\mu\mbox{K}-\mbox{arcmin}$\xspace}
\newcommand{\bea}{\begin{eqnarray}}
\newcommand{\eea}{\end{eqnarray}}

\newcommand{\sze}{Sunyaev-Zel'dovich effect}
\newcommand{\eg}{\textit{e.g.}}
\newcommand{\ie}{\textit{i.e.}}
\newcommand{\simleq}{{\raise.0ex\hbox{$\mathchar"013C$}\mkern-14mu \lower1.2ex\hbox{$\mathchar"0218$}}}
\newcommand{\simgeq}{{\raise.0ex\hbox{$\mathchar"013E$}\mkern-14mu \lower1.2ex\hbox{$\mathchar"0218$}}}

\begin{document}

\title{Measurements of the temperature and E-mode polarization of the CMB from 500 square degrees of SPTpol data}

\def\KICPChicago{1}
\def\AAUChicago{2}
\def\ColoradoAPS{3}
\def\Melbourne{4}
\def\Cardiff{5}
\def\FNAL{6}
\def\NIST{7}
\def\ArgonneHEP{8}
\def\PhysicsUChicago{9}
\def\EFIChicago{10}
\def\UKZN{11}
\def\SLAC{12}
\def\TAPIRCaltech{13}
\def\Caltech{14}
\def\Berkeley{15}
\def\McGill{16}
\def\CIFAR{17}
\def\HarveyMudd{18}
\def\ESO{19}
\def\ColoradoPhys{20}
\def\illast{21}
\def\illphy{22}
\def\UChicago{23}
\def\Stanford{24}
\def\KIPAC{25}
\def\Davis{26}
\def\LBNL{27}
\def\Michigan{28}
\def\Dunlap{29}
\def\ArgonneMSD{30}
\def\Minnesota{31}
\def\CaseWestern{32}
\def\ArtInstChicago{33}
\def\ThreeSpeedLogic{34}
\def\CfA{35}
\def\UToronto{36}
\def\Maryland{37}
\def\UCLA{38}

\shortauthors{J.~W.~Henning, J.T.~Sayre, C.~L.~Reichardt, et al.}
\author{
  J.~W.~Henning\altaffilmark{\KICPChicago,\AAUChicago},
  J.T.~Sayre\altaffilmark{\ColoradoAPS},
  C.~L.~Reichardt\altaffilmark{\Melbourne},
  P.~A.~R.~Ade\altaffilmark{\Cardiff},
  A.~J.~Anderson\altaffilmark{\FNAL},
  J.~E.~Austermann\altaffilmark{\NIST},
  J.~A.~Beall\altaffilmark{\NIST},
  A.~N.~Bender\altaffilmark{\KICPChicago,\ArgonneHEP},
  B.~A.~Benson\altaffilmark{\KICPChicago,\AAUChicago,\FNAL},
  L.~E.~Bleem\altaffilmark{\KICPChicago,\ArgonneHEP},
  J.~E.~Carlstrom\altaffilmark{\KICPChicago,\AAUChicago,\ArgonneHEP,\PhysicsUChicago,\EFIChicago},
  C.~L.~Chang\altaffilmark{\KICPChicago,\AAUChicago,\ArgonneHEP},
  H.~C.~Chiang\altaffilmark{\UKZN},
  H-M.~Cho\altaffilmark{\SLAC},
  R.~Citron\altaffilmark{\KICPChicago},
  C.~Corbett~Moran\altaffilmark{\TAPIRCaltech},
  T.~M.~Crawford\altaffilmark{\KICPChicago,\AAUChicago},
  A.~T.~Crites\altaffilmark{\KICPChicago,\AAUChicago,\Caltech},
  T.~de~Haan\altaffilmark{\Berkeley},
  M.~A.~Dobbs\altaffilmark{\McGill,\CIFAR},
  W.~Everett\altaffilmark{\ColoradoAPS},
  J.~Gallicchio\altaffilmark{\KICPChicago,\HarveyMudd},
  E.~M.~George\altaffilmark{\Berkeley,\ESO},
  A.~Gilbert\altaffilmark{\McGill},
  N.~W.~Halverson\altaffilmark{\ColoradoAPS,\ColoradoPhys},
  N.~Harrington\altaffilmark{\Berkeley},
  G.~C.~Hilton\altaffilmark{\NIST},
  G.~P.~Holder\altaffilmark{\CIFAR,\illast,\illphy},
  W.~L.~Holzapfel\altaffilmark{\Berkeley},
  S.~Hoover\altaffilmark{\KICPChicago,\PhysicsUChicago},
  Z.~Hou\altaffilmark{\KICPChicago},
  J.~D.~Hrubes\altaffilmark{\UChicago},
  N.~Huang\altaffilmark{\Berkeley},
  J.~Hubmayr\altaffilmark{\NIST},
  K.~D.~Irwin\altaffilmark{\SLAC,\Stanford},
  R.~Keisler\altaffilmark{\Stanford,\KIPAC},
  L.~Knox\altaffilmark{\Davis},
  A.~T.~Lee\altaffilmark{\Berkeley,\LBNL},
  E.~M.~Leitch\altaffilmark{\KICPChicago,\AAUChicago},
  D.~Li\altaffilmark{\NIST,\SLAC},
  A.~Lowitz\altaffilmark{\KICPChicago},
  A.~Manzotti\altaffilmark{\KICPChicago,\AAUChicago},
  J.~J.~McMahon\altaffilmark{\Michigan},
  S.~S.~Meyer\altaffilmark{\KICPChicago,\AAUChicago,\PhysicsUChicago,\EFIChicago},
  L.~Mocanu\altaffilmark{\KICPChicago,\AAUChicago},
  J.~Montgomery\altaffilmark{\McGill,\CIFAR},
  A.~Nadolski\altaffilmark{\illphy},
  T.~Natoli\altaffilmark{\Dunlap},
  J.~P.~Nibarger\altaffilmark{\NIST},
  V.~Novosad\altaffilmark{\ArgonneMSD},
  S.~Padin\altaffilmark{\KICPChicago,\AAUChicago,\Caltech},
  C.~Pryke\altaffilmark{\Minnesota},
  J.~E.~Ruhl\altaffilmark{\CaseWestern},
  B.~R.~Saliwanchik\altaffilmark{\UKZN},
  K.~K.~Schaffer\altaffilmark{\KICPChicago,\EFIChicago,\ArtInstChicago},
  C.~Sievers\altaffilmark{\KICPChicago},
  G.~Smecher\altaffilmark{\McGill,\ThreeSpeedLogic},
  A.~A.~Stark\altaffilmark{\CfA},
  K.~T.~Story\altaffilmark{\Stanford,\KIPAC},
  C.~Tucker\altaffilmark{\Cardiff},
  K.~Vanderlinde\altaffilmark{\Dunlap,\UToronto},
  T.~Veach\altaffilmark{\Maryland},
  J.~D.~Vieira\altaffilmark{\illast,\illphy},
  G.~Wang\altaffilmark{\ArgonneHEP},
  N.~Whitehorn\altaffilmark{\UCLA,\Berkeley},
  W.~L.~K.~Wu\altaffilmark{\Berkeley},
  and
  V.~Yefremenko\altaffilmark{\ArgonneHEP}
}

\altaffiltext{\KICPChicago}{Kavli Institute for Cosmological Physics, University of Chicago, 5640 South Ellis Avenue, Chicago, IL, USA 60637}
\altaffiltext{\AAUChicago}{Department of Astronomy and Astrophysics, University of Chicago, 5640 South Ellis Avenue, Chicago, IL, USA 60637}
\altaffiltext{\ColoradoAPS}{Center for Astrophysics and Space Astronomy, Department of Astrophysical and Planetary Sciences, University of Colorado, Boulder, CO, USA 80309}
\altaffiltext{\Melbourne}{School of Physics, University of Melbourne, Parkville, VIC 3010, Australia}
\altaffiltext{\Cardiff}{School of Physics and Astronomy, Cardiff University CF24 3AA, UK}
\altaffiltext{\FNAL}{Fermi National Accelerator Laboratory, MS209, P.O. Box 500, Batavia, IL 60510}
\altaffiltext{\NIST}{NIST Quantum Sensors Group, 325 Broadway Mailcode 687.08, Boulder, CO, USA 80305}
\altaffiltext{\ArgonneHEP}{High Energy Physics Division, Argonne National Laboratory, 9700 S. Cass Avenue, Argonne, IL, USA 60439}
\altaffiltext{\PhysicsUChicago}{Department of Physics, University of Chicago, 5640 South Ellis Avenue, Chicago, IL, USA 60637}
\altaffiltext{\EFIChicago}{Enrico Fermi Institute, University of Chicago, 5640 South Ellis Avenue, Chicago, IL, USA 60637}
\altaffiltext{\UKZN}{School of Mathematics, Statistics \& Computer Science, University of KwaZulu-Natal, Durban, South Africa}
\altaffiltext{\SLAC}{SLAC National Accelerator Laboratory, 2575 Sand Hill Road, Menlo Park, CA 94025}
\altaffiltext{\TAPIRCaltech}{TAPIR, Walter Burke Institute for Theoretical Physics, California Institute of Technology, 1200 E California Blvd, Pasadena, CA 91125, USA}
\altaffiltext{\Caltech}{California Institute of Technology, MS 249-17, 1216 E. California Blvd., Pasadena, CA, USA 91125}
\altaffiltext{\Berkeley}{Department of Physics, University of California, Berkeley, CA, USA 94720}
\altaffiltext{\McGill}{Department of Physics, McGill University, 3600 Rue University, Montreal, Quebec H3A 2T8, Canada}
\altaffiltext{\CIFAR}{Canadian Institute for Advanced Research, CIFAR Program in Cosmology and Gravity, Toronto, ON, M5G 1Z8, Canada}
\altaffiltext{\HarveyMudd}{Harvey Mudd College, 301 Platt Blvd., Claremont, CA 91711}
\altaffiltext{\ESO}{European Southern Observatory, Karl-Schwarzschild-Str. 2, 85748 Garching}
\altaffiltext{\ColoradoPhys}{Department of Physics, University of Colorado, Boulder, CO, USA 80309}
\altaffiltext{\illast}{Astronomy Department, University of Illinois at Urbana-Champaign, 1002 W. Green Street, Urbana, IL 61801, USA}
\altaffiltext{\illphy}{Department of Physics, University of Illinois Urbana-Champaign, 1110 W. Green Street, Urbana, IL 61801, USA}
\altaffiltext{\UChicago}{University of Chicago, 5640 South Ellis Avenue, Chicago, IL, USA 60637}
\altaffiltext{\Stanford}{Dept. of Physics, Stanford University, 382 Via Pueblo Mall, Stanford, CA 94305}
\altaffiltext{\KIPAC}{Kavli Institute for Particle Astrophysics and Cosmology, Stanford University, 452 Lomita Mall, Stanford, CA 94305}
\altaffiltext{\Davis}{Department of Physics, University of California, One Shields Avenue, Davis, CA, USA 95616}
\altaffiltext{\LBNL}{Physics Division, Lawrence Berkeley National Laboratory, Berkeley, CA, USA 94720}
\altaffiltext{\Michigan}{Department of Physics, University of Michigan, 450 Church Street, Ann  Arbor, MI, USA 48109}
\altaffiltext{\Dunlap}{Dunlap Institute for Astronomy \& Astrophysics, University of Toronto, 50 St George St, Toronto, ON, M5S 3H4, Canada}
\altaffiltext{\ArgonneMSD}{Materials Sciences Division, Argonne National Laboratory, 9700 S. Cass Avenue, Argonne, IL, USA 60439}
\altaffiltext{\Minnesota}{School of Physics and Astronomy, University of Minnesota, 116 Church Street S.E. Minneapolis, MN, USA 55455}
\altaffiltext{\CaseWestern}{Physics Department, Center for Education and Research in Cosmology and Astrophysics, Case Western Reserve University, Cleveland, OH, USA 44106}
\altaffiltext{\ArtInstChicago}{Liberal Arts Department, School of the Art Institute of Chicago, 112 S Michigan Ave, Chicago, IL, USA 60603}
\altaffiltext{\ThreeSpeedLogic}{Three-Speed Logic, Inc., Vancouver, B.C., V6A 2J8, Canada}
\altaffiltext{\CfA}{Harvard-Smithsonian Center for Astrophysics, 60 Garden Street, Cambridge, MA, USA 02138}
\altaffiltext{\UToronto}{Department of Astronomy \& Astrophysics, University of Toronto, 50 St George St, Toronto, ON, M5S 3H4, Canada}
\altaffiltext{\Maryland}{Department of Astronomy, University of Maryland College Park, MD 20742-2421}
\altaffiltext{\UCLA}{Department of Physics and Astronomy, University of California, Los Angeles, CA, USA 90095}

\begin{abstract}

We present measurements of the $E$-mode polarization angular auto-power spectrum ($EE$) and temperature-$E$-mode cross-power spectrum ($TE$) of the cosmic microwave background (CMB) using 150 GHz data from three seasons of SPTpol observations.
We report the power spectra over the spherical harmonic multipole range $50 < \ell \leq 8000$, and detect nine acoustic peaks in the $EE$ spectrum with high signal-to-noise ratio.
These measurements are the most sensitive to date of the $EE$ and $TE$ power spectra at $\ell > 1050$ and $\ell > 1475$, respectively. 
The observations cover 500\,\sqdeg, a fivefold increase in area compared to previous SPTpol analyses, which increases our sensitivity to the photon diffusion damping tail of the CMB power spectra enabling tighter constraints on \LCDM model extensions.
After masking all sources with unpolarized flux $>50$\,mJy we place a 95\% confidence upper limit on residual polarized point-source power of $D_\ell = \ell(\ell+1)C_\ell/2\pi <0.107\,\muksq$ at $\ell=3000$,
suggesting that the $EE$ damping tail dominates foregrounds to at least $\ell = 4050$ with modest source masking.
We find that the SPTpol dataset is in mild tension with the \LCDM model ($2.1\,\sigma$), and different data splits prefer parameter values that differ at the $\sim 1\,\sigma$ level.
When fitting SPTpol data at $\ell  < 1000$ we find cosmological parameter constraints consistent with those for \planck temperature.
Including SPTpol data at $\ell > 1000$ results in a preference for a higher value of the expansion rate ($\ho = 71.3 \pm 2.1\,\mbox{km}\,s^{-1}\mbox{Mpc}^{-1}$ ) and a lower value for present-day density fluctuations ($\sigma_8 = 0.77 \pm 0.02$).

\end{abstract}

\maketitle

\section{Introduction}
\setcounter{footnote}{0}

Studies of the temperature fluctuations of the cosmic microwave background (CMB) have transformed our understanding of the early universe and how it has evolved over cosmic time.
From the largest angular scales to scales of roughly seven arcminutes, satellite-based measurements over the full sky of the angular power spectrum of CMB temperature anisotropies are now cosmic variance limited \citep{bennett13, hinshaw13, planck15-11}.
Ground-based experiments have measured the temperature power spectrum of small patches of the sky to arcminute scales with high precision \citep{story13, das14, george15}.

The CMB is also polarized at the 10\% level by local radiation quadrupole fluctuations during the epoch of recombination \citep{hu97d}.
CMB polarization is often decomposed into even-parity $E$ modes and odd-parity $B$ modes \citep{seljak97, kamionkowski97b}. 
As with temperature anisotropies, $E$ modes are sourced by both scalar (density) and tensor (gravitational wave) fluctuations, and are therefore partially correlated with CMB temperature resulting in a nonzero $TE$ cross-power spectrum.

Foreground emission at typical CMB frequencies is also partially polarized. 
At low multipole $\ell$ (large angular scales), the main sources of foreground emission are Galactic dust and synchrotron, which are both expected to be polarized at roughly the same fractional level as the CMB \citep[\eg,][]{planck14-30,planck15-25}.
The contamination to $E$-mode measurements from these foregrounds is expected to be at a similar level as the contamination to low-$\ell$ temperature measurements. 
At high $\ell$ (small angular scales), however, $E$-mode measurements are expected to be fractionally less contaminated by foregrounds than temperature measurements, because the typical fractional polarization of high-$\ell$ foregrounds such as radio galaxies, dusty galaxies, and the cosmic infrared background is expected to be much less than 10\% \citep{seiffert07,battye11}.
Recent measurements indicate that the $E$-mode auto-power spectrum is free of significant foregrounds at intermediate scales to at least multipole $\ell = 3600$ with modest masking of extragalactic sources \citep[][hereafter C15]{crites14}.

The relative lack of foreground contamination in the $E$-mode auto-power spectrum and 
temperature-$E$-mode correlation makes multipoles in the so-called ``damping tail" of the CMB, where anisotropy power is damped by photon diffusion during recombination \citep{silk68}, available for more precise cosmological study.
While the \LCDM model is well constrained using multipoles at $\ell < 2000$ \citep{planck15-11},
fitting cosmological models to additional acoustic peaks in the $TE$ and $EE$ damping tails can act as a consistency test for the \LCDM paradigm.
Furthermore, the polarization damping tails are sensitive to additional physics not tested by the six-parameter \LCDM model.
By including the effects of these phenomena in models, one can check for consistency with theoretical expectations or search for signs of tension that could hint at new physics.

In recent years, great effort has gone in to measurements of CMB polarization anisotropies.
Several experiments have now made high-fidelity measurements of $E$ modes and temperature-$E$-mode correlation \citep[\eg,][C15]{planck15-11, bicep2keck15, naess14, polarbear2014b, louis16}.
While measurements of the polarized angular power spectra have yet to reach the sensitivity of those of the temperature power spectrum, constraints on cosmological models from polarization data are so far consistent with the standard \LCDM cosmological model \citep[][C15]{planck15-13,naess14,louis16}.
With sufficient sky coverage and sensitivity, however, the constraining power of CMB polarization measurements are expected to surpass that of temperature measurements \citep{galli14, louis16}.  

In this paper, we report improved measurements of the $E$-mode angular auto-power spectrum ($EE$) and the temperature-$E$-mode cross-power spectrum ($TE$) using three seasons of 150 GHz data taken with the SPTpol instrument \citep{austermann12}.
Measurements of the SPTpol $B$-mode auto-power spectrum are the subject of a separate ongoing analysis.
The data cover 500\,\sqdeg, a fivefold increase in survey area from that used in C15.
We extend the multipole range from $500 < \ell \leq 5000$ to $50 < \ell \le 8000$, which significantly improves constraints on polarized extragalactic point-source power.
We also use the expanded multipole coverage and increased sensitivity in the CMB damping tails to fit several extensions to the \LCDM model.

The paper is structured as follows:  
In Section \ref{sec:instrument}, we briefly describe the SPTpol instrument. 
We discuss the low-level data processing and mapmaking pipeline in Section \ref{sec:obs_and_redux}. 
In Section \ref{sec:ps}, we outline the procedure for estimating unbiased CMB polarization angular power spectra and their covariance from biased measurements. 
We describe a suite of systematics tests in Section \ref{sec:systematics}.
In Section \ref{sec:bandpowers}, we present our measurements of the $EE$ and $TE$ spectra.
We describe our methodology for placing cosmological constraints using these data in Section \ref{sec:cosmology} and report and interpret our results in Section \ref{sec:lcdm}. 
Finally, we state our conclusions and look toward the future in Section \ref{sec:conclusion}.

\section{The SPTpol Instrument}
\label{sec:instrument}

We installed the SPTpol receiver on the South Pole Telescope (SPT) during the austral summer of 2011-2012.
The SPT is a 10 m off-axis Gregorian telescope located at the Amundsen-Scott South Pole station that we designed for dedicated measurements of the CMB \citep{padin08, carlstrom11}.
To complement the polarization-sensitive receiver, we modified the telescope to reduce ground pickup by installing a 1 m guard ring around the 10 m primary and a small ``snout" near prime focus at the top of the receiver cabin in 2012.
In 2013, before the second SPTpol observing season commenced, we also installed larger ``side shields" that reach from either side of the guard ring to the front edge of the telescope.

The SPTpol focal plane is composed of 1536 feedhorn-coupled transition edge sensor (TES) detectors:
360 detectors in 180 polarization-sensitive pixels at 95\,GHz, and 1176 detectors in 588 polarization-sensitive pixels at 150\,GHz.
More details about the design and fabrication of the 95 and 150\,GHz pixels can be found in \cite{sayre12} and \cite{henning12}, respectively.
The detectors are operated in their superconducting transitions at $\sim\,500$\,mK and we use superconducting quantum interference device (SQUID) amplifiers and a digital frequency-domain multiplexing readout system \citep{dobbs12b, dehaan12} to record detector time-ordered data.
In this analysis, we use data from the 150\,GHz detectors.
We include data from 95\,GHz only when defining point sources to mask during data processing.
The full 5 yr, two-frequency SPTpol dataset will be used in future work.

\section{Observations and Data Reduction}
\label{sec:obs_and_redux}

In this section, we describe the observations of the SPTpol $500$\,\sqdeg survey field.
We follow with a description of the data processing pipeline that starts with detector time-ordered data and ends with a set of 125 maps we use to estimate the CMB temperature and polarization power spectra.

\subsection{Observations}
\label{sec:obs}

The SPTpol survey field is a $500$\,\sqdeg patch of sky spanning 4 hr of right ascension, from 22 hr to 2 hr, and 15 degrees of declination, from $-65^\circ$ to $-50^\circ$.
The field also overlaps the survey region of the BICEP/Keck series of experiments \citep[\eg,][] {bicep2keck15}.
We include measurements from three seasons of dedicated CMB observations during which the Sun was below the horizon or far from our observing field: 2013 April 30 --- 2013 November 27, 2014 March 25 --- 2014 December 12, and 2015 March 27 --- 2015 October 26.
Over 9087 hr of dedicated observations, the field was independently mapped 3491 times.
A fourth season of observations on this field ended in 2016 September, but these data are currently under study and are not included in this analysis.

A single observation of the field consists of either 106 or 109 constant-elevation raster scans depending on the observing strategy discussed below, with the telescope first scanning right and then left.
After each right/left scan pair, the telescope makes a step in elevation of either $9.2^\prime$ or $9.0^\prime$ before making another set of paired scans.
This process repeats until the field is completely mapped once, and we define the corresponding set of scans as a single ``observation." 

Over the observation period covered here, we used two strategies to observe the field, ``lead-trail" and ``full-field" observing.  
From the beginning of observations through 2014 May 29, we mapped the field using an azimuthal lead-trail strategy, similar to that described in C15.
In this observing mode, the field is split into two equal halves in right ascension, a ``lead'' half-field and a ``trail'' half-field.
The lead half-field is observed first over a period of 2 hr, followed immediately by a 2 hr trail half-field observation, with the scan speed and elevation steps defined such that the lead and trail observations occur over the same azimuth range.  
For lead-trail observations, we scan the half-fields at a rate of 1.09 degrees per second in azimuth, or 0.59 degrees per second on the sky at the central declination of the field.  
To increase sensitivity to larger scales on the sky, on 2014 May 29 we switched to mapping the field with a full-field strategy, where constant-elevation scans are made across the entire range of right ascension of the field over a 2 hr period.
To reduce noise at low multipoles, corresponding to larger scales on the sky, we increased the scanning speed to 2 degrees per second in azimuth, or 1.1 degrees per second on the sky.
Higher scanning speeds move sky signals of interest to higher temporal frequencies, away from instrumental $1/f$ noise.

In addition to CMB field observations, we also routinely take a series of measurements for calibration and data quality control.  See \citet{schaffer11} and C15 for more details.

\subsection{Time stream Processing}
\label{sec:timestream_filtering}

The raw data are composed of digitized, time-ordered  data, or ``time streams," for each detector in the focal plane.
These time streams are filtered before making maps to remove low-frequency signal from the atmosphere, instrumental $1/f$ noise, and scan-synchronous structure, as well as to reduce high-frequency signals that could alias down into the signal band when binning.

We Fourier-transform the time streams to apply a low-pass filter and to downsample the data by a factor of 2 to reduce computational requirements.
Three harmonics of two spectral lines originating from the pulse tube coolers, needed to cool the instrument to cryogenic temperatures, are notch-filtered at this time.
We calculate the filtered detector power spectral densities (PSDs) to determine inverse-noise-variance weights for mapmaking.

On a scan-by-scan basis, we subtract Legendre polynomial modes from each detector's time stream.
For lead-trail observations, we perform a fifth-order polynomial subtraction, while we use a ninth-order polynomial subtraction on full-field observations since the observations are twice as long in right ascension.
This filtering step performs an effective high-pass filter on the data at $\ell \sim 50$ in the scan direction, which sets the lower multipole bound for this analysis.
(Since the telescope is located at one of the geographic poles and each scan is performed at a constant elevation, time stream filtering only removes modes in the direction of the scan on the sky, \ie, from right ascension.)
Additionally, if during a scan a detector passes within $5^\prime$ of an extragalactic source with unpolarized flux $> 50$\,mJy at either 95 or 150\,GHz, the relevant time stream samples are masked during polynomial filtering.
To remove power from higher multipoles that would alias into the signal band through map pixelization, we also perform a temporal frequency low pass on the time streams that corresponds to $\ell = 11000$ in the telescope scan direction.

\subsection{Cross-talk}
\label{sec:crosstalk}

SPTpol detectors are read out using a digital frequency-domain multiplexing system.
Each detector is part of an $LCR$ resonant circuit with 12 resonant channels, which we refer to as a ``resonant comb."
The finite width and spacing of these resonances, as well as inductive coupling between readout elements,  cause cross-talk between detectors read out on the same resonant comb, and to a lesser degree different resonant combs: when a detector scans over a bright source, another detector with a neighboring resonant frequency sees a negative scaled copy of the source, typically at the $1\%$ level.
Averaging over many detectors, the effect of cross-talk is a multipole-dependent multiplicative correction to the measured power spectra.
In C15, we estimated this cross-talk correction from simulations and applied the correction to the power spectra at the end of the analysis.

In this analysis, we choose to correct cross-talk at the detector time stream level.
We measure the detector-detector cross-talk matrix $\mathbf{X}$ using frequent calibration measurements of the Galactic \textsc{HII} region \textsc{RCW38}.
For each full observation we make single-detector maps and compare them to detector-specific templates we generate by offsetting and scaling a focal-plane-averaged template of \textsc{RCW38}.
We fit each single-detector map as a linear combination of templates from possibly cross-talking detectors.
The coefficients of these fits populate our cross-talk matrix.
We find excellent temporal stability in the cross-talk matrix across an observing season and use its average when cleaning time streams.

As the first step in time stream processing for all observations, we reconstruct the cross-talk-corrected time streams $\vec{d}$,
\begin{equation}
\vec{d} = \mathbf{X}^{-1} \hat{\vec{d}}.
\end{equation}
Before correcting for cross-talk, we observe high signal-to-noise ratio copies of \textsc{RCW38} in maps.
After the above correction, we see no evidence of negative-cross-talking artifacts, demonstrating at least an order of magnitude suppression of cross-talk.
As this is a $O(1\%)$ effect and we suppress it by at least a factor of 10, we neglect any uncertainty on the correction and proceed in the processing steps described in this section assuming that the time streams are clean of cross-talk.

\subsection{Data Quality Cuts}
\label{sec:datacuts}
We cut data based on both the performance of detectors and the overall observation quality, which we discuss below.

\subsubsection{Detector Cuts}
We make the same series of detector data cuts as those made in C15.  
We refer the reader to that work for more details and summarize here.  
For each scan of each detector, time streams are flagged and removed if a ``glitch" is detected.
Glitches could be sudden spikes caused by cosmic-ray hits or discrete DC jumps attributed to changes in SQUID bias point.
A detector's time stream of a given scan is also removed from the analysis if the scan's rms noise is $5\,\sigma$ above or below the median of all detectors during an observation.
We cut 8.9\% of all scans in this manner.

A second round of cuts is performed at the detector level, removing all data from a single detector for an entire observation.
Detectors with anomalously high or low noise in the 1--3\,Hz frequency band are flagged for removal.
Additionally, any detectors with low signal-to-noise ratio in either of two regular calibration observations -- elevation dips ($100\,\sigma$ minimum) and response to an internal source ($10\,\sigma$ minimum) -- are cut.
Finally, we remove data from the polarized-pixel partner of any detector that is itself cut.
On average, 864 of the 1176 150 GHz detectors survive detector cuts, which includes unavoidable cuts from fabrication and readout yield.

\subsubsection{Observation Cuts}
\label{sec:lowl_noise}
We apply an additional round of data cuts in this analysis to single-observation maps in order to reduce polarized noise at large angular scales in the map power spectra.
Rather than de-weight high-noise maps, we conservatively choose to cut them to avoid possible systematic contamination.
We construct a statistic $\xi$ that quantifies excess power at $\ell < 300$ in Stokes $Q$ and $U$ maps from a given observation,
\begin{equation}
\xi = \log_{10}{\left(\frac{\left< N_\ell^{XX}\right>_{\ell < 300}}{\mathrm{median}\left(N_\ell^{XX}\right)}\right)},
\end{equation}
where $XX \in \{QQ,UU\}$ and $N_\ell^{XX}$ are auto-power noise spectra, created by differencing $Q$ and $U$ maps made from left-going scans and from right-going scans.
By cutting maps with anomalously high low-$\ell$ power, we reduce low-$\ell$ noise at the expense of slightly increasing the overall noise level.
We choose to cut an observation if $\xi > 1.0$ for either the $Q$ or $U$ map.
Out of 4127 total observations (lead half-field maps, trail half-field maps, and full-field maps), 501 are removed from the dataset, which increases noise at $\ell > 1000$ by $\sim10\%$ but reduces noise at $\ell < 100$ by $\sim 60\%.$

\subsection{Pre-map Calibration}
\label{sec:cal}

We apply a series of calibrations to the data to transform from raw detector time stream units to thermodynamic temperature units $\mu\mbox{K}_{\mbox{\tiny{CMB}}}$, indicating the equivalent intensity fluctuations for a 2.73\,K blackbody.
We also apply in-pixel calibration between detectors within a polarization-sensitive pair, as well as a polarization calibration to define detector polarization angles and efficiency.
A temperature calibration step is also applied to maps, but we save its discussion for Section \ref{sec:abscal}.

\subsubsection{Relative Calibration}
\label{sec:relcal}

We first convert detector time streams to measured on-sky power using the recorded detector voltages while in operation.
We then calibrate the detector response amplitudes, or gains, to $\mu\mbox{K}_{\mbox{\tiny{CMB}}}$ using measurements of an internal calibrator and RCW38 following the procedure described in detail in \cite{schaffer11}.  We refer the reader to that work for more details.

\subsubsection{In-pixel Gain Calibration}

We also perform a relative gain correction between two detectors in the same polarization-sensitive pixel.
This step is meant to reduce noise in the differenced detector time stream, particularly at low temporal frequencies, which in turn decreases polarization noise at large angular scales on the sky.
For each right/left scan pair, we calculate the Fourier-transform amplitudes between 0.1 and $0.3\,$Hz of the right-going minus left-going time streams.
For two transformed detector time streams in a polarization-sensitive pair, $\widetilde{X}$ and $\widetilde{Y}$, we calculate the relative pixel gain factors $a$ and $b$ that minimize the differenced power while keeping the total power the same:
\begin{equation}
\min{(a\widetilde{X} - b\widetilde{Y})^2}, \quad \mathrm{where}\, (a\widetilde{X} + b\widetilde{Y})^2 = (\widetilde{X}+\widetilde{Y})^2.
\end{equation}
These per-scan gain factors are averaged across an entire observation and then applied to the detector time streams.

\subsubsection{Polarization Calibration\label{sec:polcal}}

To reconstruct maps of Stokes $Q$ and $U$ polarization, we require accurate measurements of each detector's polarization angle $\theta$ and polarization efficiency $\eta_\mathrm{p}$.
Let us assume that the time stream $d$ of a detector can be written as
\begin{equation}
d = G(I + \eta_Q Q + \eta_U U),
\end{equation}
where $G$ is an overall normalization or gain, $I$, $Q$, and $U$ are the Stokes polarization parameters, and $\eta_Q$ and $\eta_U$ are the fractional responses of a detector to Stokes $Q$ and $U$, respectively.
Then we define the polarization angle $\theta$ and efficiency $\eta_\mathrm{p}$ for that detector as
\begin{equation}
\theta = \frac{1}{2}\arctan{\left(\frac{\eta_U}{\eta_Q}\right)}, \qquad \eta_\mathrm{p} = \frac{2\sqrt{\eta_Q^2+\eta_U^2}}{1+\sqrt{\eta_Q^2+\eta_U^2}}.
\end{equation}
We take measurements of an external polarized calibrator to fit for $\eta_Q$ and $\eta_U$ and determine $\theta$ and $\eta_\mathrm{p}$ for each detector.  See C15 and \cite{keisler15} for more details.
The median statistical error in detector angle per detector is $0.5^{\circ}$, and across all 150\,GHz detectors $\eta_\mathrm{p}$ averages $97\%\pm2\%$.
Additionally, while fitting cosmological models to the data, we allow the average polarization efficiency to vary.  See Section \ref{sec:nuisance} for more details.

\subsection {Maps}
We combine detector time streams into maps with square $0.5^\prime$ pixels using the oblique Lambert azimuthal equal-area projection.
Forming maps from detector time streams follows the same procedures discussed in C15 and \cite{keisler15}, which are similar to those described in \cite{couchot99} and \cite{jones07}.
We review the process here.

First, inverse-variance detector time stream weights $w$ are constructed from detector polarization efficiency $\eta_\mathrm{p}$ and the rms noise amplitude $n$ in the 1-3\,Hz range during an observation: $w \propto (\eta_\mathrm{p}/n)^2$.
If weights were calculated for each detector independently, the in-pixel calibration would be nullified.
For this reason, we assign the same weight, calculated as one over the average noise power in the 1-3\,Hz band, $w \propto 1/<(n_X^2,n_Y^2)>$, to both detectors within a pixel.  

We then combine detector time streams using telescope pointing, as well as detector polarization angles $\theta$ and weights $w$.
For the $i^\textrm{th}$ detector with time stream $d_i$ and polarization angle $\theta_i$, the contribution to pixel $\alpha$ of the weighted $T$, $Q$, and $U$ maps is
\begin{eqnarray}
T^W_{i \alpha} &=& \sum_t A_{t i \alpha} \ w_i  \ d_{ti}  \\
\nonumber Q^W_{i \alpha} &=& \sum_t A_{t i \alpha} \ w_i \ d_{ti} \ \cos{2 \theta_i} \\
\nonumber U^W_{i \alpha} &=& \sum_t A_{t i \alpha} \ w_i \ d_{ti} \ \sin{2 \theta_i},
\end{eqnarray}
where $t$ indexes time stream samples and $A_{t i \alpha}$ is a matrix that encodes during which time samples $t$ detector $i$ was pointing at pixel $\alpha$.

We also construct a $3 \times 3$ weight matrix $W_{\alpha}$ for each pixel by summing over the weight contributions from each detector.
The weight matrices encode the $T$, $Q$, and $U$ weights, as well as correlations between the three measurements.
See \cite{keisler15} for more details.
We obtain an estimate of the unweighted maps by inverting the weight matrix for each pixel:
\begin{equation}
\{T,Q,U\}_\alpha = W^{-1}_\alpha \{T^W_{\alpha},Q^W_{\alpha},U^W_{\alpha}\}.
\end{equation}

After unweighting the maps, we redefine $Q$ and $U$ by rotating polarization angles by $\psi(\alpha)$ set by the chosen map projection,
\begin{equation}
(Q'+iU') = e^{-i \psi(\alpha)} (Q+iU),
\end{equation}
where the polarization angles for the primed maps are defined on flat skies and the angles for unprimed maps are defined on curved skies.
This procedure ensures that the definition of the Stokes parameters is consistent across a map regardless of its projection.

Lastly, we combine the Stokes $Q$ and $U$ maps in Fourier space to generate Fourier maps of $E$-mode polarization \citep{zaldarriaga01},
\begin{equation}
E_{\pmb{\ell}} = Q_{\pmb{\ell}}\cos{2\phi_{\pmb{\ell}}} + U_{\pmb{\ell}}\sin{2\phi_{\pmb{\ell}}},
\label{eq:QUtoE}
\end{equation}
where $\pmb{\ell} = (\ell_x,\ell_y)$, $\ell = |\pmb{\ell}|$, and $\phi_{\pmb{\ell}}= \arctan{\ell_y/\ell_x}$.
This equation assumes the flat-sky approximation, in which we replace spherical harmonic transforms with Fourier transforms by assuming $\ell = 2\pi|\mathbf{u}|$, and where $\mathbf{u}$ is the Fourier conjugate of small angles on the sky.

\subsubsection{Map Bundles} 
\label{sec:bundles}

Individual maps of the field, from either one lead-trail pair or a single full-field observation, have nonuniform coverage due to elevation steps and cut time stream data.
The resulting map pixel weight matrix $W_\alpha$ can be ill behaved during inversion and nonuniform across observations.
To regularize the weight, we choose to combine the dataset into 125 map ``bundles."
Lead-trail pairs and full-field observations are each grouped separately into 125 bundles.
The lead-trail bundle set is generated by ordering the observations chronologically and adding single maps until the combined weight is 1/125th of the total lead-trail observation weight, at which point the next bundle is started.
The full-field bundle set is calculated similarly.
Using this procedure, we find only a $2.3\%$ rms variation in bundle weights.
The lead-trail and full-field bundles are then matched sequentially to form 125 total bundles, each with contributions from lead-trail and full-field observations that span the 3 yr observing period.  

\begin{figure*}
\begin{center}
\includegraphics[width=1.0\textwidth]{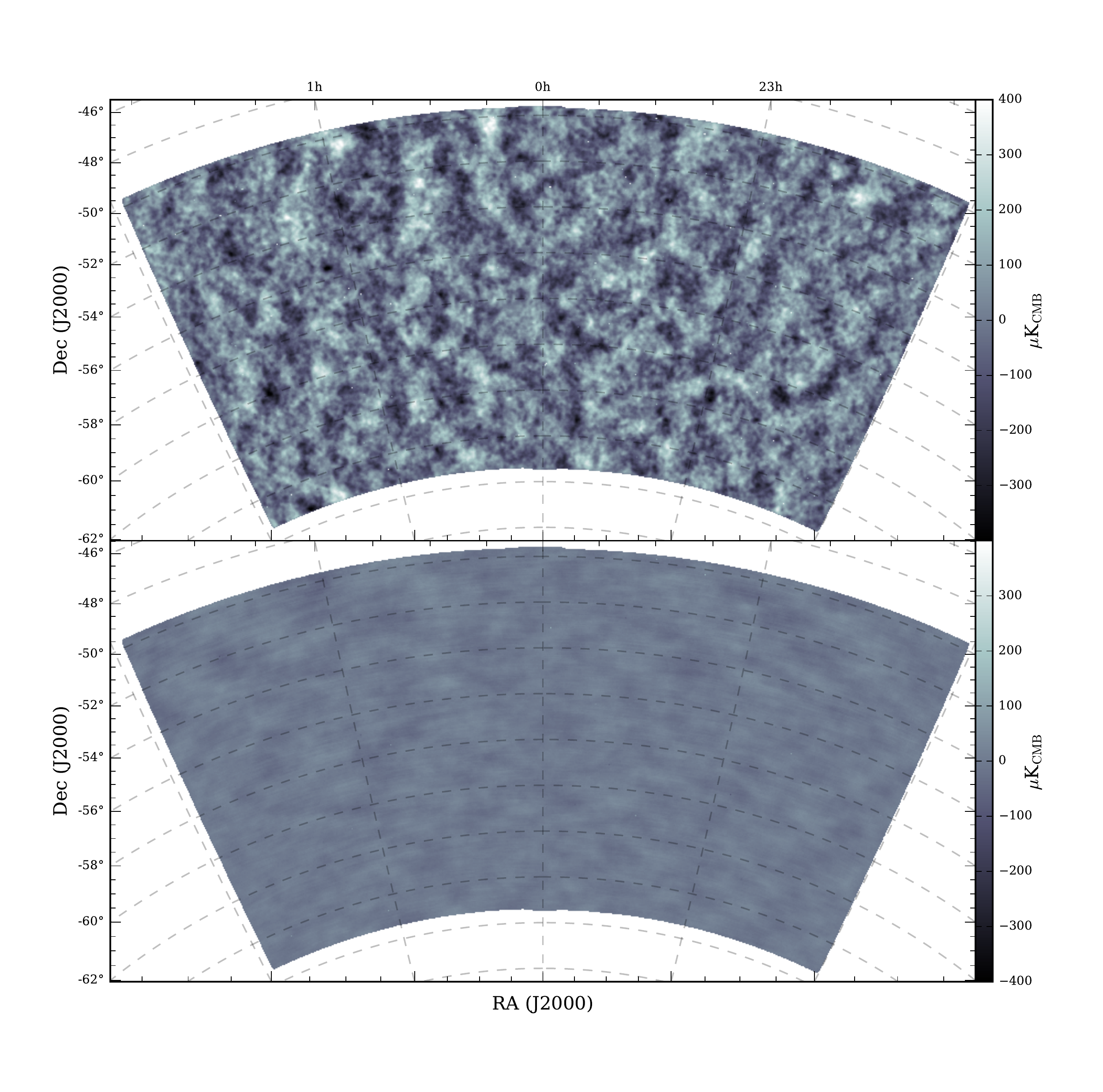}
\caption{SPTpol 500\,\sqdeg $T$ signal (top) and noise (bottom) maps.  
The noise maps are obtained by subtracting data of the first half from data of the second half of the set of bundles and dividing by 2 to reflect the effective noise level of the entire dataset.}
\label{fig:coadd_t}
\end{center}
\end{figure*}

\begin{figure*}
\begin{center}
\includegraphics[width=1.0\textwidth]{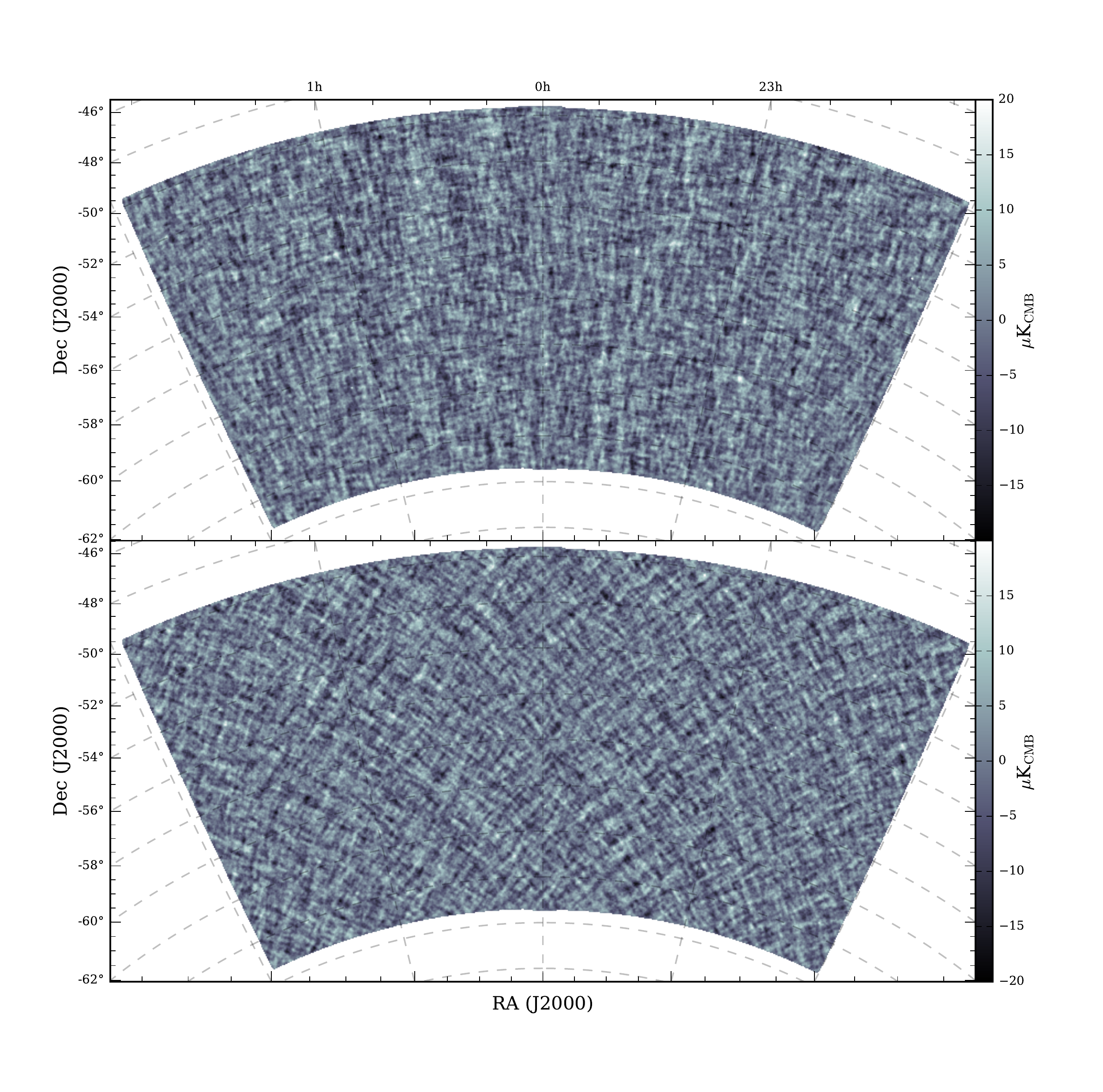}
\caption{Top: map of Stokes $Q$.  Bottom: map of Stokes $U$.  
The clear striping along lines of constant right ascension and declination in $Q$ and $\pm45^{\circ}$ striping in $U$ are indicative of high signal-to-noise ratio $E$ modes.
The maps have been smoothed by a $4^\prime$ FWHM Gaussian.}
\label{fig:coadd_qu}
\end{center}
\end{figure*}

\begin{figure*}
\begin{center}
\includegraphics[width=1.0\textwidth]{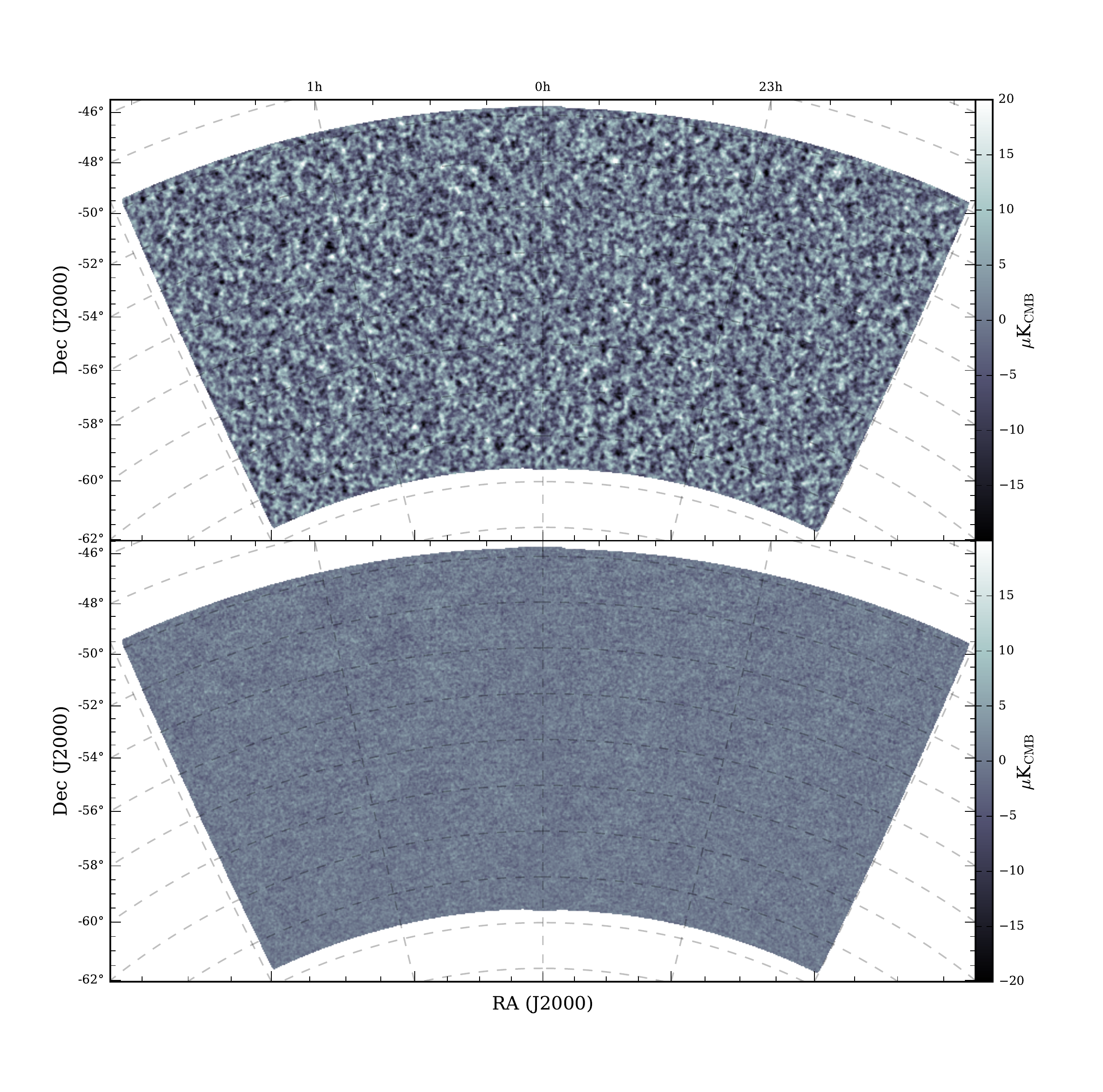}
\vspace{-0.25in}
\caption{SPTpol 500\,\sqdeg $E$-mode signal (top) and noise (bottom) maps.  
The Fourier transforms of the $Q$ and $U$ maps shown in Figure \ref{fig:coadd_qu} are combined to form $E$ modes, which are inverse Fourier transformed to generate an $E$-mode map.
Both maps have been smoothed by a $4^\prime$ FWHM Gaussian. }
\label{fig:coadd_e}
\end{center}
\end{figure*}

The resulting 125 bundles are the basic input to the power spectrum analysis described in Section \ref{sec:ps}.
However, for illustration purposes we can combine them to form a single map that contains all cuts.
Figures \ref{fig:coadd_t}-\ref{fig:coadd_e} show the $T$, $Q$ and $U$, and $E$-mode maps for the 500\,deg$^2$ survey field, respectively.
We have smoothed the polarization maps by a $4^\prime$ FWHM Gaussian.
Note that in Figure \ref{fig:coadd_qu} we have not accounted for polarization rotation caused by the map projection, while we have in Figure \ref{fig:coadd_e}.
The $Q$ map shows a clear stripe pattern along lines of constant right ascension and declination, while the $U$ map shows similar striping $\pm45^\circ$ from the coordinate lines.
These patterns are indicative of $E$-mode polarization, which we measure with high signal-to-noise ratio.
To demonstrate this, we also plot noise maps for temperature and $E$ modes, which we generate by splitting the bundles into two sets chronologically, subtracting them, and dividing by 2 to show the effective noise level in the combined dataset.

\begin{figure}
\begin{center}
\includegraphics[width=0.475\textwidth]{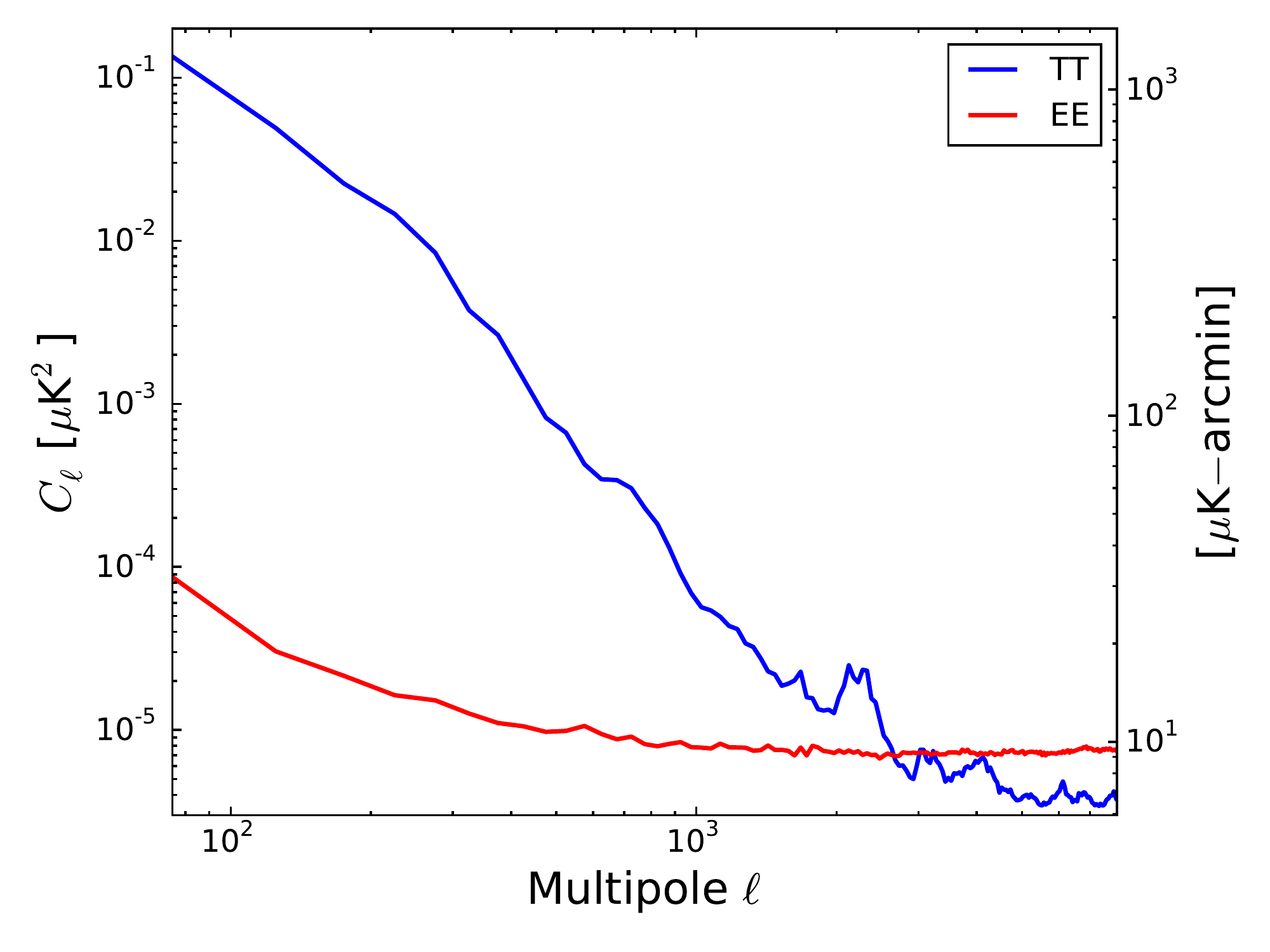}
\caption{SPTpol 500\,\sqdeg $TT$ (blue) and $EE$ (red) noise spectra.  
The left-hand labels give the noise in units of \muksq while the right-hand labels give the equivalent map depth in \mukarcmin.}
\label{fig:noise_spectra}
\end{center}
\end{figure}

After data and map cuts, the SPTpol 500\,\sqdeg field reaches an average polarization map depth of $9.4\,\microK$-arcmin in the multipole range $1000 < \ell < 3000$.
This is similar in depth to the 100\,\sqdeg polarization analysis of C15 but covers five times more sky, which decreases power spectrum uncertainties by more than a factor of 2.
We plot the temperature and $E$-mode polarization noise spectra in Figure \ref{fig:noise_spectra} after correcting for filtering in the analysis pipeline (see Section \ref{sec:transfer_function}) and calibration (see Sections \ref{sec:abscal} and \ref{sec:nuisance}).
The polarization noise is white at $\ell > 1000$ but rises at larger scales, by an order of magnitude at $\ell = 50$.
At these noise levels our $TE$ and $EE$ polarization spectra are dominated by sample variance at $\ell \lesssim 1000$, so we do not pursue further noise improvements in this analysis.

We use minimal time stream filtering to maintain sensitivity at large scales in the polarization power spectra.
This analysis choice leads to the temperature noise spectrum being dominated by atmospheric noise at large scales.
Aliasing of the atmospheric noise by our scan strategy also contaminates the temperature noise at higher $\ell$.
An analysis designed to measure the temperature power spectrum with low noise requires more aggressive time stream filtering, which would restrict the useful multipole range of the analysis.
Here we choose to focus on the $TE$ and $EE$ power spectra and recover modes at larger scales at the cost of higher temperature noise.

\section{Power Spectrum}
\label{sec:ps}

We now discuss how we calculate $TE$ and $EE$ angular power spectra from the map bundles constructed in the previous section.
First, we describe a bundle-bundle cross-spectrum formulation we use to avoid noise bias.
Second, we outline a pseudo-$C_\ell$ procedure we use to clean the spectra of bias introduced by our observation and analysis procedures.
Third, we outline the calculation of each source of bias in the pseudo-$C_\ell$ framework.
Fourth, we discuss some additional cleaning procedures we apply to the unbiased spectra.
Finally, we describe the calculation of the bandpower covariance matrix.

\subsection{Cross-spectra}

As in C15 and other SPT analyses, we choose to calculate power spectra with a cross-spectrum approach to avoid noise bias \citep{polenta05, tristram05} and follow the framework laid out in \cite{lueker10}.
The map bundles $m^X_i$, where $X\in\{T,E\}$ and $i$ indexes bundle number, contain true sky signal and noise.
Since each bundle has an independent realization of noise, calculating cross-spectra between pairs of bundles $(m_i, m_j)$, where $i \ne j$ eliminates the noise bias one would incur calculating the auto-spectrum of a map containing the entire dataset. While each bundle has higher noise than a single combined map, and thus each bundle-pair cross-spectrum is noisier than the combined-map auto-spectrum, for a sufficient number of bundles the average over all possible bundle-pair cross-spectra approaches the sensitivity of the combined-map auto-spectrum. 
Additionally, the noise penalty from ignoring the bundle auto-spectra is negligible, as they represent a small fraction of the total available bundle-pair spectra.

When calculating cross-spectra $\widehat{C}^{XY}_\ell$, we use the flat-sky approximation.
Here the overhat denotes a biased, or ``pseudo," quantity.
We multiply the bundles by an apodization and point-source mask and zero-pad the maps before calculating their Fourier transforms $\widehat{m}^X_{\pmb{\ell},i}$.

Next, we calculate the average cross-spectra between two bundles $i$ and $j$ within $\ell$-bins b,
\begin{equation}
\widehat{D}^{XY}_b = \left< \frac{\ell(\ell+1)}{2\pi}\mathrm{Re}\left[\widehat{m}^X_{\pmb{\ell},i}\widehat{m}^{Y*}_{\pmb{\ell},j}\right]\right>_{\ell \in b}.
\end{equation}
With 125 bundles there are 7750 independent cross-spectra with $i \ne j$, which we average to obtain one-dimensional estimates of the binned pseudo-power spectrum $\widehat{D}^{XY}_b$, which we refer to as ``bandpowers."

\subsection{Pseudo-spectra to Spectra}
\label{sec:pseudo_spectra}

Finite instrument resolution, time stream filtering, map resolution, and masking all make the measured spectra $\widehat{D}^{XY}_{b}$ biased estimates of the true $XY$ spectra.
As in previous SPT analyses, we follow the pseudo-$C_\ell$ MASTER method of \cite{hivon02} to estimate the unbiased and binned power spectra.
We relate $\widehat{D}^{XY}_{b}$ to the unbiased bandpower estimates $D^{XY}_{b}$ by
\begin{equation}
\widehat{D}^{XY}_{b} = K_{bb^{\prime}}D^{XY}_{b^{\prime}}.\label{eq:biased_D}
\end{equation}
The kernel $K_{bb^{\prime}}$ encodes a series of operations performed on the true spectra during observations and analysis.
It can be expanded into constituent operations as 
\begin{align}
  K_{bb^{\prime}} &= P_{b\ell} \mathcal{M}_{\ell\ell^{\prime}}Q_{\ell^{\prime}b^{\prime}}\nonumber \\
 &= P_{b\ell}\left(M_{\ell\ell^{\prime}}[\pmb{\mathrm{W}}]F_{\ell^{\prime}}B^2_{\ell^{\prime}}\right)_{\ell\ell^{\prime}}Q_{\ell^{\prime}b^{\prime}},\label{eq:K}
 \end{align}
where we perform element-wise multiplication over $\ell^\prime$ when calculating $\mathcal{M}_{\ell\ell^{\prime}}$ and follow Einstein summation notation otherwise.
Here $M_{\ell\ell^{\prime}}[\pmb{\mathrm{W}}]$ accounts for coupling between Fourier modes due to the sky mask $\pmb{\mathrm{W}}$, $F_\ell$ is the filter transfer function that accounts for time stream processing and map pixelization, and $B_\ell$ is the Fourier transform of the SPTpol instrument $\delta$-function response or ``beam."
As defined in \cite{hivon02}, $P_{b\ell}$ is the binning operator, which takes independent multipoles $\ell$ and bins them into bandpowers $b$, while $Q_{\ell b}$ is its reciprocal operator.
We find the unbiased estimates of the true spectra by inverting $K_{bb^\prime}$,
\begin{equation}
D^{XY}_{b} = K^{-1}_{bb^{\prime}}\widehat{D}^{XY}_{b^{\prime}}.
\end{equation}
In Sections \ref{sec:mode_coupling}, \ref{sec:transfer_function}, and \ref{sec:beams} we describe the calculation of the mode-coupling matrix $M_{\ell\ell^{\prime}}[\pmb{\mathrm{W}}]$, the filtering transfer function $F_\ell$, and the beam $B_\ell$, respectively.

\subsection{Map Apodization and Mode Coupling}
\label{sec:mode_coupling}

To reduce ringing in Fourier space from sharp edges at the survey boundary, we apply an apodization mask $\pmb{\mathrm{W}}$ before Fourier-transforming the bundles.
We also use this step to mask bright point sources in the survey region. 
For each bundle, we find the region where the weight is greater than 30\% of the median weight.
The intersection of these areas for all bundles is then apodized with a $15^\prime$ cosine taper to define the apodization mask.
We also mask all point sources with unpolarized flux $>50\,\mbox{mJy}$ at 95 or 150\,GHz with a $10^\prime$ disk and $10^\prime$ cosine taper.
The effective area of the mask $\pmb{\mathrm{W}}$ is 490.2 \sqdeg.

Masking the full sky couples otherwise-independent Fourier modes.
We analytically calculate the mode-coupling matrix $M_{\ell\ell^{\prime}}[\pmb{\mathrm{W}}]$ following the description in Appendix A of \cite{hivon02} and the Appendix of C15.
Mode-coupling matrices are calculated independently for the $TT$, $TE$, and $EE$ spectra.
Mode coupling can also leak $B$ to $E$ modes; however, we ignore this term under the assumption that power in $EE \gg BB$. 
Finally, to conserve Fourier-space power when applying the apodization and point-source mask, the mode-coupling matrices are normalized by the second moment of the mask,
\begin{equation}
\sum_{\ell^{\prime}} M_{\ell\ell^{\prime}}[\pmb{\mathrm{W}}] = \frac{1}{\Omega} \int d^2 r \pmb{\mathrm{W}}^2  \equiv w_2,
\end{equation}
where $\Omega$ is the area of a map in steradians.

We test the fidelity of our analytic mode-coupling matrix calculation using full-sky simulations. 
In particular, we want to test the effect of the flat-sky approximation on large angular scales.
These simulations will test for any errors in the analytic calculation, however.
We generate a HEALPix realization \citep{gorski05} of the full sky from spectra limited to a small range of input multipole $\Delta\ell=5$.
We then multiply the sky realization by our apodization mask $\pmb{\mathrm{W}}$ before calculating the power spectrum using spherical harmonic transforms.
The ratio of the input spectrum to the output spectrum reveals to what multipoles $\ell$ the power from the limited $\Delta\ell=5$ input range is mixed by masking the map.
This process is repeated for each $\Delta\ell=5$ input range from $0 < \ell < 500$ to construct one realization of the mode-coupling matrix.
We make 400 realizations of the mode-coupling matrix in this way and compare their average to the result of the flat-sky analytical calculation at $0 < \ell < 500$.
We find that the two calculations are in good agreement, so we proceed in using only the flat-sky analytical solution for the mode coupling when unbiasing bandpowers.

\subsection{Transfer Function}
\label{sec:transfer_function}

Our mapmaking procedure is a lossy process that does not recover all modes of the true sky.
We lose information during time stream filtering, as well as when we bin data into map pixels.
In order to obtain an unbiased estimate of the on-sky power spectrum, we must determine what the loss is and account for it.
In the MASTER formalism, this loss is quantified by the filtering transfer function $F_\ell^{X}$, where $X\in\{TT,EE,TE\}$.
We calculate $F_\ell^{X}$ by creating 300 simulated skies, which we process into spectra, replicating each step in the analysis pipeline.
We generate the sky realizations using the best-fit theory of the \textsc{plikHM\_TT\_lowTEB\_lensing} \planck dataset to define the CMB spectra \citep{planck15-15}.
Gaussian realizations of foreground power are also added to the simulated skies.
We define the foreground parameters and summarize their priors in Section \ref{sec:foregrounds}.
After convolving the skies with the SPTpol beam, we ``mock-observe" each realization, generating noiseless time streams by ``scanning" the skies using the recorded pointing information for each of the 3626 observations that pass data quality and map cuts.
The resulting time streams are then processed identically to the real data to generate 300 sets of simulated map bundles.

Following the prescription of \cite{hivon02}, we calculate $F_\ell^{X}$ from the mock-observed spectra using an iterative approach.
The first iteration is the ratio of the mean simulated spectra over the input theoretical spectrum,
\begin{equation}
F^X_{\ell,1} = \frac{\left< \widehat{C}^X_{\ell,\mathrm{sim}}\right>}{w_2 C^X_{\ell,\mathrm{th}} B^2_\ell}.
\end{equation}
We then iteratively remove mode coupling:
\begin{equation}
F^X_{\ell,i+1} = F^X_{\ell,i} + \frac{M_{\ell\ell^{\prime}} F^X_{\ell^{\prime},i} C_{\ell^{\prime},th} B^2_{\ell^{\prime}} }{w_2 C_{\ell,\mathrm{th}} B^2_\ell}.
\end{equation}
We find that the calculation converges after two iterations.

\begin{figure*}[t!]
\begin{center}
\includegraphics[width=1.0\textwidth]{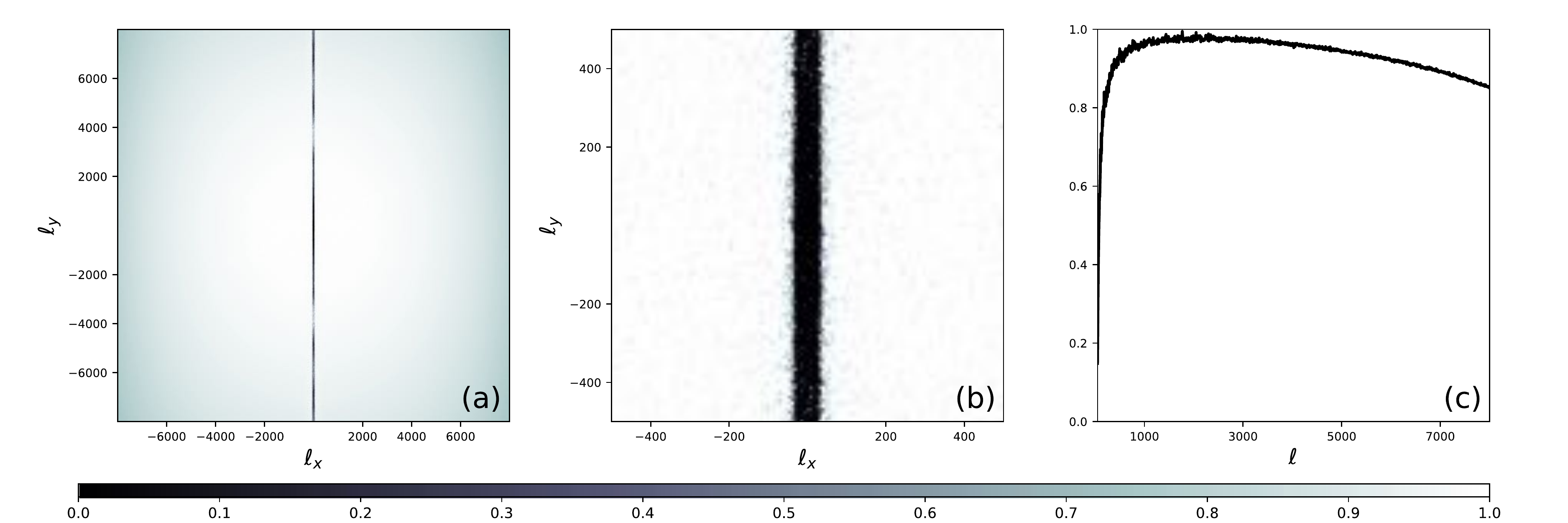}
\end{center}
\caption{
(a) 2D Fourier-space filtering transfer function, which we calculate from simulated maps with the Sanson-Flamsteed projection.
Note that we have not corrected the 2D transfer function for mode coupling.
(b) Zoom-in of the 2D transfer function at low $\ell_x$ and $\ell_y$.
(c) Geometric mean of the $TT$ and $EE$ 1D filtering transfer functions corrected for mode coupling, which we calculate from maps using the oblique Lambert azimuthal equal-area projection.}
\label{fig:tf}
\end{figure*}

We plot the 1D and 2D Fourier-space transfer functions in Figure \ref{fig:tf}.
Since time stream filtering only removes modes along the scan direction in a map, we calculate the 2D transfer function from mock-observed simulated maps in the Sanson-Flamsteed projection.
In this projection, the Fourier conjugate of the scan direction is purely $\ell_x$.
We see in Figures \ref{fig:tf} (a) and (b) that filtering predominantly removes modes at $\ell_x < 50$.
Additionally, the low-pass filter and map pixelization remove some information at higher $\ell_x$.
The same information can be read off from the geometric mean of the azimuthally averaged 1D $TT$ and $EE$ transfer functions in Figure \ref{fig:tf} (c), now properly calculated from simulated maps in the oblique Lambert azimuthal equal-area projection and corrected for mode coupling.
Finally, we note that while our filtering is nonisotropic, we are ultimately unbiasing azimuthally averaged 1D power spectra, and the 1D transfer function captures the mean loss of modes in a given azimuthal bin.

To avoid the numerical complications introduced by zero-crossings in the $TE$ spectrum, we set $F_\ell^{TE}$ to the geometric mean of the $TT$ and $EE$ transfer functions.
Unlike a transfer function constructed to recover a $TE$ spectrum with zero crossings at specific multipoles, this approximation is applicable to any cosmology.
We find that this approximation introduces a bias to our constraints on the angular scale of the sound horizon $\thetaMC$ as discussed in Section \ref{sec:consistency}, although it is small compared to our parameter errors.

\subsection{Beam Function and Map Calibration}
\label{sec:beams}
To properly calibrate the measured angular power spectra, we must understand the optical response of the system. 
We need to know both the differential response as a function of angle from boresight, otherwise known as the beam, and the absolute response, or the absolute calibration.
In this section, we describe how we measure the beam using Venus.
We cross-check the beam  on small angular scales by fitting radio sources in the field and on large scales by comparing to \textit{Planck} data. 
In the process, we also determine a map calibration factor that matches SPTpol temperatures to that of \textit{Planck} in the SPTpol survey region.

\begin{figure*}[t!]
\begin{center}
\includegraphics[width=1.0\textwidth]{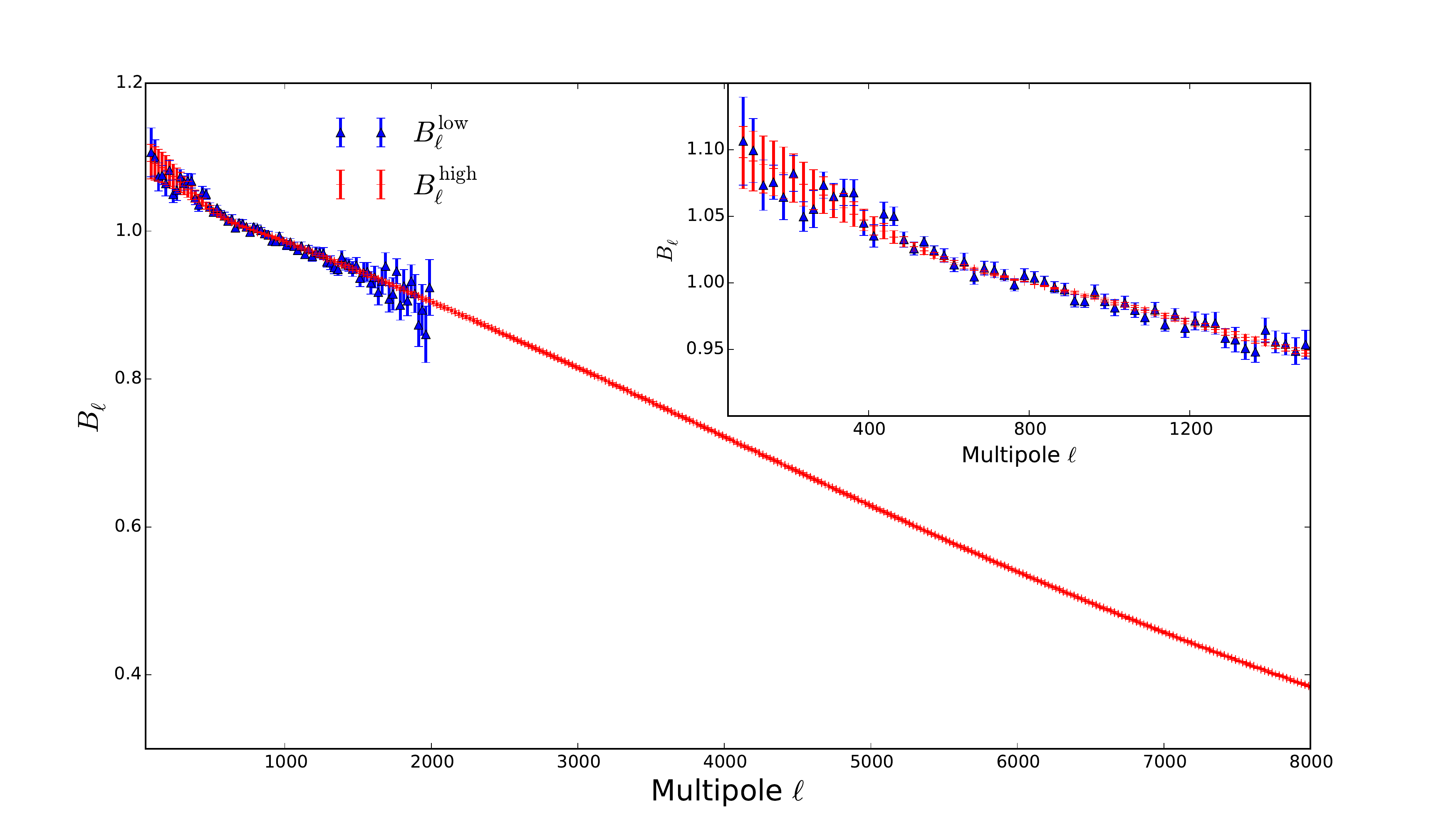}
\end{center}
\caption{
The SPTpol low-$\ell$ and high-$\ell$ beams, calculated using a \planck-SPTpol $TT$ cross-spectrum and SPTpol observations of Venus, respectively.
The inset highlights the agreement between the beams at low $\ell$.
We use the Venus-derived beam when unbiasing bandpowers over the entire multipole range of this analysis.
}
\label{fig:beam}
\end{figure*}

\subsubsection{Beam Measurement from Venus}
\label{sec:highell_beam}

We estimate the beam from observations of Venus. 
After upgrading the focal plane in late 2012, we made seven observations of Venus that pass data quality cuts in 2013 January. 
We measure the beam $B_\ell$ by averaging the 21 independent cross-spectra of these seven observations. 
We estimate the beam covariance from the noise variance of these 21 independent cross-spectra. 
While the noise variance does not include all sources of statistical uncertainty in the beam, it is the dominant source.
Furthermore, we find in Section \ref{sec:lcdm_sptpol} that increasing the beam covariance defined in this way by a factor of 100 has negligible impact on cosmological constraints.

The beam in the CMB field will be broadened by the rms pointing error or jitter but not have the broadening caused by the finite size of Venus (the angular diameter of Venus at the time was $\sim 11^{\prime\prime}$). 
To estimate the net result of these two competing effects, we fit a Gaussian to a combined map of Venus and to maps of the brightest point source in the SPTpol survey field made with data from 2013, 2014, and 2015 separately, as well as all years combined. 
We find that the broadening owing to pointing jitter dominates. 
The quadrature difference in width between Venus and the point source is $9.9^{\prime\prime}$, $14.0^{\prime\prime}$, and $23.4^{\prime\prime}$ for 2013, 2014, and 2015 observations, respectively, and $13.8^{\prime\prime}$ for the complete dataset. 
We convolve the Venus beam with a 32.5" FWHM Gaussian to capture the effective pointing jitter during the observation period of this analysis, and we use this convolved beam to unbias pseudo-spectra.

After the bulk of this work was complete, we discovered that a data cut threshold was mistakenly set too loosely when calculating pointing solutions, which caused poor-performing detectors to degrade the overall pointing.
The beam for the full three-season dataset is nevertheless well fit by a $1.22^\prime$ FWHM Gaussian, compared to a $1.18^\prime$ FWHM Gaussian in C15.
The increased beam width has a minimal impact on the sensitivity of the data, and only on small scales.
We have since corrected the pointing error, and future analyses will use the updated pointing solutions.

\subsubsection{Absolute Temperature Calibration from Planck}
\label{sec:abscal}
We get an absolute temperature calibration by comparing the SPTpol 150\,GHz maps with the 143\,GHz \planck{} maps over the angular multipole range $600 < \ell < 1000$.
Specifically,  we calculate the ratio of the SPTpol 150\,GHz auto-spectrum to the cross-spectrum of SPTpol with the \planck 143\,GHz temperature map. 
This ratio can be expressed as
\begin{equation}
\label{eq:beam_lowell}
\frac{\left<\mathrm{Re}\left[\tilde{m}_i^{T_\mathrm{S}} \tilde{m}_j^{*T_\mathrm{S}}\right]\right>_{i \ne j}}{\left<\mathrm{Re}\left[\tilde{m}^{T_\mathrm{P}} \tilde{m}_i^{*T_\mathrm{S}}\right]\right>_i} = \frac{C_\ell^{TT} M_{\ell\ell^\prime}F_{\ell^\prime}^\mathrm{S}(\epsilon_T B_{\ell^\prime}^\mathrm{S})^2}{C_\ell^{TT} M_{\ell\ell^\prime}\sqrt{F_{\ell^\prime}^\mathrm{P}} F_{\ell^\prime}^\mathrm{S}B_{\ell^\prime}^\mathrm{P}(\epsilon_T B_{\ell^\prime}^\mathrm{S})},
\end{equation}
where superscript S and P denote SPTpol and \planck, respectively, and $\epsilon_T$ is a map calibration factor that matches the scale of SPTpol temperature measurements to that of \planck.
We mock-observe the \planck map in an identical fashion to the SPTpol simulations and use the same apodization mask applied to the SPTpol data.
In this way we compare exactly the same modes on the sky, which have been identically processed, so that the true sky spectrum and the effects of filtering and mode coupling fall out of the ratio.
The filtering transfer function $\sqrt{F_\ell^\mathrm{P}}$ for the \planck map we mock-observe is unity aside from an effective low-pass filtering from the size of the \planck map pixels.
Rearranging terms yields
\begin{equation}
\label{eq:beam_lowell}
\epsilon_T B_{b}^\mathrm{S} = \sqrt{F_{b}^\mathrm{P}} B_{b}^\mathrm{P}  \frac{\left<\mathrm{Re}\left[\tilde{m}_i^{T_\mathrm{S}} \tilde{m}_j^{*T_\mathrm{S}}\right]\right>_{i \ne j, b}}{\left<\mathrm{Re}\left[\tilde{m}^{T_\mathrm{P}} \tilde{m}_i^{*T_\mathrm{S}}\right]\right>_{i, b}},
\end{equation}
where the subscript $b$ refers to a quantity averaged over $\ell$-bin $b$.

We estimate the calibration factor $\epsilon_T$ by averaging across the multipole range $600 < \ell < 1000$, where both \planck and SPTpol have similar sensitivity on a 500 deg$^2$ patch of sky.
(Above $\ell \sim 1000$, \planck becomes less sensitive than SPTpol on a patch of this size.)
We estimate the uncertainty from the standard deviation of the ratio over each multipole bin across the bundle cross-spectra. 
To account for the contribution of \planck{} noise in the beam uncertainty, we use noisy simulations of \planck{} and SPTpol maps when calculating the cross-spectra in Equation \ref{eq:beam_lowell}.
We generate SPTpol noise realizations by combining our map bundles with random signs, and we use publicly available realizations of \planck{} noise.
Since the beam function is normalized to unity at $\ell=800$, the beam and calibration uncertainties are effectively independent.
We find that the preliminary RCW38-based calibration discussed in \ref{sec:relcal} must be scaled by $\epsilon_T = 0.9088$, with an uncertainty in temperature of 0.34\%.
This calibration is similar to that found in C15 where we compared SPT-SZ and SPTpol maps on the same patch of sky, and where SPT-SZ was calibrated to \planck{}.
We marginalize over $\epsilon_T$ when fitting cosmological parameters (see Section \ref{sec:nuisance}).

\subsubsection{Beam cross-check}

We are able to independently confirm the Venus beam function at $\ell<2000$ by comparing the SPTpol and \planck{} maps and at $\ell>3000$ by looking at the brightest radio source in the survey region. 
The \planck{} comparison on large angular scales uses the same framework as the absolute calibration analysis, but  instead of averaging across $600<\ell<1000$, it looks for variation in the ratio as a function of multipole. 
Due to the \planck{} beam size, this yields a strong test of the beam function for  $\ell<2000$. 
As shown in Figure \ref{fig:beam}, the \planck{} and Venus beams agree very well over these multipoles.

We can calculate the high-$\ell$ beam directly from a 3 yr combined map of the brightest radio source in the survey field, which automatically includes the effects of pointing jitter.
The point source is significantly dimmer than Venus, so a reliable measurement of the beam is only available at scales $\lesssim 3.5'$ corresponding to multipoles $ \ell > 3000$.
At these multipoles, the Venus-derived and point-source-derived Fourier-space beam functions agree.  
We find no evidence for deviations away from the Venus-derived beam at large or small angular scales.

\subsection{$T\rightarrow P$ Deprojection}
\label{sec:leakage}

A variety of effects can leak total intensity $T$ into measurements of polarization, known as $T\rightarrow P$ leakage.
For example, a difference in relative gains in a detector pair will produce a scaled ``monopole" copy of temperature in the $Q$ and $U$ maps.
Higher-order effects can also leak $T$ into $P$, such as differential detector pointing and beam ellipticity, which add copies of the first and second derivatives of $T$ into polarization, respectively \citep{hu03}.
Given the low $\sim 10\%$ polarization fraction of the CMB and the corresponding factor of 100 reduction in amplitude between the $TT$ and $EE$ power spectra, $T\rightarrow P$ leakage is a serious systematic contaminant we must address.

We characterize and deproject a monopole leakage term from all bundles, quantifying false polarization signal that scales with $T$ as $P^{\prime} = \epsilon^PT$ for $P\in \{Q,U\}$.
To estimate the degree of monopole leakage, we take a weighted average of half-data-set cross-correlated $T$ and $P$ maps,
\begin{equation}
\epsilon^P = \frac{\sum_{\ell=50}^{2500}w_\ell \frac{C_\ell^{TP}}{C_\ell^{TT}}}{\sum_{\ell=50}^{2500}w_\ell }.
\end{equation}
Here $w_l$ is a weighting function designed to minimize the uncertainty of $\epsilon^P$.
We measure $\epsilon^Q = 0.016\pm0.001$ and $\epsilon^U = 0.009\pm0.001$, where the uncertainties are the error in the mean of $\epsilon^P$ from 125 cross-spectra of left-going-scan and right-going-scan subsets of the map bundles.
We remove the monopole leakage by subtracting scaled copies of the temperature map from each bundle according to $\epsilon^P$,
\begin{equation}
P =  \hat{P} - \epsilon^P T
\label{eq:monopole}
\end{equation}
As in C15, we ignore additional uncertainty caused by uncertainties in $\epsilon^P$.

We implicitly assume that $\epsilon^P$ are zero in the absence of systematics.
Any isotropic correlations between $T$ and $E$ modes or $B$ modes are averaged out when converted to Stokes $Q$ and $U$.
This is true for all sources of power in the maps, including foregrounds.
To demonstrate this, we calculate $\epsilon^P$ for the average of our mock-observed simulations.
We find $\epsilon^Q_\textrm{sims} = 5.3 \times 10^{-5} \pm 2.3 \times 10^{-5}$ and $\epsilon^U_\textrm{sims} = 1.0 \times 10^{-6} \pm 2.3 \times 10^{-5}$, which are orders of magnitude below the values measured for the SPTpol data.

While monopole $T\rightarrow P$ leakage is dominant in the data, we find non-negligible leakage in power spectra from higher-order effects that we call the ``leakage beam" $G_\ell$.
Using measurements of Venus, which we assume is unpolarized, we calculate $G_\ell$ as the ratio of the $TE$ and $EE$ spectra to the $TT$ spectrum of Venus,
\begin{equation}
G_\ell^{XY} = \frac{\sum_{\phi_\ell} \left( C^{XY}_{\ell, \phi_\ell} \right)_{\mathrm{Venus}} }{\sum_{\phi_\ell} \left( C^{TT}_{\ell, \phi_\ell} \right)_{\mathrm{Venus}} }. 
\end{equation}
The shape and amplitude of $G_\ell$ are well matched to the expected leakage beam for $\sim 1\%$ differential beam ellipticity.
Contributions from the leakage beam are removed by subtracting a copy of the measured $C_\ell^{TT}$ spectrum scaled by the leakage beam,
\begin{equation}
C_{\ell,\mathrm{corrected}} ^{XY} = C_{\ell,\mathrm{uncorrected}} ^{XY} - G_\ell^{XY} C_\ell^{TT}.
\end{equation}
$G_\ell^{XY}$ is everywhere less than $5\%$ and generally much smaller; in particular, at $\ell < 3500$ $G_\ell^{TE}$ and $G_\ell^{EE}$ are less than 1\% and 0.02\%, respectively.
We neglect additional uncertainty from this correction.

\subsection{Bandpower Window Functions}
\label{subsec:bpwf}

To constrain cosmological parameters, we calculate bandpowers from unbinned theoretical spectra $D^{th}_{\ell}$ for a given model.
We define bandpower window functions $W_\ell^b$ that transform unbiased theoretical spectra from unbinned to binned bandpower space:
\begin{equation}
\label{eq:bp}
D_b^{th} = W^b_\ell D^{th}_\ell.
\end{equation}
Once binned, we can directly compare the theoretical spectra to our measured bandpowers $D_{b}$ to calculate cosmological model likelihoods.

We derive the bandpower window functions from the biasing kernel $K_{bb^\prime}$ and from the fact that binned and unbinned bandpowers are related via
\begin{equation}
D_{b} = P_{b\ell} D_{\ell}
\end{equation} 
for each bin $b$:
\begin{equation}
W^b_\ell = K^{-1}_{bb^{\prime}} \left( P_{b^{\prime}\ell^{\prime}}M_{\ell^{\prime}\ell}F_{\ell}B^2_{\ell}\right).
\end{equation}
In Section \ref{sec:consistency} we describe the validation tests we perform on the bandpower window functions.

\subsection{Bandpower Covariance Matrix}
\label{sec:cov}

The bandpower covariance matrix $C_{bb^\prime}$ quantifies the uncertainties and correlations between bandpowers $b$ and $b^\prime$ and accounts for correlations between different spectra.
Sample variance from limited sky coverage and noise variance from the instrument and atmosphere contribute to the bandpower covariance matrix.
For the $TE$ and $EE$ spectra we include in this analysis, the covariance matrix has a $2\times2$ block structure.
The ``on-diagonal" blocks are auto-covariance ($TE\times TE$ and $EE\times EE$), while the two ``off-diagonal" blocks encode cross-covariance ($TE\times EE$).
As we discuss in Section \ref{sec:nuisance}, we treat the absolute calibration and beam uncertainties separately during parameter estimation and therefore do not include them in the covariance matrix.

Unlike in C15, we calculate sample and noise variance simultaneously using noisy simulations.
We add realizations of SPTpol map noise, which we generate in the same way as in Section \ref{sec:abscal}, to the 300 mock-observed noiseless simulated maps we use to calculate the transfer function.
We calculate the total bandpower covariance from the resulting set of $TE$ and $EE$ power spectra we generate from the mock-observed noisy simulations.
In another change from C15, all three independent covariance blocks ($TE\times TE$, $EE\times EE$, and $TE\times EE$) are calculated in this way, as opposed to algebraically constructing the $TE\times EE$ covariance from the $TE$ and $EE$ auto-covariances.

As in C15, covariance elements are noisy owing to a finite number of simulations.
However, given mode coupling from our map apodization, we expect elements far from the diagonal of a covariance block to also have near-zero mean.
We therefore ``condition" each block in the covariance matrix to conform to these expectations.
We calculate the bandpower correlation matrix $\rho_{bb^\prime}$ for each covariance block and average elements the same distance from the diagonal,  
\begin{equation}
\rho_{bb^\prime}=\frac{\sum_{b_1-b_2=b-b^\prime} \widehat{\rho}_{b_1b_2}}
{\sum_{b_1-b_2=b-b^\prime} 1},
\end{equation}
where the hat denotes an unconditioned matrix.
We then reconstruct the covariance blocks from the conditioned correlation matrices.
In the auto-covariance blocks, all elements greater than $\Delta\ell= 400$ from the diagonal are set to zero where the bin-bin correlations are expected to be negligible.
Because of high noise in the $TE\times EE$ covariance from just 300 realizations, we condition this block more aggressively, keeping only its diagonal elements.
While bin-bin correlations are in practice nonzero, we find no evidence of bias during cosmological fitting using this conditioning scheme.
We discuss this and other tests for bias and systematics in the following section.

\section{Tests for Systematics and Pipeline Consistency}
\label{sec:systematics}

\subsection{Null Tests}
\label{sec:jackknives}
We perform a set of null tests to look for potential systematics contaminating our maps. 
In each test, we split the data into two halves based on a metric related to the systematic in question. 
For instance, we look for time variation in the instrument by splitting the data in half temporally. 
We pair and difference observations with maximally different values of the metric to create bundles that should (nearly) null out any true sky signal. 
In practice, the nulling is slightly imperfect because of differences in the coverage and filtering of observations. 
We handle this by looking at the null spectra in signal-only simulations and subtracting their expectation values from the real null spectra. 
Any deviation from zero signal would suggest the existence of a systematic signal in the maps. 

We quantify the consistency with zero by calculating a $\chi^2$ with respect to zero for each null spectrum. 
The null spectra are binned to the same resolution as the bandpowers reported in Section \ref{sec:bandpowers}.
Note that since we are comparing to zero, there is no reason to unbias the spectra as is done in Section \ref{sec:ps}, and we do not apply the unbiasing step to the null spectra. 
In the absence of systematic errors, we expect the probabilities to exceed (PTEs) for the null test $\chi^2$ to follow a uniform distribution. 


\begin{table}
\begin{center}
\caption{Jackknife PTEs}
\small
\begin{tabular}{|c|c|c|}
\hline
\hline
Jackknife & $TE$ & $EE$ \\ 
\hline
Left-Right & 0.85 & 0.36\\ 
1st Half-2nd Half & 0.60 & 0.04\\ 
Sun & 0.95 & 0.20\\ 
Moon & 0.08 & 0.32\\ 
Azimuth & 0.79 & 0.41\\ 
\hline
\end{tabular}
\label{tab:jacks}
\end{center}
\end{table}

We perform five null tests on the SPTpol $TE$, and $EE$ spectra:
\begin{enumerate}
\item Left-Right:  We split data according to left-going or right-going telescope scans.
This test searches for scan-dependent effects from, \eg, telescope movements.  
\item 1st Half-2nd Half: This tests for time-dependent errors, such as might be induced by a drift in detector responsivity. 
Additionally, as we swapped observing strategies from lead-trail to full-field observations approximately midway through the data taking, this null test would be sensitive to any effects related to the scan strategy.
\item Sun: We test for systematics from beam sidelobe pickup by splitting data by whether they were observed with the sun above or below the horizon.
This test was degenerate with the 1st half-2nd half test in C15, but the degeneracy is mostly broken in this analysis through the inclusion of data from multiple observing seasons.
\item Moon: We test for additional beam sidelobe pickup by splitting data by whether the Moon is above or below the horizon.
\item Azimuth: We probe for contamination from stationary objects or ground features by combining maps in azimuth-elevation coordinates over the entire 3 yr observation period.
We then use the rms noise from this map as a function of azimuth as a metric to sort the standard CMB field observations.
Data are split according to whether the field azimuth during an observation was ``high noise" or ``low noise." 
\end{enumerate}

The PTEs for each test are summarized in Table \ref{tab:jacks}, and each null spectrum is plotted in the Appendix.
None are exceedingly close to zero or unity, and we conclude that our maps and resulting power spectra are free of significant systematic bias from the sources tested here.
We also find reasonable PTEs when restricting the multipole range to $50 < \ell < 500$.
We note that while performing these null tests including data up to $\ell=10,000$, we find anomalously negative values at high multipole in the $EE$ spectrum, resulting in a PTE for the $EE$ 1st Half-2nd Half test of 0.003.
Cutting data above $\ell = 8000$ improves the PTE to the value quoted in Table \ref{tab:jacks}.
To avoid a potential systematic, we limit the multipole range of the analysis to $\ell < 8000$. 
Given the current polarization map noise, imposing this limit has negligible impact on the cosmological constraints discussed below.

\subsection{Consistency Tests}
\label{sec:consistency}

Our analysis pipeline relies on accurately removing sources of bias introduced by observing, data reduction, and analysis.
To search for biases in the pipeline and resulting data products, we perform several consistency tests.
To test the self-consistency of the conditioned bandpower covariance matrix and binning operations, we check that the ensemble of unbiased mock-observed simulated bandpowers is in statistical agreement with the average values of the simulated bandpowers.
We look for potential bias in the filtering transfer function caused by generating simulations using spectra that differ from the true spectra on the sky.
Finally, we verify that the likelihood used to calculate parameter constraints recovers the input cosmological values for unbiased simulated bandpowers while using our constructed bandpower window functions and covariance matrix.

First, we test the self-consistency of the conditioned bandpower covariance matrix and binning operations.
We unbias the set of 300 noisy simulated spectra bandpowers using $K_{bb^\prime}$, the same unbiasing matrix used on the data, and we calculate the resulting set of average simulated bandpowers.
For each realization we calculate $\chi^2$ between the unbiased simulated bandpowers and the average simulated bandpowers using the conditioned covariance matrix calculated in Section \ref{sec:cov} and we consider the resulting distributions of $\chi^2$ and their probabilities to exceed.
If the conditioned covariance sufficiently captures bin-bin correlations, then in the limit of infinite simulations the resulting distribution of PTEs should be uniform.
For the set of simulations we find that $\chi^2 = 113.1 \pm 14.8$ for 112 degrees of freedom (dof).
The distribution of PTEs is consistent with being uniform, with a median value of 0.47.
The $\chi^2$ and PTE values are reasonable and show no significant evidence for bias.

Second, we check the dependence of the filtering transfer function and the process of unbiasing pseudo-spectra on the assumed cosmological model used in simulations.
We generate and mock-observe 100 sets of simulated spectra that use an alternate input cosmology from that used to generate the filtering transfer functions in the standard pipeline and add noise realizations.
The standard simulations use as input the best-fit \LCDM model to the \textsc{plikHM\_TT\_lowTEB\_lensing} \planck dataset along with foregrounds as defined in Table \ref{tab:foreground_priors}.
For this test we use a contrived \LCDM model meant to test sensitivity to spectral tilt, expansion rate, and the sound horizon: \{$\omb=0.018$, $\omc=0.14$, $\thetaMC=1.079$, $\tau=0.058$, $\As=2.2\times10^{-9}$, $\ns=0.92$\}.
Foregrounds in the alternate cosmology are also doubled compared to their values in Table \ref{tab:foreground_priors}.
See section \ref{sec:fitting} for a description of each parameter.
We unbias the pseudo-spectra of the alternate-cosmology simulations using the standard biasing kernel $K_{bb^\prime}$, which we note is calculated with the original set of simulations.
We then perform another $\chi^2$ test on the distribution of unbiased noise-free simulated bandpowers, comparing them to their binned input theory spectra.
If the transfer function and the process of unbiasing the pseudo-spectra are insensitive to small changes in input cosmology, we would expect the unbiased alternate-cosmology simulated bandpowers to be in good agreement with their input theory bandpowers.
We find that $\chi^2 = 116.7 \pm 15.5$ for 112 dof, and the median PTE of the distribution is 0.41.
If we instead use the sample covariance for the standard cosmology simulations in the calculation, we find $\chi^2 = 107.4 \pm 14.7$ with a median PTE of 0.62.
Therefore, we find no statistically significant evidence for biases in the filtering transfer function.

Finally, we verify that we can recover the input cosmological parameters to our simulations by fitting the simulated bandpowers using our likelihood, which is described in Section \ref{sec:cosmology}. 
We fit for cosmological parameters using the mean unbiased bandpowers of our standard set of 300 noiseless mock-observed sky realizations.
By using the mean bandpowers, we force the inputs to parameter estimation to be as similar as possible to the known theory.
We also use the bandpower covariance matrix and bandpower window functions calculated for the SPTpol data bandpowers to test these products during parameter estimation.
After finding mean marginalized parameters and the parameter covariance using the likelihood discussed in Section \ref{sec:cosmology}, we find the following shifts in values compared to the input cosmology, where in each case $\sigma$ is the standard deviation of a given parameter when fitting noisy SPTpol data alone: \{$\Delta\omb=-0.24\,\sigma$, $\Delta\omc=0.11\,\sigma$, $\Delta\thetaMC=-0.34\,\sigma$, $\Delta\tau=-0.03\,\sigma$, $\Delta\As=-0.10\,\sigma$, $\Delta\ns=0.23\,\sigma$\}.
We see some degeneracy between \LCDM and nuisance parameters in the likelihood related to Galactic dust foregrounds.
(See Section \ref{sec:foregrounds} for more details on the parameters.)
Fixing these foreground parameters at their input values and recalculating parameter shifts, we find  \{$\Delta\omb=0.02\,\sigma$, $\Delta\omc=0.10\,\sigma$, $\Delta\thetaMC=-0.32\,\sigma$, $\Delta\tau=-0.02\,\sigma$, $\Delta\As=0.01\,\sigma$, $\Delta\ns=-0.07\,\sigma$\}.
All input cosmological parameters are recovered within small fractions of the parameter errors, the largest difference being a $0.3\,\sigma$ shift caused by the approximate $TE$ transfer function we discuss in Section \ref{sec:transfer_function}.

We conclude from these tests that there are no significant biases in the analysis pipeline or data products and proceed to report bandpowers and resulting cosmological parameter constraints.

\section{Bandpowers}
\label{sec:bandpowers}

\begin{figure*}
\begin{center}
\includegraphics[width=1.0\textwidth]{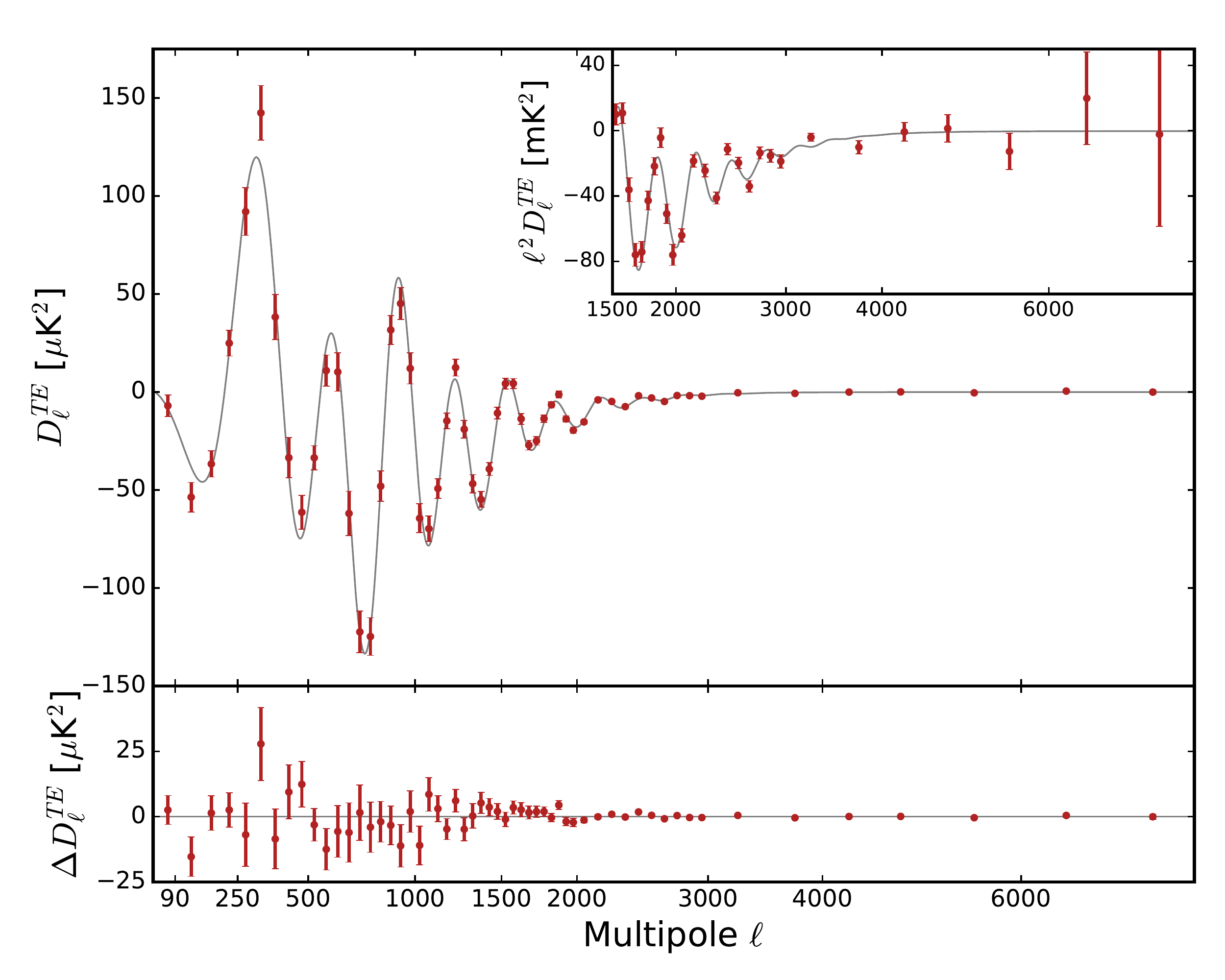}
\end{center}
\caption{SPTpol 500\,deg$^2$ $TE$ cross-correlation angular power spectrum.  
The solid gray lines are the best-fit \LCDM{} model to the \textsc{plikHM\_TT\_lowTEB} dataset.
The x-axis is scaled to $\ell^{0.6}$.
The inset plot has bandpowers scaled by an additional $\ell^2$ to highlight features at smaller angular scales.
Error bars include sample and noise variance. 
We plot residuals $\Delta D_{\ell}$ to the \textsc{plikHM\_TT\_lowTEB} model in the subpanel.
}
\label{fig:TE}
\vspace{0.1in}
\end{figure*}

The primary data products of this analysis are the $TE$ and $EE$ bandpowers and their covariance.
Using nearly 3 yr of observations on 500\,\sqdeg, we extend the measured multipole range of C15 to $50 < \ell \leq 8000$.
The low-$\ell$ cutoff is defined by our time stream filtering, while the high-$\ell$ cutoff is informed by jackknife tests as discussed in Section \ref{sec:jackknives}.
We bin the spectra to several multipole resolutions to reduce the total number of bandpowers and therefore computational complexity while maintaining sensitivity to spectral features.
The increased sky coverage results in high signal-to-noise ratio measurements of the first nine acoustic peaks of the $EE$ spectrum at $50 < \ell < 3000$ --- each peak is measured with at least three bandpowers, each with signal-to-noise ratio greater than 3.5.

\begin{figure*}
\begin{center}
\includegraphics[width=1.0\textwidth]{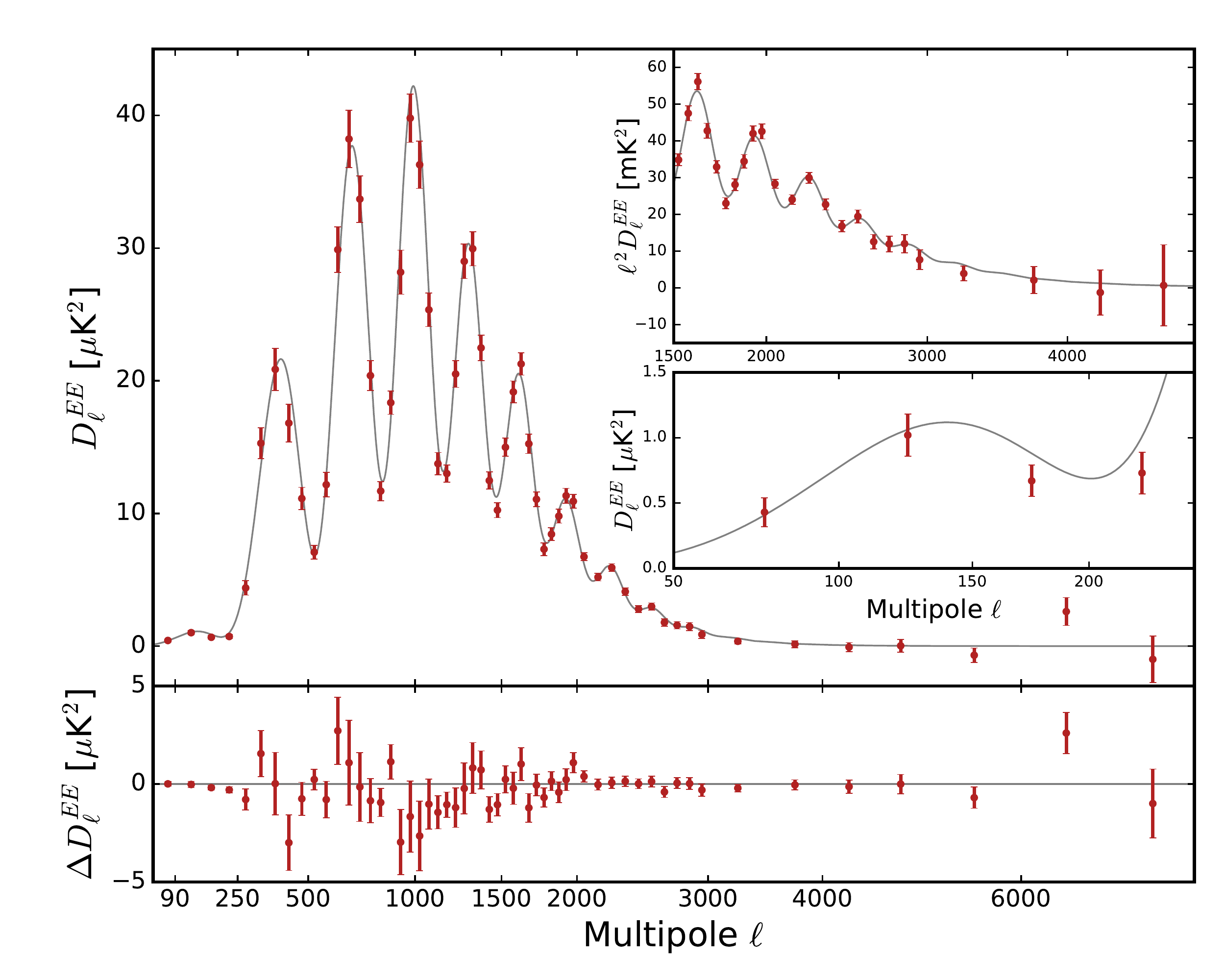}
\end{center}
\caption{SPTpol 500\,deg$^2$ $EE$ auto-correlation angular power spectrum.  
The solid gray lines are the best-fit \LCDM{} model to the \textsc{plikHM\_TT\_lowTEB} dataset.
The x-axis is scaled to $\ell^{0.6}$.
The top right inset has bandpowers scaled by an additional $\ell^2$ to highlight features at smaller angular scales.
The lower inset highlights features at low multipole without the additional scaling.
Error bars include sample and noise variance. 
We plot residuals $\Delta D_{\ell}$ to the \textsc{plikHM\_TT\_lowTEB} model in the subpanel.
}
\label{fig:EE}
\vspace{0.1in}
\end{figure*}

We plot the SPTpol bandpowers in Figures \ref{fig:TE} and \ref{fig:EE}.
Error bars include contributions from sample and noise variance.
Residuals to the best-fit \LCDM{} model to the \textsc{plikHM\_TT\_lowTEB} dataset are also plotted for each spectrum.
We present the bandpowers in Table \ref{tab:bandpowers}.
While we do not use SPTpol $TT$ bandpowers to constrain cosmology, we do use them to clean the $TE$ and $EE$ spectra of multipole-dependent beam leakage as discussed in Section \ref{sec:leakage}.
Therefore, we also include the measured $TT$ bandpowers in Table \ref{tab:bandpowers}.  

\begin{figure*}
\begin{center}
\includegraphics[width=1.0\textwidth]{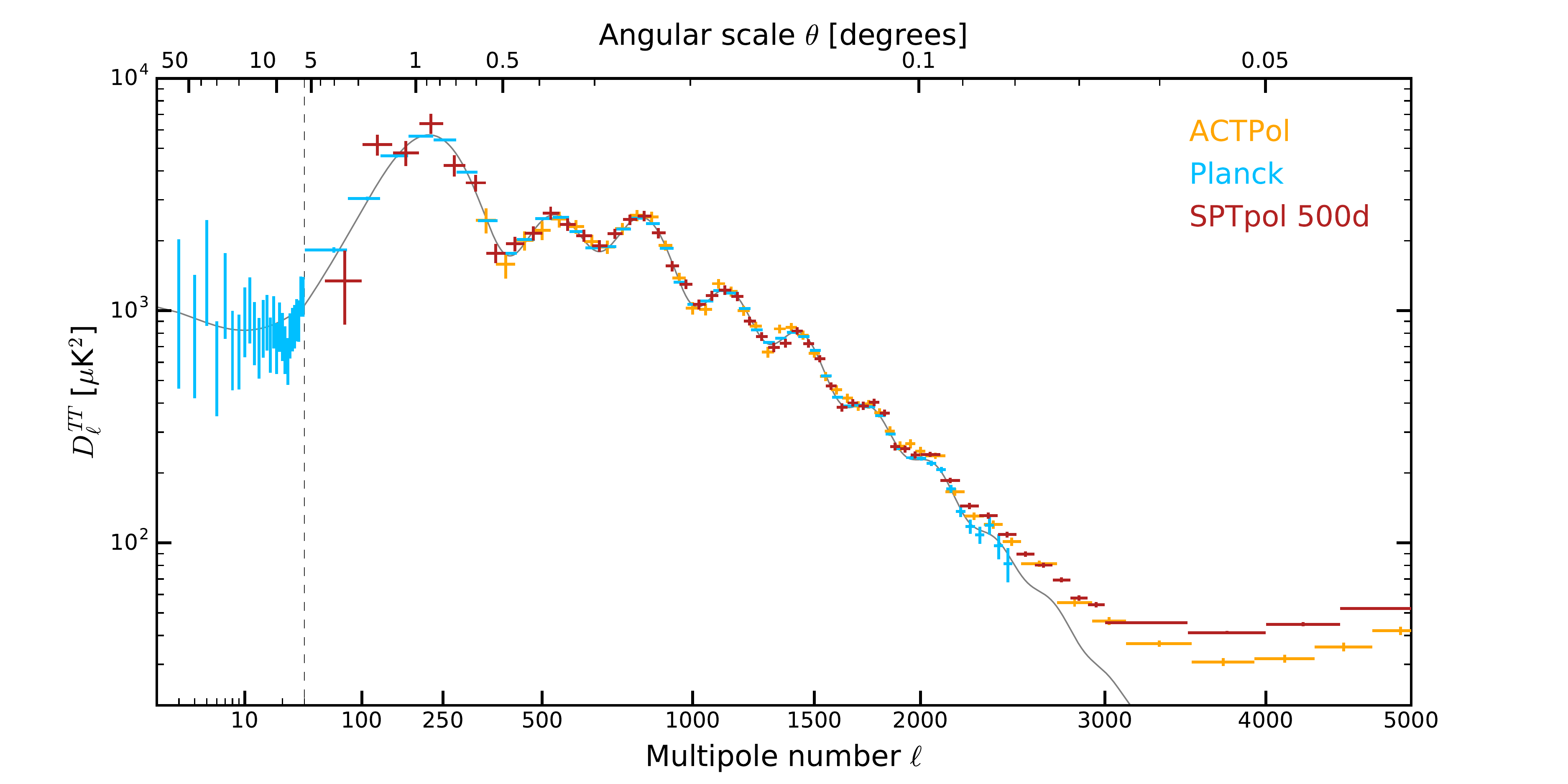}
\end{center}
\caption{Summary of recent $TT$ measurements \citep{planck15-11, louis16} with the results of this work.
The spectrum is plotted on a log scale at $\ell < 30$ (vertical dashed line) and otherwise scaled by $\ell^{0.6}$.
The solid gray line is the best-fit \LCDM model to the \planck \textsc{plikHM\_TT\_lowTEB} dataset.
Differences in power between experiments at high $\ell$ are caused by varying levels of foreground masking and/or component fitting in the respective analyses.}
\label{fig:spectra_TT_summary}
\vspace{0.1in}
\end{figure*}

\begin{figure*}
\begin{center}
\includegraphics[width=1.0\textwidth]{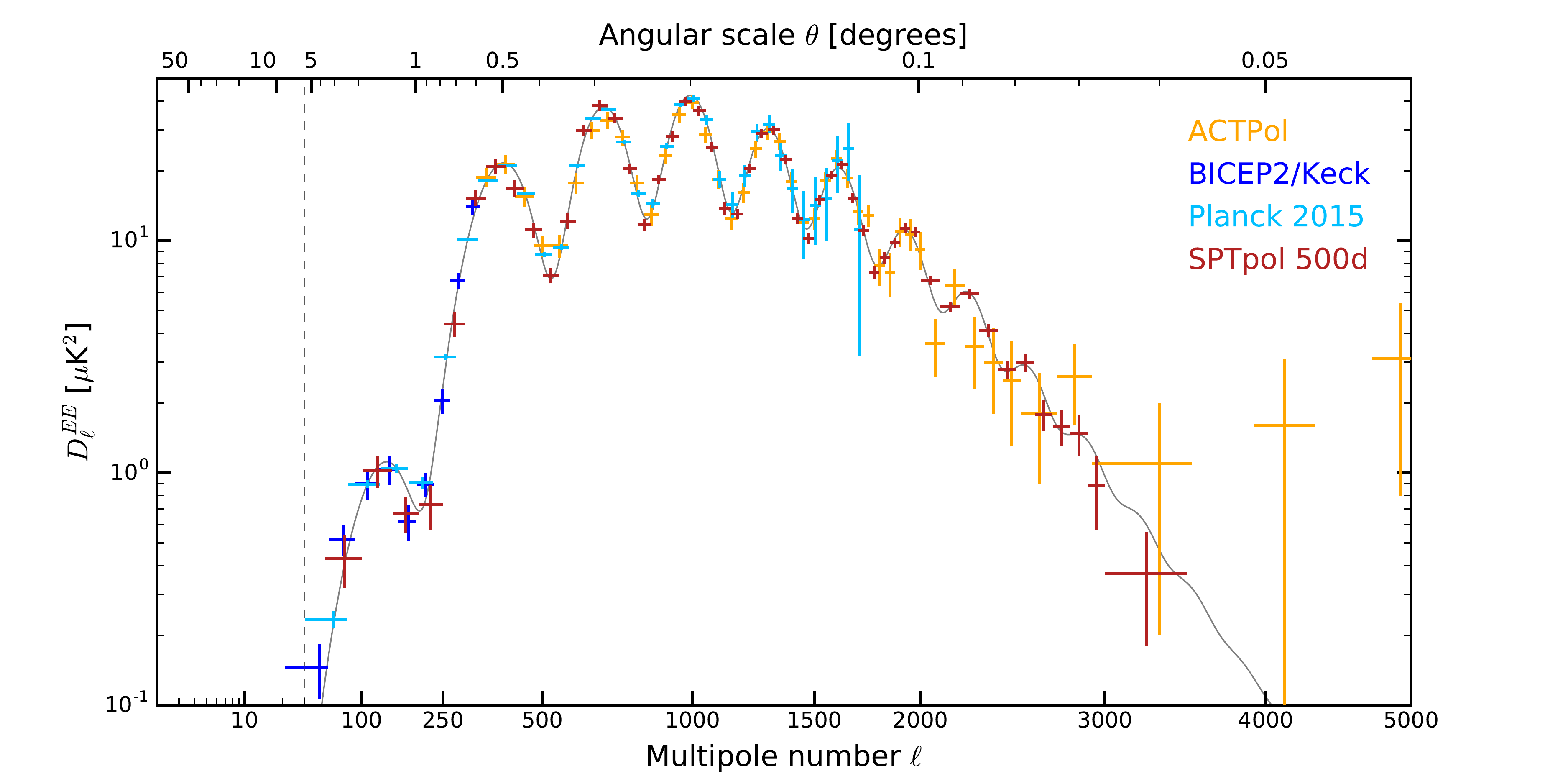}
\end{center}
\caption{Summary of recent $EE$ measurements \citep{planck15-11, bicep2keck15,louis16} with the results of this work.
The spectrum is plotted on a log scale at $\ell < 30$ (vertical dashed line) and otherwise scaled by $\ell^{0.6}$.
The solid gray line is the best-fit \LCDM model to the \planck \textsc{plikHM\_TT\_lowTEB} dataset.
Differences in power at high $\ell$ between ACTPol and SPTpol data are caused by varying levels of foreground masking.
\planck{} data are restricted to $\ell < 1750$.}
\label{fig:spectra_EE_summary}
\vspace{0.1in}
\end{figure*}

\begin{figure*}
\begin{center}
\includegraphics[width=1.0\textwidth]{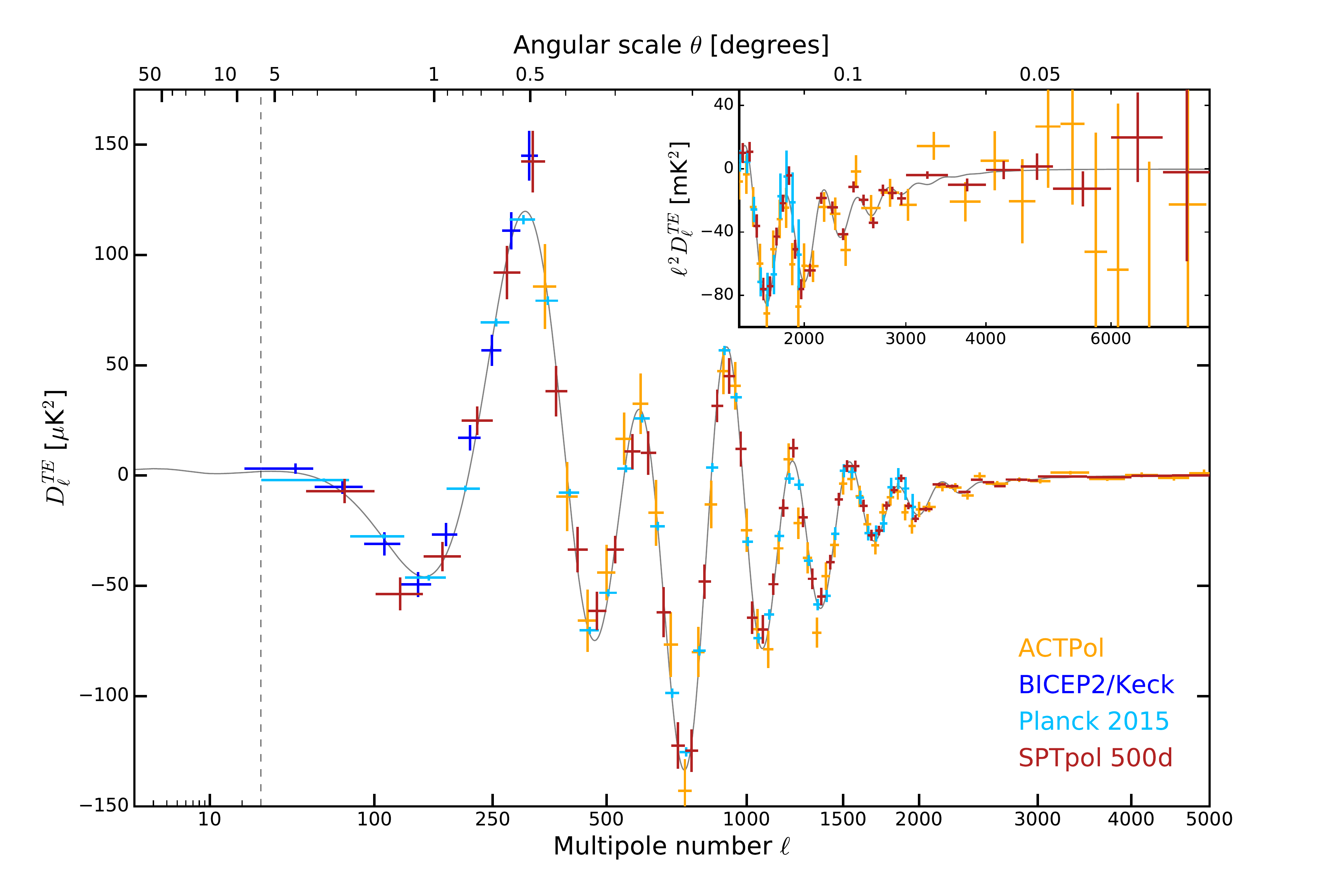}
\end{center}
\caption{Summary of recent $TE$ measurements \citep{planck15-11, bicep2keck15,louis16} with the results of this work.
The spectrum is plotted on a log scale at $\ell < 30$ (vertical dashed line).
In both panels the spectrum is scaled by $\ell^{0.3}$.
The solid gray line is the best-fit \LCDM model to the \planck \textsc{plikHM\_TT\_lowTEB} dataset.}
\label{fig:spectra_TE_summary}
\vspace{0.1in}
\end{figure*}

Recent bandpower measurements by several polarization-sensitive experiments are compiled with the results of this work in Figures \ref{fig:spectra_TT_summary}, \ref{fig:spectra_EE_summary}, and \ref{fig:spectra_TE_summary} \citep{planck15-11, bicep2keck15,louis16}.
The SPTpol $TE$ spectrum is sample variance limited at $\ell < 2050$, while the $EE$ spectrum is sample variance limited at $\ell < 1750$.
These data are the most sensitive to date of the $EE$ and $TE$ angular power spectra at $\ell > 1050$ and $\ell > 1475$, respectively.

\begin{table*}
\begin{center}
\caption[SPTpol Bandpowers and Bandpower Errors]{SPTpol Bandpowers and Bandpower Errors}

\begin{tabular}{|c| ccc | ccc | ccc |}
\hline\hline
\rule[-2mm]{0mm}{6mm}
$\ell$ Range&$\ell^{TT}_{\rm eff}$ & $D_{\ell}^{TT}$ & $\sigma^{TT}$ & $\ell^{EE}_{\rm eff}$ & $D_{\ell}^{EE}$ & $\sigma^{EE}$ & $\ell^{TE}_{\rm eff}$ & $D_{\ell}^{TE}$ & $\sigma^{TE}$\\
\hline

50  -  100  &  75.8  &  1341.6  &  469.8  &  75.8  &  0.4  &  0.1  &  76.0  &  -7.0  &  5.5 \\
100  -  150  &  124.5  &  5187.0  &  530.8  &  124.8  &  1.0  &  0.2  &  124.7  &  -53.6  &  7.5 \\
150  -  200  &  174.7  &  4785.2  &  588.8  &  174.7  &  0.7  &  0.1  &  174.7  &  -36.7  &  6.6 \\
200  -  250  &  224.6  &  6397.1  &  639.4  &  224.6  &  0.7  &  0.2  &  224.6  &  24.9  &  6.5 \\
250  -  300  &  274.5  &  4226.3  &  447.3  &  274.6  &  4.4  &  0.5  &  274.6  &  92.0  &  12.1 \\
300  -  350  &  324.5  &  3551.0  &  298.2  &  324.5  &  15.3  &  1.2  &  324.5  &  142.4  &  14.0 \\
350  -  400  &  374.5  &  1767.6  &  167.4  &  374.5  &  20.9  &  1.6  &  374.5  &  38.2  &  11.5 \\
400  -  450  &  424.4  &  1941.2  &  142.4  &  424.5  &  16.8  &  1.4  &  424.5  &  -33.5  &  10.3 \\
450  -  500  &  474.5  &  2157.1  &  151.3  &  474.5  &  11.1  &  0.8  &  474.5  &  -61.4  &  8.7 \\
500  -  550  &  524.5  &  2634.1  &  164.8  &  524.4  &  7.1  &  0.5  &  524.5  &  -33.6  &  6.2 \\
550  -  600  &  574.5  &  2349.0  &  144.5  &  574.6  &  12.2  &  0.9  &  574.5  &  10.9  &  7.9 \\
600  -  650  &  624.4  &  2102.6  &  125.3  &  624.5  &  29.9  &  1.7  &  624.5  &  10.2  &  9.9 \\
650  -  700  &  674.5  &  1903.2  &  108.7  &  674.5  &  38.2  &  2.2  &  674.5  &  -62.0  &  11.3 \\
700  -  750  &  724.5  &  2143.2  &  109.3  &  724.5  &  33.7  &  1.8  &  724.5  &  -122.4  &  10.6 \\
750  -  800  &  774.6  &  2466.3  &  128.9  &  774.5  &  20.4  &  1.1  &  774.5  &  -124.8  &  9.6 \\
800  -  850  &  824.5  &  2555.5  &  131.7  &  824.4  &  11.7  &  0.7  &  824.5  &  -48.1  &  7.8 \\
850  -  900  &  874.5  &  2160.0  &  111.1  &  874.5  &  18.4  &  0.9  &  874.5  &  31.6  &  7.4 \\
900  -  950  &  924.5  &  1556.4  &  81.2  &  924.5  &  28.2  &  1.7  &  924.5  &  45.2  &  8.1 \\
950  -  1000  &  974.5  &  1301.6  &  61.6  &  974.5  &  39.8  &  1.8  &  974.5  &  12.0  &  7.9 \\
1000  -  1050  &  1024.5  &  1062.9  &  53.9  &  1024.5  &  36.3  &  1.8  &  1024.5  &  -64.4  &  7.4 \\
1050  -  1100  &  1074.5  &  1161.7  &  56.9  &  1074.5  &  25.4  &  1.3  &  1074.5  &  -69.7  &  6.4 \\
1100  -  1150  &  1124.5  &  1228.0  &  58.0  &  1124.4  &  13.8  &  0.8  &  1124.5  &  -49.2  &  5.0 \\
1150  -  1200  &  1174.6  &  1152.2  &  51.6  &  1174.4  &  13.0  &  0.6  &  1174.5  &  -14.7  &  4.0 \\
1200  -  1250  &  1224.5  &  902.7  &  39.8  &  1224.5  &  20.5  &  1.0  &  1224.5  &  12.5  &  4.3 \\
1250  -  1300  &  1274.5  &  773.8  &  33.0  &  1274.5  &  29.0  &  1.3  &  1274.5  &  -19.0  &  4.4 \\
1300  -  1350  &  1324.4  &  695.6  &  32.2  &  1324.5  &  30.0  &  1.3  &  1324.5  &  -46.8  &  4.7 \\
1350  -  1400  &  1374.4  &  724.8  &  31.4  &  1374.5  &  22.5  &  1.0  &  1374.5  &  -54.8  &  4.0 \\
1400  -  1450  &  1424.5  &  818.2  &  30.2  &  1424.5  &  12.5  &  0.6  &  1424.5  &  -39.3  &  3.4 \\
1450  -  1500  &  1474.5  &  722.0  &  30.2  &  1474.5  &  10.2  &  0.6  &  1474.5  &  -10.8  &  2.9 \\
1500  -  1550  &  1524.5  &  622.1  &  22.3  &  1524.5  &  15.0  &  0.7  &  1524.5  &  4.3  &  2.8 \\
1550  -  1600  &  1574.5  &  473.5  &  17.6  &  1574.5  &  19.2  &  0.8  &  1574.5  &  4.3  &  2.5 \\
1600  -  1650  &  1624.4  &  383.4  &  16.4  &  1624.4  &  21.3  &  0.8  &  1624.4  &  -13.7  &  2.8 \\
1650  -  1700  &  1674.5  &  401.0  &  15.0  &  1674.4  &  15.2  &  0.7  &  1674.5  &  -27.1  &  2.5 \\
1700  -  1750  &  1724.5  &  389.3  &  15.8  &  1724.4  &  11.1  &  0.6  &  1724.5  &  -25.0  &  2.1 \\
1750  -  1800  &  1774.4  &  403.5  &  14.9  &  1774.5  &  7.3  &  0.5  &  1774.4  &  -13.6  &  1.8 \\
1800  -  1850  &  1824.5  &  362.7  &  12.2  &  1824.5  &  8.4  &  0.5  &  1824.5  &  -6.5  &  1.6 \\
1850  -  1900  &  1874.5  &  260.4  &  10.3  &  1874.5  &  9.8  &  0.5  &  1874.5  &  -1.2  &  1.7 \\
1900  -  1950  &  1924.5  &  254.2  &  9.0  &  1924.5  &  11.3  &  0.6  &  1924.5  &  -13.7  &  1.6 \\
1950  -  2000  &  1974.5  &  239.5  &  8.1  &  1974.5  &  10.9  &  0.5  &  1974.5  &  -19.5  &  1.6 \\
2000  -  2100  &  2049.5  &  240.6  &  6.2  &  2049.5  &  6.7  &  0.3  &  2049.5  &  -15.3  &  1.0 \\
2100  -  2200  &  2149.4  &  186.1  &  4.9  &  2149.5  &  5.2  &  0.3  &  2149.5  &  -4.0  &  0.8 \\
2200  -  2300  &  2249.5  &  144.5  &  4.5  &  2249.5  &  5.9  &  0.3  &  2249.5  &  -4.8  &  0.8 \\
2300  -  2400  &  2349.5  &  131.2  &  3.9  &  2349.5  &  4.1  &  0.3  &  2349.5  &  -7.5  &  0.7 \\
2400  -  2500  &  2449.5  &  108.8  &  3.1  &  2449.5  &  2.8  &  0.2  &  2449.5  &  -1.9  &  0.6 \\
2500  -  2600  &  2549.5  &  89.6  &  2.2  &  2549.5  &  3.0  &  0.3  &  2549.5  &  -3.0  &  0.5 \\
2600  -  2700  &  2649.5  &  80.2  &  2.0  &  2649.5  &  1.8  &  0.3  &  2649.5  &  -4.8  &  0.5 \\
2700  -  2800  &  2749.5  &  69.4  &  1.8  &  2749.5  &  1.6  &  0.3  &  2749.5  &  -1.8  &  0.5 \\
2800  -  2900  &  2849.5  &  58.0  &  1.5  &  2849.5  &  1.5  &  0.3  &  2849.5  &  -1.9  &  0.5 \\
2900  -  3000  &  2949.5  &  54.3  &  1.4  &  2949.5  &  0.9  &  0.3  &  2949.5  &  -2.2  &  0.5 \\
3000  -  3500  &  3249.5  &  45.4  &  0.6  &  3249.5  &  0.4  &  0.2  &  3249.5  &  -0.4  &  0.2 \\
3500  -  4000  &  3749.5  &  41.1  &  0.7  &  3749.5  &  0.2  &  0.3  &  3749.5  &  -0.7  &  0.3 \\
4000  -  4500  &  4249.5  &  44.7  &  0.9  &  4249.5  &  -0.1  &  0.3  &  4249.5  &  -0.0  &  0.3 \\
4500  -  5000  &  4749.5  &  52.2  &  0.7  &  4749.5  &  0.0  &  0.5  &  4749.5  &  0.1  &  0.4 \\
5000  -  6000  &  5499.5  &  66.0  &  0.6  &  5499.5  &  -0.7  &  0.5  &  5499.5  &  -0.4  &  0.4 \\
6000  -  7000  &  6499.5  &  89.4  &  1.1  &  6499.5  &  2.6  &  1.0  &  6499.5  &  0.5  &  0.7 \\
7000  -  8000  &  7499.5  &  115.7  &  1.4  &  7499.5  &  -1.0  &  1.8  &  7499.5  &  -0.0  &  1.0 \\

\hline
\end{tabular}
\label{tab:bandpowers}
\tablecomments{
The $\ell$ range, bandpower window function-weighted multipole $\ell_{\rm eff}$, bandpowers $D_{\ell}^{XY}$, and associated bandpower uncertainties, $\sigma^{XY}$, of the SPTpol 150\,GHz $TT$, $EE$, and $TE$ power spectra.  Bandpowers and errors are given in units of $\microKsq$.
The errors are the square root of the diagonal elements of the covariance matrix and do not include beam or calibration uncertainties.
}
\end{center}
\end{table*}


\section{Fitting Methodology and Likelihood}
\label{sec:cosmology}

In this section we describe the methodology we use for cosmological parameter fitting.
We discuss our chosen \LCDM parameterization, treatment of foregrounds, and instrument nuisance parameters.
Finally, we describe additional corrections in our likelihood to account for biases related to measuring a small patch of sky.

\subsection{Fitting Methodology}
\label{sec:fitting}

We calculate constraints on cosmological parameters with the 2016 November version of the Markov Chain Monte Carlo (MCMC) package \textsc{CosmoMC} \citep{lewis02b}.
Unlike in C15, where we used \textsc{PICO} \citep{fendt07b,fendt07a} trained with the Boltzmann code \textsc{CAMB} \citep{lewis99}, we have configured \textsc{CosmoMC} to use \textsc{CAMB} directly.
We continue to use the SPTpol likelihood discussed in C15 to constrain cosmology with these bandpower measurements; however, the likelihood is better integrated into \textsc{CosmoMC}, and several additional nuisance parameters have been introduced, which we discuss below.
Details on how to install and use the SPTpol likelihood and dataset are available on the SPT website.\footnote{http://pole.uchicago.edu/public/data/henning17/}

For this analysis, we choose the following parameterization of the \LCDM model: the content of baryons and cold dark matter, \omb and \omc, respectively; \thetaMC, an internal \textsc{CosmoMC} variable that is a proxy for the angular scale of the sound horizon at decoupling \thetas; the amplitude of primordial scalar fluctuations \As; the spectral tilt of primordial scalar fluctuations \ns, defined at a pivot scale of $k_0 = 0.05\,\mbox{Mpc}^{-1}$; and the optical depth to reionization $\tau$.
We report \As and $\tau$ as a single combined amplitude parameter \clamp.
In addition to these six base parameters, we also report constraints on \ho, the expansion rate today, and \sigmaeight, the present amplitude of matter fluctuations at $8/h$\,Mpc scales, where $h = \ho / 100\,\mbox{km}\,\mbox{s}^{-1}\,\mbox{Mpc}^{-1}$.

We fit the SPTpol $TE$ and $EE$ bandpowers from this work, hereafter the \sptpolEETE dataset, to several \LCDM models both independently and simultaneously with the \textsc{plikHM\_TT\_lowTEB} \planck dataset \citep{planck15-11}, hereafter \textsc{PlanckTT}.
We allow the \planck and SPTpol likelihoods to treat foregrounds independently.
The SPTpol survey region covers a small fraction of the total sky, so we also neglect correlations between experiments resulting from shared sky signal.

We choose not to fit SPTpol $TT$ bandpowers simultaneously with $TE$ and $EE$ for several reasons.
First, as discussed in Section \ref{sec:bundles}, the analysis is tailored to the recovery of large-scale polarization modes leaving significant atmospheric contamination in the $TT$ spectrum.
The $TT$ spectrum does not, in general, pass our null tests, and further data cuts or analysis techniques to clean the $TT$ spectrum would reduce our polarization sensitivity.
Second, including $TT$ in the SPTpol likelihood necessitates the inclusion of many more foreground terms and nuisance parameters, some with temperature-polarization correlations.
To properly account for these new terms, the likelihood would need significant updates that are beyond the scope of this work. 
Efforts are ongoing, however, to update the likelihood module of \citet{story13}, as well as to include TT measurements in the SPTpol likelihood.

\subsection{Foreground Parameterization}
\label{sec:foregrounds}

The primordial CMB $EE$ and $TE$ power spectra are expected to be less contaminated by foreground power than the temperature spectrum at small scales. 
For example, C15 did not see any evidence of contamination from polarized extragalactic source power after masking the brightest $\sim$10 sources over $\sim100\,$deg$^2$, and the level of $EE$ power from Galactic dust expected in our sky patch based on \citet{planck14-30} is a factor of $\sim$20 below our measured $EE$ power in the lowest $\ell$ bin. Nevertheless, we add parameters to our cosmological model to account for these two potential sources of polarized power. We do not attempt to model contributions from Galactic synchrotron emission because we expect the polarized Galactic foreground power to be dominated by dust at 150 GHz.

We introduce four parameters to model contributions to the $TE$ and $EE$ spectra from polarized Galactic dust.
We assume the angular power spectrum of Galactic dust follows the model of \citet{planck14-30},
\begin{equation}
D_{\ell,\mathrm{dust}}^{XY} = A_{80}^{XY} \left(\frac{\ell}{80}\right)^{\alpha_{XY}+2}.
\end{equation}
Here $ A_{80}^{XY}$ is the amplitude of the spectrum in units of $\mu$K$^2$ at $\ell=80$ and $\alpha_{XY}$ is the angular power dust spectral index.
As the SPTpol survey field overlaps the \textsc{BICEP2} field, we use the \textsc{Planck} constraints over the \textsc{BICEP2} patch corrected for the SPTpol 150\,GHz bandpass to define priors on $A_{80}^{XY}$ and $\alpha_{XY}$ for generating the simulations discussed in Section \ref{sec:transfer_function}, which we summarize in Table \ref{tab:foreground_priors}.
We obtain a pessimistic expectation for the $TE$ dust amplitude by assuming that the temperature and $E$-mode dust spectra are 100\% correlated and taking the geometric mean of their amplitudes.
During cosmological fitting, we apply flat priors between 0 and $2\,\mu$K$^2$ on $A_{80}^{XY}$ and Gaussian priors on $\alpha_{XY}$ centered on -2.42 with standard deviations of 0.02, motivated by the findings of \citet{planck14-30}.

While masking extragalactic sources removes most point-source power, we parameterize the level of residual polarized power in the $EE$ spectrum by fitting a component $D_\ell \propto \ell^2$.
(There is also a clustering component to the point-source power, but this component is a modulation of the mean power from all sources and is thus effectively unpolarized.)
We include a single additional foreground nuisance term, $D^{\mathrm{PS_{EE}}}_{3000}$, which is the amplitude of residual power $\propto \ell^2$ at $\ell=3000$ after masking all sources above 50\,mJy in unpolarized flux at 95 and 150\,GHz.
We apply a uniform prior between 0 and $2.5\,\mu$K$^2$ on $D^{\mathrm{PS_{EE}}}_{3000}$ in all cosmological fits.


\begin{table*}
\begin{center}
\caption{Foreground and Nuisance Parameter Priors}
\small
\begin{tabular}{|c|c|c|c|c|c|c|c|c|c|}
\hline
\hline

& $D^{\mathrm{PS_{EE}}}_{3000}$& $A_{80}^{TT}$ & $A_{80}^{EE}$ & $A_{80}^{TE}$& $\alpha_{XX}$ & $T_\mathrm{cal}$ & $P_\mathrm{cal}$ & $\kappa$ & $A^n_{\mathrm{beam}}$\\ 
& [$\mu\mbox{K}^2$] & [$\mu\mbox{K}^2$] & [$\mu\mbox{K}^2$] & [$\mu\mbox{K}^2$] & - & - & - & - & - \\
\hline
Value for simulations & $0.21$ & $1.15$ & $0.0236$ & $0.1647$ & -2.42  & 1.0 & 1.0 & 0.0 & 0.0 \\
Prior & $0-2.5$ & $ 0-2$ & $ 0-2$ & $ 0-2$ & $\sigma=0.02$ & $\sigma=0.0034$ & $\sigma=0.01$ & $\sigma=0.001$ & $\sigma=1.0$\\
\hline
\end{tabular}
\tablecomments{We marginalize over six foreground parameters ($D^{\mathrm{PS_{EE}}}_{3000}$, $A_{80}^{TT}$, $A_{80}^{EE}$, $A_{80}^{TE}$, $\alpha_{XY}$, where $XY \in \{TE,EE\}$), the super-sample lensing variance $\kappa$, and four instrumental calibration and beam uncertainty terms ($T_\mathrm{cal}$, $P_\mathrm{cal}$, and $A^n_{\mathrm{beam}}$, $n \in \{1,2\}$). 
We impose flat priors on the first four of these and Gaussian priors on the remainder. 
For Gaussian prior central values, we use the values chosen when generating the simulated skies discussed in Section \ref{sec:transfer_function}.}
\label{tab:foreground_priors}
\end{center}
\end{table*}

\subsection{Nuisance Parameters}
\label{sec:nuisance}

In addition to the five foreground terms we discuss above, the SPTpol likelihood includes four additional parameters.
As in C15, $T_\mathrm{cal}$ and $P_\mathrm{cal}$ represent the map-space temperature and polarization calibration.
That is, we scale the theoretical spectra to which we are comparing our data by $1/(T_\mathrm{cal}^2P_\mathrm{cal}$) for $TE$ and $1/(T_\mathrm{cal}^2P_\mathrm{cal}^2$) for $EE$.
Note that with spectrum calibration parameterized in this way, we do not include calibration covariance in the bandpower covariance matrix.
As discussed in section \ref{sec:beams}, we obtain an absolute temperature calibration when cross-correlating the SPTpol survey map with the \planck{} 143 GHz map.
We correct for this calibration when unbiasing bandpowers; therefore, when fitting cosmology, $T_\mathrm{cal}$ has an expectation of unity.
We apply a Gaussian prior to $T_\mathrm{cal}$ centered on one with a standard deviation of 0.0034, the calibration error obtained from matching amplitudes of low-$\ell$ and high-$\ell$ beams we discuss above.

$P_\mathrm{cal}$ can be interpreted as the inverse of the effective polarization efficiency of the instrument.
In C15, where detector cross-talk was treated as a multiplicative bias partially degenerate with polarization efficiency, we used a flat prior on $P_\mathrm{cal}$.
We now correct cross-talk at the time stream level before we generate maps, so one would expect that the physical interpretation of $P_\mathrm{cal}$ is more clearly defined.
Furthermore, as we correct for nonunity polarization efficiency when adding detector time streams into maps, we expect $P_\mathrm{cal}$ to be unity.
However, we find $P_\mathrm{cal} = 1.06 \pm 0.01$ when calculating the ratio of an $EE$ cross-spectrum between \planck and SPTpol $E$-mode maps over the $EE$ spectrum from SPTpol over the multipole range $500 < \ell < 1500$.
This result disagrees with the expected polarization efficiency from measurements of our polarization calibration source, the error for which was 2\%.
The source of this discrepancy is the subject of ongoing study, but we note that $P_\mathrm{cal}$ is a constant multiplicative correction to the polarized spectra and does not alter their shape as a function of multipole.
For the purposes of this analysis, we choose to apply the additional calibration factor of 1.06 to our $E$-mode maps and use a Gaussian prior for $P_\mathrm{cal}$ centered on 1.0 with a standard deviation of 0.01 motivated by the \planck-SPTpol $EE$ cross-spectrum result when fitting cosmology.
We provide the priors for $P_\mathrm{cal}$ and $T_\mathrm{cal}$ in Table \ref{tab:foreground_priors}.

We continue to marginalize our cosmological constraints over the effects of so-called ``super-sample lensing" variance \citep{manzotti14}.
When studying small regions of the sky, gravitational lensing occurring over scales larger than the region itself can dilate or contract scales across the entire patch, leading to a bias in the constraint on $\theta_\mathrm{s}$.
As in C15, we modify the theoretical spectrum returned from \textsc{CAMB} for parameter vector $\mathbf{p}$ at every step in a Markov chain,
\begin{equation}
\hat{C}_\ell^{XY} \left( \mathbf{p};\kappa \right) = C_\ell^{XY} \left( \mathbf{p}\right) - \frac{\partial \ell^2 C_\ell^{XY} \left( \mathbf{p}\right)}{\partial \ln \ell}\frac{\kappa}{\ell^2},
\end{equation}
where $\kappa$ is the mean lensing convergence in a field.
As demonstrated by \citet{manzotti14}, the uncertainty on $\kappa$ decreases with increasing survey area.
We apply a Gaussian prior centered on zero with a standard deviation of $\sigma_{\kappa} = 1.0\times 10^{-3}$, which is more than a factor of two tighter than the prior applied in C15 where the field was only 100 deg$^2$.

Finally, we must incorporate SPTpol beam uncertainty into the MCMC. 
We first calculate the beam correlation matrix from the beam covariance matrix discussed in Section \ref{sec:highell_beam}.
This matrix is singular, as we would expect since the number of dof in the beam uncertainties is less than the number of bandpowers. 
To resolve this, we only keep the eigenvalues (and associated eigenvectors) that are at least 0.01 times the maximum eigenvalue. 
We define a beam error $C_{\ell,n}^\mathrm{beam}$ for both of the two surviving eigenvectors of the beam correlation matrix, where $n$ indexes the eigenvector $H_\ell^n$, 
\begin{equation}
C_{\ell,n}^\mathrm{beam} = A^n_\mathrm{beam} H_\ell^n.
\end{equation}
We perturb the theoretical spectrum $C_\ell$ to which we compare according to
\begin{equation}
C_\ell \rightarrow C_\ell \left( 1 + \sum_{n=1}^2 C_{\ell,n}^\mathrm{beam}\right).
\end{equation} 
The two beam error eigenmode amplitudes $A^n_\mathrm{beam}$ are treated as nuisance parameters and sampled in the MCMC.
We apply a Gaussian prior with unit width centered on zero to both of these parameters.

\subsection{Zero-parameter Corrections}
\label{sec:aberration}

Recently, \citet{louis16} accounted for aberration owing to relative motion with respect to the CMB dipole when fitting ACTPol data for cosmological parameters.  
They find that not accounting for this effect amounts to a $0.5\,\sigma$ bias in \thetas.  
The effect of aberration is well approximated by a simple formula \citep{jeong14} and is completely determined by the input spectrum and the separation angle between the observation patch and the direction of the CMB dipole.

We also correct for aberration in our likelihood, requiring zero extra nuisance parameters.
Before binning with our bandpower window functions, we adjust the theoretical spectra according to 
\begin{equation}
C_\ell \rightarrow C_\ell - C_\ell \frac{d\ln C_\ell}{d\ln \ell}\beta\left<\cos\theta\right>,
\end{equation} in a similar fashion to \cite{louis16}.  
Here $\beta=1.23 \times 10^{-3}$ for the CMB dipole, and $\left<\cos\theta\right>=-0.40$ for the SPTpol survey field.
Note that the correction has the opposite sign to that found in \cite{louis16}, as we apply it to the theoretical spectrum instead of our data.
We find that turning on the aberration correction shifts our value of \thetas from SPTpol-only fits by $-0.4\,\sigma$.
We note that this shift is comparable to the increase in uncertainty on \thetas caused by the effects of super-sample lensing variance on a $500$\,\sqdeg patch \citep{manzotti14}.

\section{Constraints}
\label{sec:lcdm}

We now discuss cosmological constraints calculated using the \sptpolEETE dataset, both independently and combined with \planckTT.
First, we fit the standard \LCDM model to \sptpolEETE over several multipole ranges.
We next consider the implications for current and future experiments from our constraints on polarized foregrounds.
Finally, we calculate constraints for several one- and two-parameter extensions to \LCDM that probe additional physics sensitive to power in the damping tail.

\subsection {\LCDM}
\label{sec:lcdm_sptpol}

\begin{figure*}[t]
\begin{center}
\includegraphics[width=1.0\textwidth]{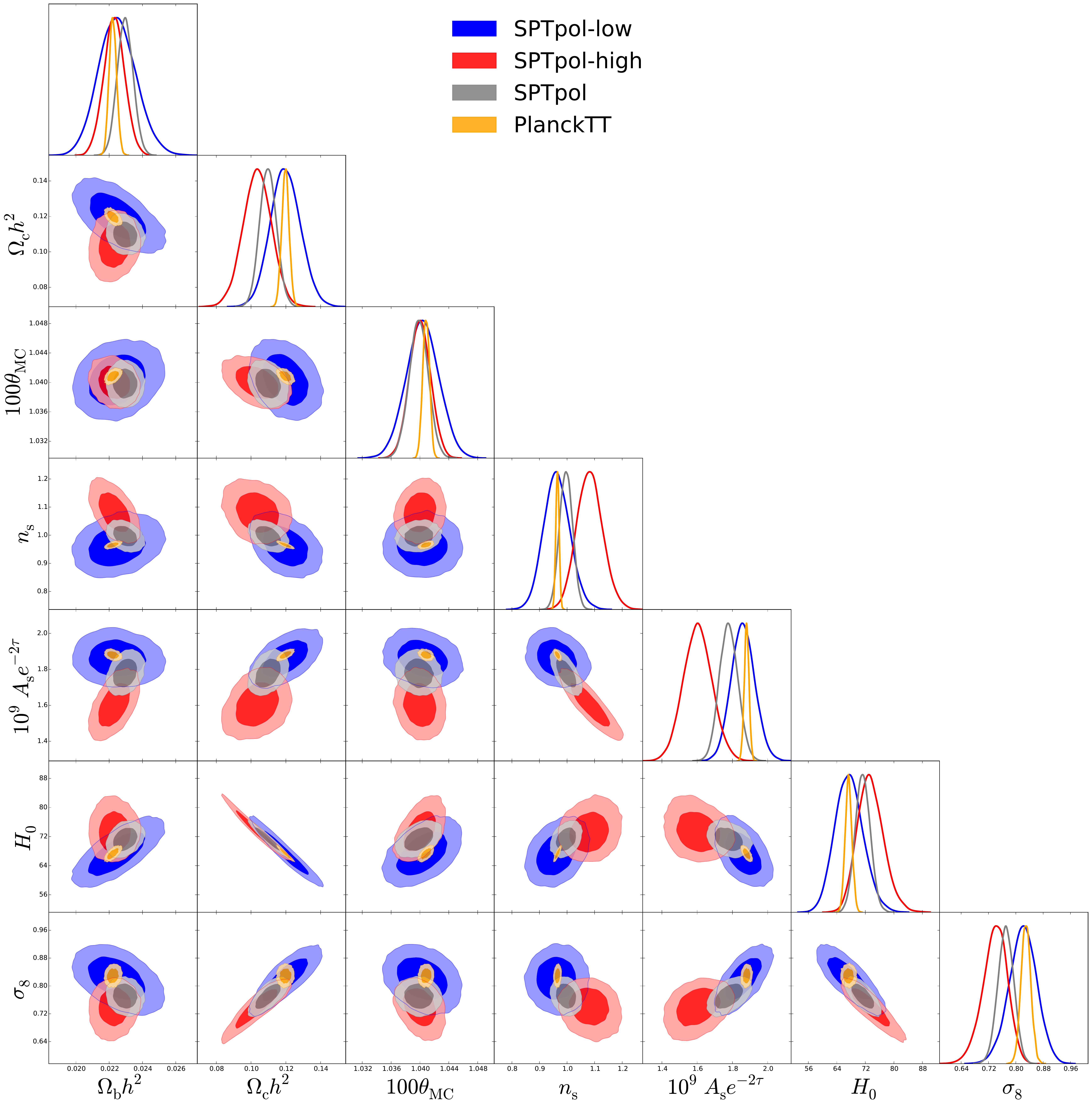}
\end{center}
\caption{Marginalized parameter constraints for the \LCDM model using \sptpolEETE.
We explore constraints over three multipole ranges: $50 < \ell \le 8000$ (\sptpolEETE), $50 < \ell \le 1000$ (\sptpolEETE-low), and $1000 < \ell \le 8000$ (\sptpolEETE-high).
For comparison, we include constraints from \planckTT as well.
}
\label{fig:TEEE_lcdm}
\end{figure*}

We first place constraints on the standard \LCDM model with the SPTpol $TE+EE$ dataset, which we present in Figure \ref{fig:TEEE_lcdm} and Table \ref{tab:lcdm_ellCut}, over several multipole ranges: $50 < \ell \le 8000$ referred to as the ``full" dataset (\sptpolEETE), $50 < \ell \le 1000$ referred to as the ``low-$\ell$" dataset (\sptpolEETE-low), and $1000 < \ell \le 8000$ referred to as the ``high-$\ell$" dataset (\sptpolEETE-high).
This cut in multipole space is roughly where the sensitivity of the SPTpol dataset surpasses that of \planck polarization.
Figure \ref{fig:TEEE_lcdm} shows 1D and 2D marginalized posterior probabilities for \LCDM parameters, as well as the derived quantities \ho and \sigmaeight.
Table \ref{tab:lcdm_ellCut} shows 68\% marginalized constraints for each parameter.
With a minimum multipole of 50, \sptpolEETE has little sensitivity to $\tau$, so we place on it a Gaussian prior $0.078\pm0.019$ informed by constraints from \planckTT.

\begin{table*}
\begin{center}
\caption{\LCDM Constraints}
\small

\begin{tabular}{c|c|c|c}
\hline\hline
\multicolumn{1}{c|}{Parameter} & \multicolumn{3}{c}{Data Set} \\ 
 & \multicolumn{1}{c}{\textsc{SPTpol-$\ell < 1000$}} & \multicolumn{1}{c}{\textsc{SPTpol-$\ell > 1000$}} & \multicolumn{1}{c}{\textsc{SPTpol}} \\ 
\hline 

\multicolumn{4}{l}{ Free }\\ 
$100\Omega_{\mathrm{b}}h^2$ & 2.250 $\pm$ 0.114 & 2.230 $\pm$ 0.063 & 2.296 $\pm$ 0.048\cr
$\Omega_{\mathrm{c}}h^2$ & 0.1198 $\pm$ 0.0087 & 0.1036 $\pm$ 0.0083 & 0.1098 $\pm$ 0.0048\cr
$100\theta_{\mathrm{MC}}$ & 1.0404 $\pm$ 0.0023 & 1.0400 $\pm$ 0.0015 & 1.0398 $\pm$ 0.0013\cr
$n_{\mathrm{s}}$ & 0.9635 $\pm$ 0.0478 & 1.0827 $\pm$ 0.0472 & 0.9967 $\pm$ 0.0238\cr
$10^{9}A_\mathrm{s}e^{-2\tau}$ & 1.8604 $\pm$ 0.0675 & 1.6035 $\pm$ 0.0818 & 1.7791 $\pm$ 0.0528\cr
\hline

\multicolumn{4}{l}{ Derived }\\ 
$\Omega_\Lambda$ & 0.681 $\pm$ 0.055 & 0.762 $\pm$ 0.039 & 0.736 $\pm$ 0.025\cr
$\sigma_8$ & 0.820 $\pm$ 0.041 & 0.738 $\pm$ 0.037 & 0.771 $\pm$ 0.024\cr
$H_0$ & 67.49 $\pm$ 3.99 & 73.49 $\pm$ 3.73 & 71.29 $\pm$ 2.12\cr
\hline

\multicolumn{4}{l}{ Nuisance + Foreground }\\ 
$\widehat{T}_\mathrm{cal}$ & 1.000 $\pm$ 0.003 & 1.000 $\pm$ 0.003 & 1.000 $\pm$ 0.003\cr
$\widehat{P}_\mathrm{cal}$ & 1.008 $\pm$ 0.011 & 0.994 $\pm$ 0.011 & 1.003 $\pm$ 0.010\cr
$100\widehat{\kappa}$ & 0.001 $\pm$ 0.100 & 0.000 $\pm$ 0.101 & 0.000 $\pm$ 0.102\cr
$\widehat{\alpha}^\mathrm{TE}_\mathrm{dust}$ & -2.42 $\pm$ 0.02 & -2.42 $\pm$ 0.02 & -2.42 $\pm$ 0.02\cr
$\widehat{\alpha}^\mathrm{EE}_\mathrm{dust}$ & -2.42 $\pm$ 0.02 & -2.42 $\pm$ 0.02 & -2.42 $\pm$ 0.02\cr
$D_{3000}^\mathrm{PS_{EE}}$  &  $<$ 2.500$\,\mu$K$^2$ at 95\% &  $<$ 0.089$\,\mu$K$^2$ at 95\% &  $<$ 0.098$\,\mu$K$^2$ at 95\% \cr
$D_{80}^\mathrm{dust_{TE}}$ &  $<$ 2.00$\,\mu$K$^2$ at 95\% &  $<$ 1.35$\,\mu$K$^2$ at 95\% &  $<$ 0.98$\,\mu$K$^2$ at 95\% \cr
$D_{80}^\mathrm{dust_{EE}}$ &  $<$ 0.06$\,\mu$K$^2$ at 95\% &  $<$ 0.70$\,\mu$K$^2$ at 95\% &  $<$ 0.07$\,\mu$K$^2$ at 95\% \cr
\hline\hline
\end{tabular}
\label{tab:lcdm_ellCut}
\tablecomments{Parameters with hats have Gaussian priors. All other parameters have uniform priors.  See Section \ref{sec:lcdm} for details.}
\end{center}
\end{table*}

To test the consistency of \sptpolEETE with the \LCDM model, we perform a $\chi^2$ test on the SPTpol bandpowers, comparing them to binned theoretical bandpowers generated from the maximum likelihood theory curve for the \sptpolEETE-only constraints over each multipole range.
We find $\chi^2=137.0$ for 104 dof, having a PTE of 0.017 over the full dataset, which corresponds to a $2.1\,\sigma$ discrepancy with the \LCDM model.
The $\chi^2$ is high and PTE is low when calculated for \sptpolEETE-low and \sptpolEETE-high separately as well; $\chi^2=47.6$ for 28 dof (PTE of 0.012) and $\chi^2=82.9$ for 68 dof (PTE of 0.106), respectively.
We also note that the two datasets prefer slightly different cosmologies as can be seen in Table \ref{tab:lcdm_ellCut}.

\begin{figure*}[t]
\begin{center}
\includegraphics[width=1.0\textwidth]{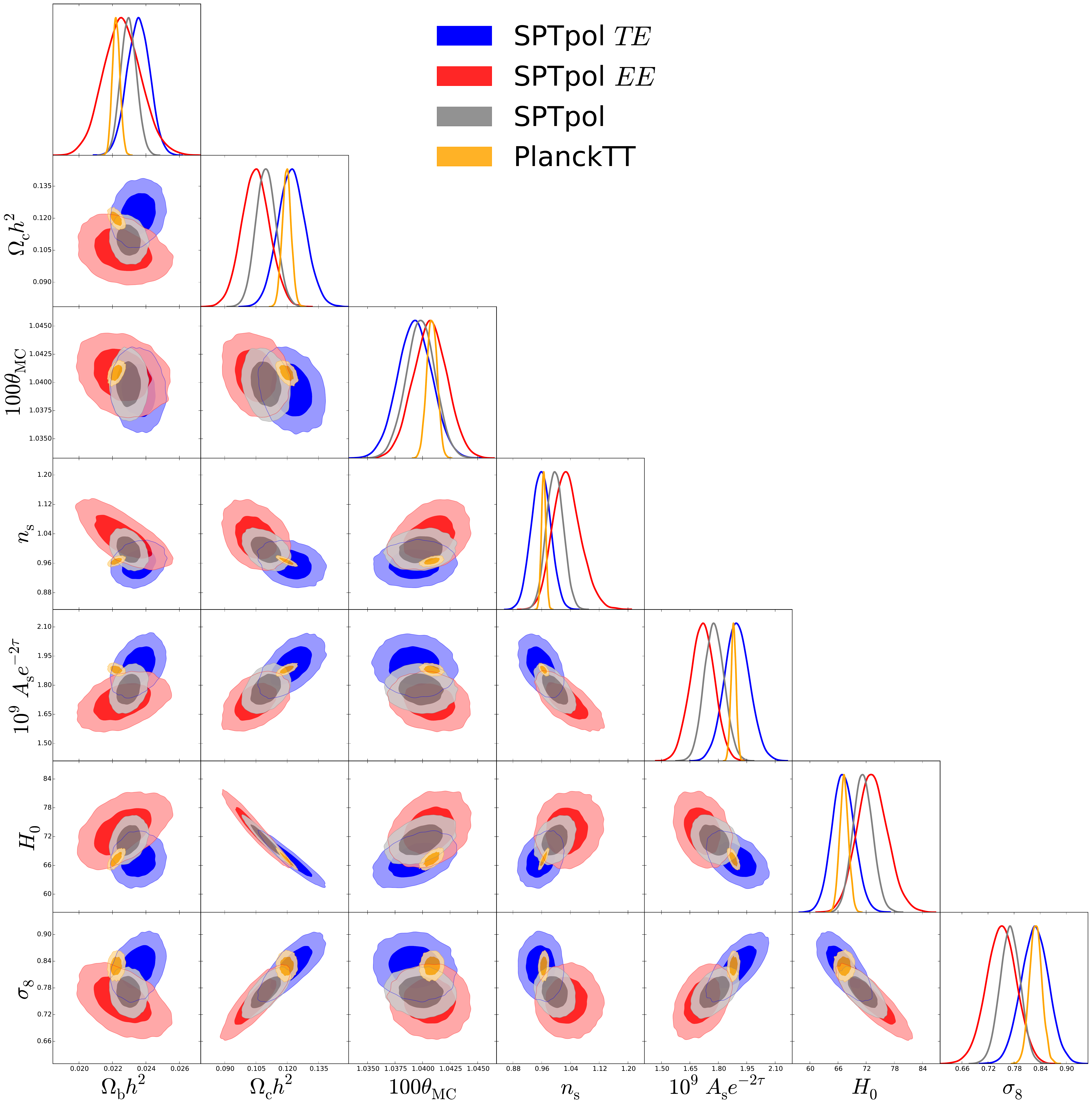}
\end{center}
\caption{Marginalized parameter constraints for the \LCDM model from several SPTpol data splits: $TE_\mathrm{full}$ (blue), $EE_\mathrm{full}$ (red), \sptpolEETE (gray).
We also plot the constraints for \planckTT (orange) for comparison.}
\label{fig:TE_EE_2D}
\vspace{0.5in}
\end{figure*}

To investigate the low PTEs, we further split \sptpolEETE-low and \sptpolEETE-high into $TE$-only and $EE$-only sets and recalculate parameter constraints for each split: $TE_\mathrm{low}$, $TE_\mathrm{high}$, $EE_\mathrm{low}$, and $EE_\mathrm{high}$.
We find the following PTEs when fitting the \LCDM model: \{$TE_\mathrm{low}$ - 0.128, $TE_\mathrm{high}$ - 0.024, $EE_\mathrm{low}$ - 0.004, $EE_\mathrm{high}$ - 0.366\}.
We also calculate constraints for $TE$ and $EE$ independently over the entire multipole range, $TE_\mathrm{full}$ and $EE_\mathrm{full}$, and find PTEs of 0.007 and 0.022, respectively.
In general, the data splits are poorly fit by the \LCDM model.
Furthermore, each data split pulls toward a slightly different \LCDM solution.
These solutions are difficult to compare directly since each split is sensitive to the \LCDM model differently and resulting parameter constraints exhibit varying degrees of degeneracy.
However, we plot marginalized parameter constraints for $TE_\mathrm{full}$, $EE_\mathrm{full}$, \sptpolEETE, and \planckTT in Figure \ref{fig:TE_EE_2D}.
For any one \LCDM parameter, there are $\sim 1\,\sigma$ shifts in the best-fit values between datasets.
While the tension between the best-fit cosmologies for SPTpol $TE$ and $EE$ is small for any single parameter, we nevertheless find this tension contributes to lowering the PTE when fitting the spectra simultaneously and over a large multipole range.

While the data pass null tests, a cross-spectrum analysis with a different dataset over the same patch of sky would further test for potential systematic contamination.
Unfortunately, the two datasets available for direct comparison over the same area as the SPTpol survey, those of \planck and of BICEP2/Keck, either are noisy or incompletely match sky coverage, making a cross-spectrum comparison difficult.
A detailed cross-data-set comparison correctly accounting for differences in coverage, filtering, and noise is beyond the scope of this work but is worth pursuing.

One possible source of discrepancies, particularly between low-$\ell$ and high-$\ell$ solutions, is the instrument beam.  
We point out, however, that we measure Venus and therefore the beam with high signal-to-noise ratio.  
Furthermore, any multipole-dependent error in the beam caused by $T\rightarrow P$ leakage is corrected via the leakage beam $G_\ell$, which is known with high precision through the high signal-to-noise ratio measurements of Venus.
Finally, if we multiply the beam covariance by a factor of 100 before calculating error eigenvectors, we find that the $\chi^2$ to the maximum-likelihood SPTpol cosmology changes by only -1.5, which suggests that neither the beam error nor any potential systematics in the beam are driving the cosmological fits. 

\subsubsection{Comparing \sptpolEETE to \planckTT}

What can we say about the preferred cosmological models of \sptpolEETE and \planckTT?
The marginalized contours for \planckTT are shown in Figure \ref{fig:TEEE_lcdm}, to illustrate that the preferred cosmological parameters for \sptpolEETE-low are in good agreement with \planckTT despite a poor goodness of fit. 
As more high-$\ell$ information is added, $\omc$ is driven lower and $\omb$ is driven higher.
Both of these changes drive up $\ho$ in order to preserve the acoustic peak scale, leading to 
\begin{equation}
\ho = 71.3\pm2.1\,\mbox{km}\,s^{-1}\mbox{Mpc}^{-1}
\end{equation} 
for \sptpolEETE.

Interestingly, this behavior is similar to that observed by \citet{aylor17} in the 2500 deg$^2$ SPT-SZ temperature power spectrum, especially since the SPTpol 500\,deg$^2$ survey is a subset of this larger field. 
\citet{aylor17} find that the \planck{} and SPT-SZ temperature data are completely consistent when restricted to the same sky modes. 
However, their density parameters shift toward a higher baryon density and lower matter density when going from the full sky to the SPT-SZ survey patch, and when adding the small-scale data. 
Together the two effects drive the Hubble constant approximately $2\,\sigma$ higher than the \planck full-sky value.
Similar analyses of \planck data with different $\ell$ splits have found that including high-$\ell$ ($>\sim800$) measurements from \planck causes shifts in the derived cosmological parameters \citep{addison16, planck16-51}, though in a different direction in $\omc$, $\ho$, and $\sigmaeight$ than we find in this work with SPTpol data.
Given the recent tension between low-$z$ and high-$z$ measurements of $\ho$ \citep{riess16}, these hints of $\ell$-dependent differences in cosmology are worth continued investigation.

The same trends in densities in the SPTpol constraints also drive a shift in $\sigmaeight$. 
\sptpolEETE-low prefers a value of $\sigmaeight = 0.820 \pm 0.041$ close to ($0.2\,\sigma$ below) the \planckTT{} value.
Adding the higher multipole data drives $\sigmaeight$ to a much lower value,
\begin{equation}
\sigmaeight= 0.771 \pm 0.024,
\end{equation}
which is $-2.5\,\sigma$ from the value preferred by \planckTT. 
As a result, $\sigmaeight$ is also the parameter that shows the most significant shift when adding \sptpolEETE{} to the \planckTT{} data. 
For the combined dataset, the preferred $\sigmaeight$ value shifts down by $0.9\,\sigma$ from \planckTT{} to
\begin{equation}
\sigmaeight = 0.817 \pm 0.014.
\end{equation}
Lower values of $\sigmaeight$ have also been inferred by other measurements of large-scale structure, including cosmic shear \citep[\eg,][]{joudaki17, hildebrandt17}, clusters of galaxies \citep[\eg,][]{dehaan16}, redshift space distortions \citep[\eg,][]{gil-marin17}, and CMB lensing \citep[\eg,][]{planck15-15}.

\subsubsection{Interpretation of Model Differences}

\begin{figure*}[t]
\begin{center}
\includegraphics[width=1.0\textwidth]{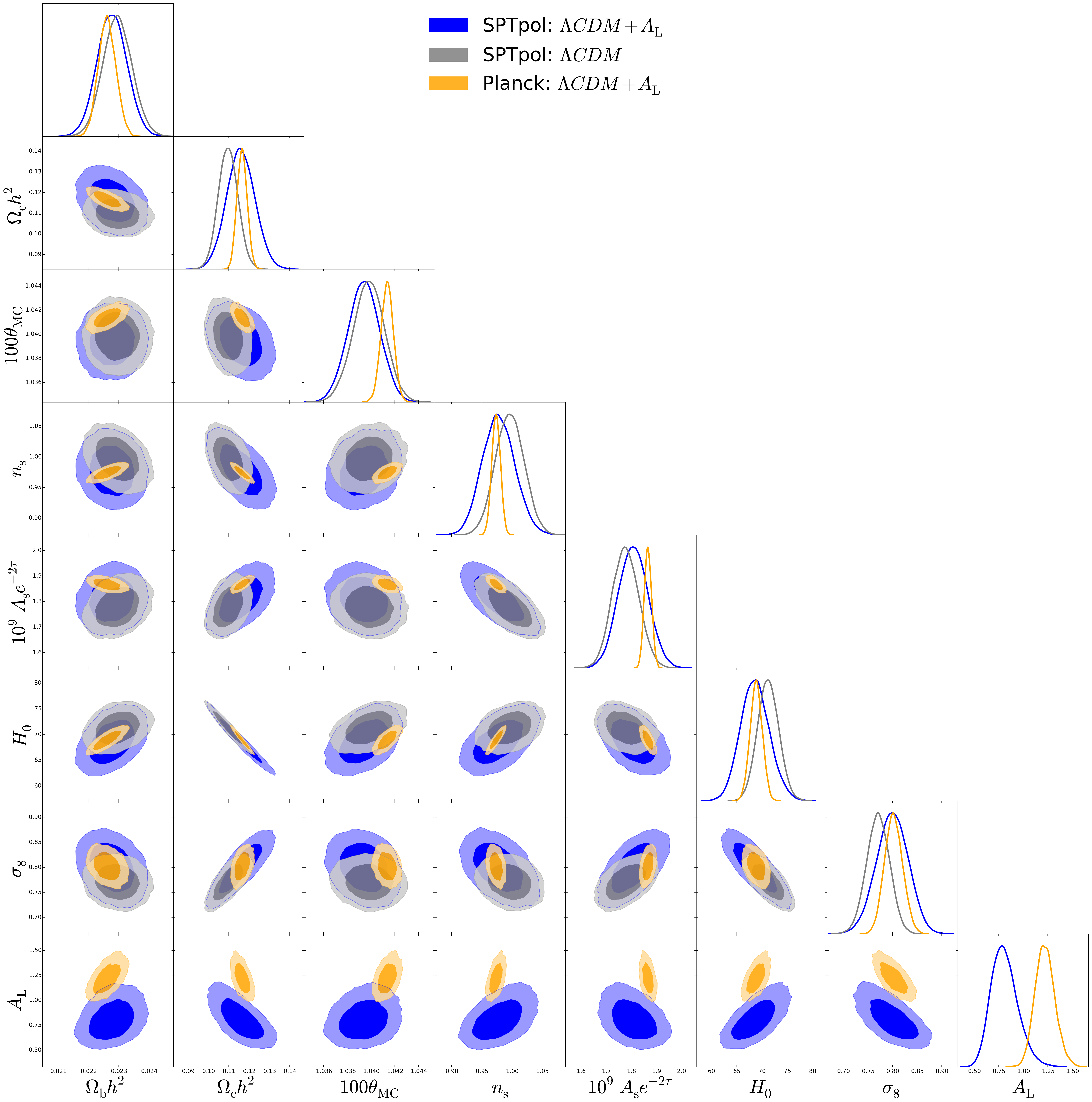}
\end{center}
\caption{Marginalized parameter constraints for the \LCDMAlens model using \sptpolEETE.
Marginalizing over $\Alens$ shifts \sptpolEETE constraints toward those preferred by \planckTT.
}
\label{fig:TEE_lcdm_Alens}
\vspace{0.5in}
\end{figure*}

Given that our data maps pass null tests, that we find no evidence for significant bias in our analysis pipeline or likelihood, and that our cosmological constraints appear insensitive to error in our beam uncertainties, we interpret the differences in cosmology over multipole range at face value.
At low $\ell$, the SPTpol dataset is in good agreement with \planckTT, but the bulk of the sensitivity of the dataset lies at higher $\ell$ and the inclusion of this new information pulls $\ho$ higher and $\sigmaeight$ lower in part because of a lower preferred value for $\omc$.

Lower matter content would imply less gravitational lensing and therefore sharper acoustic peaks.
Looking at the residuals of the SPTpol bandpowers compared to the \planckTT theory in Figures \ref{fig:TE} and \ref{fig:EE}, we see that the acoustic peaks in the high-$\ell$ data appear sharper than what the \planckTT theory prefers.
To quantify this, we fit a \LCDMAlens model to \sptpolEETE.
As in \citet{planck15-13}, $\Alens$ scales the theoretical lensing power spectrum $C_\ell^{\phi\phi}$ at every point in parameter space, and this scaled lensing spectrum is used to lens the CMB power spectra.
Using only \sptpolEETE, we find
\begin{equation}
\Alens = 0.81 \pm 0.14,
\end{equation}
which is $1.4\,\sigma$ below the \LCDM expectation of $\Alens=1.0$ and $2.9\,\sigma$ lower than the value preferred by \planckTT: $\Alens = 1.22 \pm 0.10$.
$\chi^2$ for the maximum likelihood \sptpolEETE \LCDMAlens model is 135.3, 1.7 lower than that for the \LCDM model, which is the largest improvement to $\chi^2$ of all model extensions we test in this analysis.
Furthermore, we see in Figure \ref{fig:TEE_lcdm_Alens} that marginalizing over $\Alens$ shifts the \sptpolEETE parameter constraints toward those preferred by \planckTT, suggesting that lensing is driving parameter differences between the datasets.
Measurements of CMB lensing using the CMB four-point function will be a valuable cross-check, similar to the results of  \citet{story14, planck15-15,sherwin16}, and an in-preparation analysis of the data used in this work.

\subsubsection{\sptpolEETE + \planckTT}

\begin{figure*}[t]
\begin{center}
\includegraphics[width=1.0\textwidth]{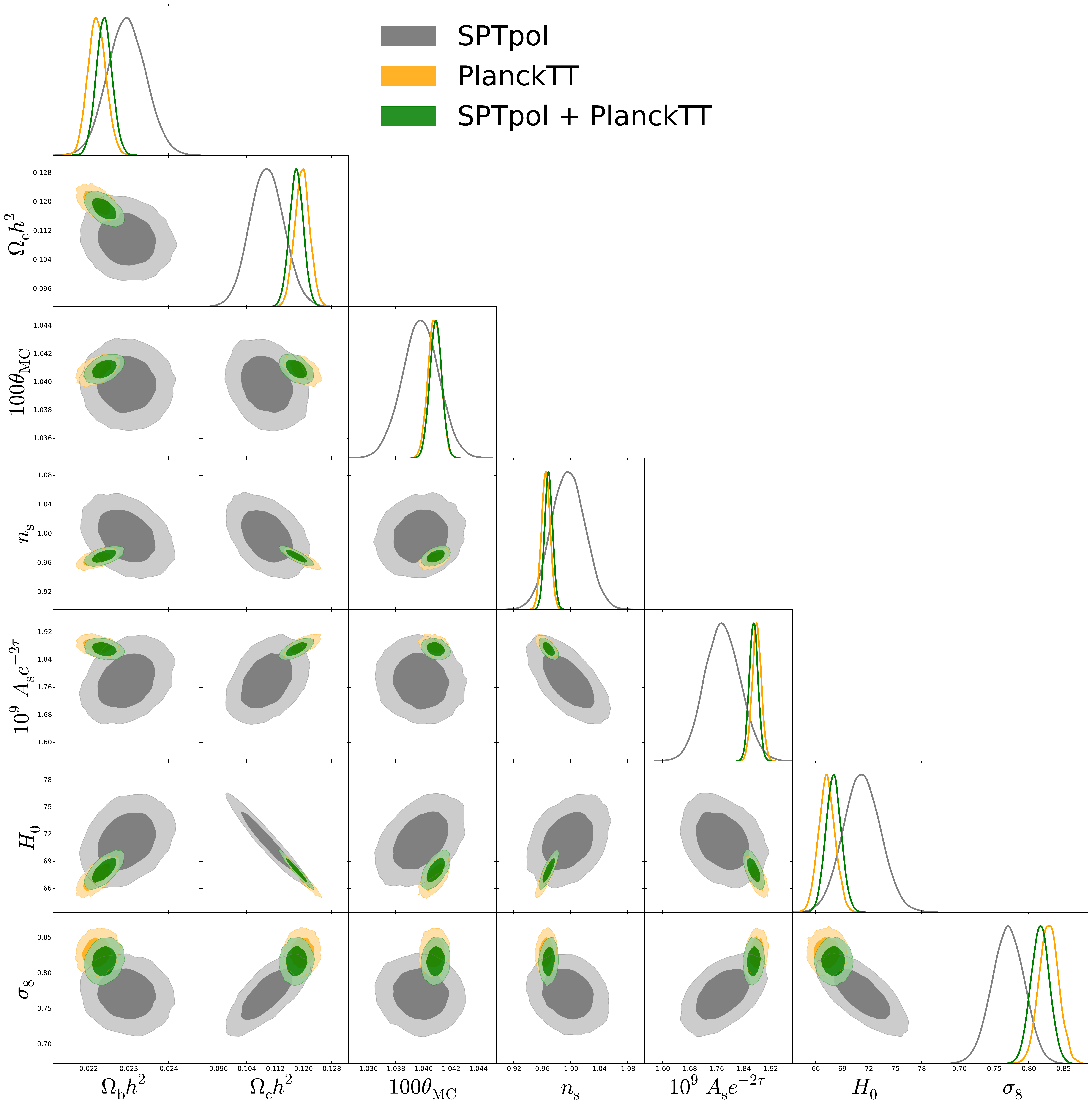}
\end{center}
\caption{Marginalized parameter constraints for the \LCDM model using \sptpolEETE, both independently and combined with \planckTT.
}
\label{fig:pTTso8_lcdm}
\end{figure*}

The SPTpol data presented here are the most sensitive in polarization to date at multipoles $\ell \gtrsim 1000$ and the inclusion of these data informs cosmological modeling.  
We consider joint fits between \sptpolEETE and \planckTT, which we include in Figure \ref{fig:pTTso8_lcdm} and Table \ref{tab:lcdm}.
While the joint 1D parameter constraints in Table \ref{tab:lcdm} are only marginally improved with the inclusion of SPTpol data, we also consider the overall improvement to the volume of 6D cosmological parameter space for the \LCDM model.
We use the determinant of the parameter covariance matrix as a metric for parameter space volume.
The ratio of covariance determinants for \cmb over \planckTT is 0.56, a factor of 1.8 reduction in the nonmarginalized parameter space.

\begin{table*}
\begin{center}
\caption{\LCDM Constraints}
\small

\begin{tabular}{c|c|c|c}
\hline\hline
\multicolumn{1}{c|}{Parameter} & \multicolumn{3}{c}{Data Set} \\ 
 & \multicolumn{1}{c}{\textsc{SPTpol}} & \multicolumn{1}{c}{\textsc{PlanckTT}} & \multicolumn{1}{c}{\textsc{PlanckTT+SPTpol}} \\ 
\hline 

\multicolumn{4}{l}{ Free }\\ 
$100\Omega_{\mathrm{b}}h^2$ & 2.296 $\pm$ 0.048 & 2.222 $\pm$ 0.023 & 2.240 $\pm$ 0.020\cr
$\Omega_{\mathrm{c}}h^2$ & 0.1098 $\pm$ 0.0048 & 0.1198 $\pm$ 0.0022 & 0.1181 $\pm$ 0.0020\cr
$100\theta_{\mathrm{MC}}$ & 1.0398 $\pm$ 0.0013 & 1.0408 $\pm$ 0.0005 & 1.0409 $\pm$ 0.0004\cr
$n_{\mathrm{s}}$ & 0.9967 $\pm$ 0.0238 & 0.9655 $\pm$ 0.0062 & 0.9693 $\pm$ 0.0057\cr
$10^{9}A_\mathrm{s}e^{-2\tau}$ & 1.7791 $\pm$ 0.0528 & 1.8805 $\pm$ 0.0138 & 1.8713 $\pm$ 0.0130\cr
\hline

\multicolumn{4}{l}{ Derived }\\ 
$\Omega_\Lambda$ & 0.736 $\pm$ 0.025 & 0.685 $\pm$ 0.013 & 0.695 $\pm$ 0.012\cr
$\sigma_8$ & 0.771 $\pm$ 0.024 & 0.830 $\pm$ 0.014 & 0.817 $\pm$ 0.014\cr
$H_0$ & 71.29 $\pm$ 2.12 & 67.30 $\pm$ 0.96 & 68.05 $\pm$ 0.89\cr
\hline

\multicolumn{4}{l}{ Nuisance + Foreground }\\ 
$\widehat{T}_\mathrm{cal}$ & 1.000 $\pm$ 0.003 & --- & 1.001 $\pm$ 0.003\cr
$\widehat{P}_\mathrm{cal}$ & 1.003 $\pm$ 0.010 & --- & 1.010 $\pm$ 0.006\cr
$100\widehat{\kappa}$ & 0.000 $\pm$ 0.102 & --- & 0.104 $\pm$ 0.066\cr
$\widehat{\alpha}^\mathrm{TE}_\mathrm{dust}$ & -2.42 $\pm$ 0.02 & --- & -2.42 $\pm$ 0.02\cr
$\widehat{\alpha}^\mathrm{EE}_\mathrm{dust}$ & -2.42 $\pm$ 0.02 & --- & -2.42 $\pm$ 0.02\cr
$D_{3000}^\mathrm{PS_{EE}}$ & 0.098$\,\mu$K$^2$ at 95\% & --- &  $<$ 0.107$\,\mu$K$^2$ at 95\% \cr
$D_{80}^\mathrm{dust_{TE}}$ &  $<$ 0.98$\,\mu$K$^2$ at 95\% & --- &  $<$ 1.17$\,\mu$K$^2$ at 95\% \cr
$D_{80}^\mathrm{dust_{EE}}$ &  $<$ 0.07$\,\mu$K$^2$ at 95\% & --- &  $<$ 0.06$\,\mu$K$^2$ at 95\% \cr
\hline\hline
\end{tabular}
\label{tab:lcdm}
\tablecomments{Parameters with hats have Gaussian priors. All other parameters have uniform priors.  $\tau$ has a Gaussian prior only when SPTpol data are fitted independently.  See Section \ref{sec:lcdm} for details.}
\end{center}
\end{table*}

\subsubsection{Polarized Power from Extragalactic Sources}
\label{sec:ptsrc}

One benefit of using the polarized CMB power spectra to constrain cosmology over the temperature spectrum is the comparative lack of foregrounds.
At current map depths, only polarized power from extragalactic sources is a potentially limiting foreground at high $\ell$.
Unmasked sources contribute power that is constant in $C_\ell$ or $\propto \ell^2$ in $D_\ell$, often referred to as ``Poisson'' power.
In C15, we placed a 95\% confidence upper limit on residual polarized Poisson power in the $EE$ spectrum of $\Dps < 0.40\,\muksq$ at $\ell=3000$ after masking all sources with unpolarized flux greater than 50\,mJy at 150\,GHz.

While the SPTpol 500\,\sqdeg dataset is of comparable map depth to the measurements of C15, the survey area is five times greater, and we include multipoles out to $\ell_\mathrm{max}=8000$, up from $\ell_\mathrm{max}=5000$ in C15.
These two factors considerably improve our sensitivity to residual Poisson power at high $\ell$.
As stated in Section \ref{sec:mode_coupling} we also mask sources with unpolarized flux greater than 50\,mJy at 95\,GHz, but this improves the constraint only negligibly.
In Table \ref{tab:lcdm}, we see that from \sptpolEETE alone we find $\Dps < 0.098\,\muksq$ at 95\% confidence.
When including \planckTT, the constraint remains virtually the same,
\begin{equation}
\Dps <0.107\,\muksq\,\mbox{at 95\% confidence},
\end{equation}
which is a factor of 4 improvement over the C15 upper limit.
The constraint corresponds to $C_\ell^{EE} < 7.4\times10^{-8}\,\muksq$, or $< 0.94\,\microK$-arcmin rms fluctuations in the SPTpol 500\,\sqdeg $E$-mode map coming from Poisson power from unmasked sources.

Following the arguments of C15, we find that this upper limit corresponds to a Poisson term that crosses the best-fit \cmb \LCDM spectrum at $\ell \sim 3800$.
As discussed in C15, this upper limit could in principle be improved by at least a factor of 2 by reducing the unpolarized flux mask threshold from $> 50\,$mJy to $> 6\,$mJy.
In this case, a residual $\ell^2$ term from Poisson sources would cross the $EE$ spectrum at $\ell  = 4050$, making up to 13 acoustic peaks potentially resolvable above the polarized foreground.

As in C15, we can interpret the upper limit on \Dps\ as an upper limit on the mean-squared polarization fraction of extragalactic sources.
The calculation here differs from C15 in two ways: (1) here we take into account the extra uncertainty in the inferred polarization fraction due to sample variance on \Dps, and (2) we correct a $\sqrt{2}$ error in the calculation from C15 (in which all the polarized Poisson power was mistakenly assigned to the $EE$ spectrum).
We take sample variance into account by creating many mock skies with point-source populations consistent with the measurements of \citet{mocanu13}, assigning each source in those mock skies a random degree of polarization drawn from an underlying distribution with a single mean-squared polarization fraction (and a random polarization angle drawn from a uniform distribution), repeating this for many values of underlying mean-squared polarization fraction.
We apply the two-frequency source cut applied to the data in this work to these mock skies, we calculate \Dps\ for each mock sky, and we compare the mock-measured \Dps\ values to the posterior distribution for \Dps\ in our cosmological fits to obtain a 95\% upper limit on polarization fraction.

As discussed in C15, with a 50\,mJy cut, point-source power in the temperature power spectrum is 
expected to be roughly equally distributed between synchrotron-dominated and dust-dominated sources, 
but the synchrotron population is expected to be much more strongly polarized.
If we assume that \Dps\ comes entirely from synchrotron sources, we infer a 95\% upper limit to the rms polarization fraction of those sources of 0.15 (15\%).
If we instead assume that all sources (synchrotron and dusty) have the same underlying polarization distribution, we infer a 95\% upper limit to the rms polarization fraction of all sources of 0.11 (11\%). After correcting for the $\sqrt{2}$ error in C15, the limits from that work were 29\% and 20\% respectively; hence, we find the expected factor of $\sim$2 improvement over that work (and find that including sample variance does not degrade the result significantly).

\subsection{High-$\ell$ Extensions to \LCDM}
\label{sec:extensions}
The low quality of fit of the \LCDM{} model to the data could reflect a need for additional physics beyond \LCDM. 
Motivated by the observation that the low-$\ell$ and high-$\ell$ data favor different cosmological parameters, in this section we consider several model extensions to \LCDM that alter the spectra at high $\ell$. 
First, we place constraints on the primordial helium fraction \yhe.
Second, we allow the effective number of relativistic species \neff to vary from the expectation of 3.046.
Finally, we consider constraints for a two-parameter extension, \LCDMyheneff.

We do not find a clear preference in the data for any of these extensions, all of which lead to minimal changes in $\chi^2$.
While we do not highlight them in this section, we also studied adding running of the scalar spectra index $\frac{d\ns}{dk}$ as well as energy injection from dark matter particle annihilation based on the work of \citet{finkbeiner12}.
These one-parameter extensions improve the best-fit $\chi^2$ by less than 0.3.
Searching for physically motivated extensions supported by the data merits further study.

\subsubsection{\LCDMyhe}
\label{sec:lcdm_yhe}

Precision CMB polarization spectra strengthen tests of the big bang nucleosynthesis (BBN) predictions for the primordial helium abundance \yhe. 
Since helium recombines at higher temperatures than hydrogen, and therefore at earlier times, an increase in \yhe would decrease the density of electrons before matter-radiation decoupling and increase the photon mean free path at the last-scattering surface. 
This shifts the damping scale to lower $\ell$. The $TE$ and $EE$ spectra approximately double the number of modes compared to $TT$ alone, and thus the SPTpol $TE$ and $EE$ bandpowers across the damping tail can tighten our constraints on the CMB damping scale.
Acoustic peaks in the $TE$ and $EE$ spectra are resolved in the SPTpol measurements to $\ell \sim 3000$, so we use this measurement of the damping tail to place constraints on \yhe in a \LCDMyhe model.
Normally \yhe is kept constant at the prediction from BBN given \omb and \neff.
By freeing \yhe, we are probing physics that alters the helium content of the universe between BBN and the epoch of recombination.

\sptpolEETE-only prefers
\begin{equation}
\yhe = 0.234 \pm 0.052,
\end{equation}
a shift of $-0.1\,\sigma$ from the BBN expectation and $-0.2\,\sigma$ from the \planckTT-only constraint.
Combining the data sets, we find
\begin{equation}
\yhe = 0.244 \pm 0.019,
\end{equation}
a 10\% reduction in marginalized uncertainty from the inclusion of \sptpolEETE.
When considering the full volume of 7D cosmological parameter space for the \LCDMyhe model, we also find that including \sptpolEETE reduces the parameter space by a factor of 2.3, which begins to demonstrate the constraining power of damping tail measurements.

\subsubsection{\LCDMneff}
\label{sec:lcdm_neff}

The energy density from non-photon-relativistic sources, \ie, neutrinos and potentially other particles, is parameterized by \neff, the effective number of relativistic species.
Standard particle theory predicts $\neff = 3.046$, where there are three neutrino species that are slightly heated compared to photons by electron-positron annihilation at early times.
A significant deviation from 3.046 could indicate additional physics, for example, additional weakly interacting relativistic species or nonstandard particle heating/cooling.

By itself, \sptpolEETE finds no evidence for additional relativistic species, with $\neff = 3.66\pm0.72$ while assuming consistency with BBN.
When combining with \planckTT, we find that
\begin{equation}
\neff = 3.18 \pm 0.28,
\end{equation}
a 12\% improvement in uncertainty in \neff over \planckTT alone.
Furthermore, including \sptpolEETE reduces the 7D parameter space of the model by a factor of 2.9.
From these datasets, we see no statistically significant evidence for additional relativistic species.

\subsubsection{\LCDMyheneff}
\label{sec:lcdm_yhe_neff}

\begin{figure*}[t]
\begin{center}
\includegraphics[width=1.0\textwidth]{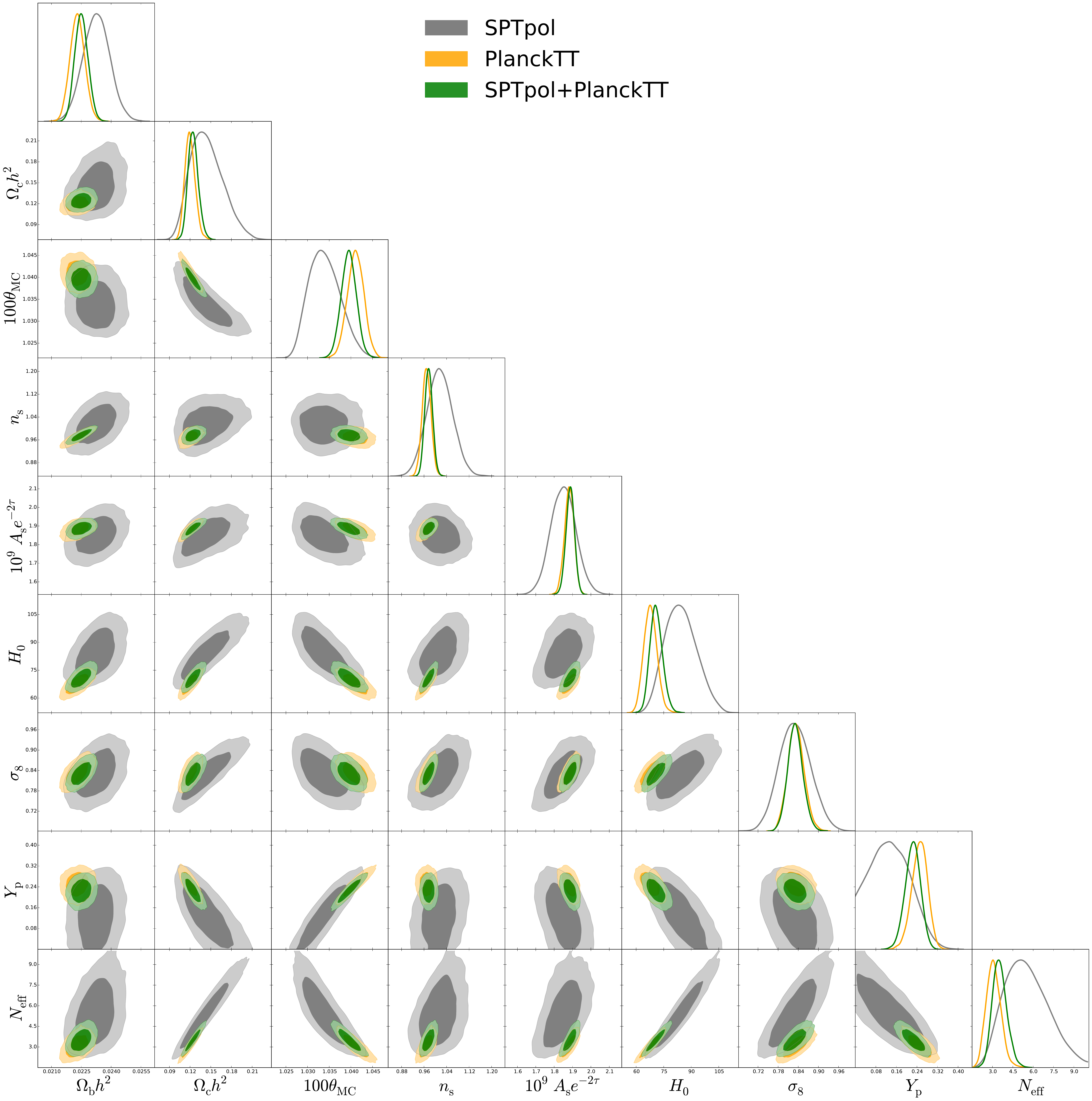}
\end{center}
\caption{Marginalized parameter constraints for the \LCDMyheneff model using \sptpolEETE, both independently and combined with \planckTT.
}
\label{fig:pTTso8_lcdm_yhe_neff}
\vspace{0.5in}
\end{figure*}

In this section we constrain a two-parameter extension model, \LCDMyheneff.
The amount of primordial helium \yhe is predicted in BBN by the baryon and neutrino content of the universe, \omb and \neff, respectively.
Allowing both to vary simultaneously probes for physics that breaks this expectation, \eg, additional energy injection after BBN but before the epoch of recombination.
While both \yhe and \neff have similar effects on the damping tail, \neff also induces an $\ell$-dependent phase shift on the acoustic peaks, which partially breaks the degeneracy.

Constraints for \LCDMyheneff are given in Table \ref{tab:lcdm_yhe_neff} and Figure \ref{fig:pTTso8_lcdm_yhe_neff}.
With bandpowers spaced by $\delta \ell=50$ through most of the acoustic oscillations in the SPTpol dataset, the phase shift from \neff is not well resolved, leaving \yhe and \neff largely degenerate.
\sptpolEETE finds $\yhe = 0.137\pm0.075$ and $\neff = 5.53\pm1.59$.
Adding more resolved acoustic peaks from \planckTT helps to break the degeneracy, and we find
\begin{equation}
\begin{split}
\yhe &= 0.224\pm0.030, \\
\neff &= 3.54\pm0.54. \\
\end{split}
\end{equation}
The joint constraints on \yhe and \neff from \planckTT alone are marginally improved with the addition of \sptpolEETE.
The constraining power of the SPTpol data manifests as a factor of 2.2 decrease in the volume of 8D cosmological parameter space for this model.
With \cmb we find that \yhe and \neff are consistent with standard particle theory and BBN.
We note, however, that the significant decrease in parameter space volume we see in all of the \LCDM extension models highlights the potential of future experiments to further probe these theories using more precise measurements of the CMB damping tail at smaller scales.

\begin{table*}
\begin{center}
\caption{\LCDMyheneff Constraints}
\small

\begin{tabular}{c|c|c|c}
\hline\hline
\multicolumn{1}{c|}{Parameter} & \multicolumn{3}{c}{Data Set} \\ 
 & \multicolumn{1}{c}{\textsc{SPTpol}} & \multicolumn{1}{c}{\textsc{PlanckTT}} & \multicolumn{1}{c}{\textsc{PlanckTT+SPTpol}} \\ 
\hline 

\multicolumn{4}{l}{ Free }\\ 
$100\Omega_{\mathrm{b}}h^2$ & 2.326 $\pm$ 0.067 & 2.231 $\pm$ 0.037 & 2.251 $\pm$ 0.033\cr
$\Omega_{\mathrm{c}}h^2$ & 0.1445 $\pm$ 0.0232 & 0.1200 $\pm$ 0.0075 & 0.1246 $\pm$ 0.0075\cr
$100\theta_{\mathrm{MC}}$ & 1.0340 $\pm$ 0.0036 & 1.0411 $\pm$ 0.0019 & 1.0395 $\pm$ 0.0017\cr
$n_{\mathrm{s}}$ & 1.0159 $\pm$ 0.0459 & 0.9691 $\pm$ 0.0159 & 0.9766 $\pm$ 0.0139\cr
$10^{9}A_\mathrm{s}e^{-2\tau}$ & 1.8503 $\pm$ 0.0681 & 1.8821 $\pm$ 0.0260 & 1.8872 $\pm$ 0.0234\cr
$Y_\mathrm{p}$ & 0.137 $\pm$ 0.075  &  0.250 $\pm$ 0.031 & 0.224 $\pm$ 0.030\cr
$N_\mathrm{eff}$ & 5.53 $\pm$ 1.59  &  3.09 $\pm$ 0.54 & 3.54 $\pm$ 0.54\cr
\hline

\multicolumn{4}{l}{ Derived }\\ 
$\Omega_\Lambda$ & 0.762 $\pm$ 0.027 & 0.688 $\pm$ 0.022 & 0.706 $\pm$ 0.018\cr
$\sigma_8$ & 0.831 $\pm$ 0.047 & 0.833 $\pm$ 0.024 & 0.831 $\pm$ 0.023\cr
$H_0$ & 84.39 $\pm$ 8.50 & 67.79 $\pm$ 3.62 & 70.98 $\pm$ 3.45\cr
\hline

\multicolumn{4}{l}{ Nuisance + Foreground }\\ 
$\widehat{T}_\mathrm{cal}$ & 1.000 $\pm$ 0.003 & --- & 1.001 $\pm$ 0.003\cr
$\widehat{P}_\mathrm{cal}$ & 1.003 $\pm$ 0.010 & --- & 1.010 $\pm$ 0.006\cr
$100\widehat{\kappa}$ & -0.000 $\pm$ 0.101 & --- & 0.100 $\pm$ 0.066\cr
$\widehat{\alpha}^\mathrm{TE}_\mathrm{dust}$ & -2.42 $\pm$ 0.02 & --- & -2.42 $\pm$ 0.02\cr
$\widehat{\alpha}^\mathrm{EE}_\mathrm{dust}$ & -2.42 $\pm$ 0.02 & --- & -2.42 $\pm$ 0.02\cr
$D_{3000}^\mathrm{PS_{EE}}$ & 0.095$\,\mu$K$^2$ at 95\% & --- &  $<$ 0.105$\,\mu$K$^2$ at 95\% \cr
$D_{80}^\mathrm{dust_{TE}}$ &  $<$ 0.98$\,\mu$K$^2$ at 95\% & --- &  $<$ 1.19$\,\mu$K$^2$ at 95\% \cr
$D_{80}^\mathrm{dust_{EE}}$ &  $<$ 0.08$\,\mu$K$^2$ at 95\% & --- &  $<$ 0.06$\,\mu$K$^2$ at 95\% \cr
\hline\hline
\end{tabular}
\label{tab:lcdm_yhe_neff}
\tablecomments{Parameters with hats have Gaussian priors. All other parameters have uniform priors.  $\tau$ has a Gaussian prior only when SPTpol data are fitted independently.  See Section \ref{sec:lcdm_yhe_neff} for details.}
\end{center}
\end{table*}


\section{Conclusion}
\label{sec:conclusion}

We have presented measurements of the $E$-mode angular auto-power and temperature-$E$-mode cross-power spectra of the CMB over the multipole range $50 < \ell \le 8000$.
These data are the most sensitive measurements to date of the $EE$ and $TE$ spectra at $\ell > 1050$ and $\ell >1475$, respectively, and demonstrate the potential of constraining cosmological parameters with information from the polarized CMB damping tail.

We have placed an upper limit on residual polarized point-source power after masking sources with unpolarized flux $> 50$\,mJy at 95 and 150\,GHz: $D_\ell < 0.107\,\muksq$ at $\ell=3000$.
This upper limit implies that with more aggressive source masking the power from polarized extragalactic sources could be reduced to a level such that the amplitude of primordial $EE$ power would be greater than the amplitude of extragalactic source power to at least $\ell=4050,$ and possibly much higher.
Compared to the $TT$ spectrum, which becomes dominated by several foregrounds including clustered and nonclustered extragalactic sources and the thermal and kinetic \sze s by $\ell \sim 3000$, the $EE$ damping tail promises to provide a much deeper look into physics at the photon diffusion scale.

The SPTpol dataset is in mild tension with the \LCDM model, discrepant at $2.1\,\sigma$.
This tension can be attributed in part to slightly different preferred cosmologies between the $TE$ and $EE$ bandpowers and between low and high $\ell$.
Interestingly, while SPTpol data at $\ell < 1000$ are in good agreement with the best-fit model of \planckTT, we see parameters pulled to new values with the addition of higher multipole polarization information, resulting in a higher $\ho$ and lower $\sigmaeight$.
This is similar to the behavior measured by \citet{aylor17} on SPT-SZ temperature data on $2500\,\mbox{deg}^2$, of which the SPTpol field is a subset.
The parameter most affected by the inclusion of \sptpolEETE is $\sigmaeight$, which decreases by $1\,\sigma$ from the value preferred by \planckTT alone.
This behavior is related to the preference for less lensing in the SPTpol data set, which prefers a value of $\Alens$ that is $2.9\,\sigma$ less than \planckTT.

Tensions between CMB and non-CMB datasets exist for some parameters.
For example, the local value of $\ho$ as measured from Type Ia supernova light curves is between $2.1\,\sigma$ and $3.4\,\sigma$ discrepant with the CMB-derived value, depending on the dataset considered \citep{riess16}.
Constraints using baryon acoustic oscillation measurements and estimates of the primordial deuterium abundance, which use no CMB measurements, are also $3\,\sigma$ discrepant with low-$z$ $\ho$ estimates \citep{addison17}.
While there is the possibility of unknown systematics in any of these datasets, this tension could hint at physics beyond \LCDM.
A higher value of \neff, for example, would help alleviate some of the discrepancy, though we do not find statistically significant preference for a higher \neff in this analysis.
Regardless, sensitivity to physics at the photon diffusion scale makes the polarized damping tail an important laboratory for studying the source of parameter tensions.

While 1D marginalized parameter constraints are modestly improved over \planckTT alone, the inclusion of SPTpol data significantly reduces the volume of nonmarginalized parameter space, a factor of 1.8 for \LCDM and 2.3, 2.9, and 2.2 for the \LCDMyhe, \LCDMneff, and \LCDMyheneff models, respectively.
As current and future high-resolution CMB polarization experiments generate deeper data sets we will continue to make dramatic progress in constraining cosmological parameters.  
For example, SPT-3G is forecasted to improve constraints on \neff over \textit{Planck} alone by nearly a factor of 2 \citep{benson14}, and CMB-S4 is expected to provide at least another factor of 2 improvement \citep{cmbs4-sb1}.

Finally, we note that such a deep high-resolution measurement of $E$-mode polarization is a key component in the search for inflationary gravitational wave-induced large-scale $B$ modes \citep[\eg,][]{abazajian15a}.
Along with polarized Galactic dust, so-called lensing $B$ modes are a significant foreground contaminant at large scales.
This additional $B$-mode signal is generated at few-arcminute scales by the gravitational lensing of primordial $E$ modes by large-scale structure \citep{zaldarriaga98}.
Improved constraints on the presence of an inflationary $B$-mode signal require \textit{delensing}, the removal of lensing $B$-mode power from either the real-space polarization maps or the total $BB$ angular power spectrum.
The SPTpol Collaboration recently released a delensing analysis of its own data \citep{manzotti17}, and is pursuing delensing jointly with the BICEP2/Keck Collaboration, whose observation field overlaps the SPTpol field.
Future constraints on inflationary $B$ modes can only benefit from these and other high-resolution measurements of $E$-mode polarization.

\acknowledgements{
The South Pole Telescope program is supported by the National Science Foundation through grant PLR-1248097.
Partial support is also provided by the NSF Physics Frontier Center grant PHY-0114422 to the Kavli Institute of Cosmological Physics at the University of Chicago, the Kavli Foundation, and the Gordon and Betty Moore Foundation through grant GBMF\#947 to the University of Chicago.  
This work is also supported by the U.S. Department of Energy. 
J.W.H. is supported by the National Science Foundation under Award No. AST-1402161.
C.R. acknowledges support from an Australian Research Council Future Fellowship (FT150100074). 
B.B. is supported by the Fermi Research Alliance LLC under contract no. De-AC02-07CH11359 with the U.S. Department of Energy.  
The Cardiff authors acknowledge support from the UK Science and Technologies Facilities Council (STFC).
The CU Boulder group acknowledges support from NSF AST-0956135.  
The McGill authors acknowledge funding from the Natural Sciences and Engineering Research Council of Canada, Canadian Institute for Advanced Research, and Canada Research Chairs program.
Work at Argonne National Lab is supported by UChicago Argonne LLC, Operator of Argonne National Laboratory (Argonne). 
Argonne, a U.S. Department of Energy Office of Science Laboratory, is operated under contract no. DE-AC02-06CH11357. 
We also acknowledge support from the Argonne Center for Nanoscale Materials.  
This work used resources made available on the Jupiter cluster, a joint data-intensive computing project between the High Energy Physics Division and the Computing, Environment, and Life Sciences (CELS) Directorate at Argonne National Laboratory.
The data analysis pipeline also uses the scientific python stack \citep{hunter07, jones01, vanDerWalt11} and the HDF5 file format \citep{hdf5}.
}
\bibliographystyle{apj}
\bibliography{spt}

\appendix
\label{sec:appendix}

\begin{figure*}
\begin{center}
\includegraphics[width=0.64\textwidth]{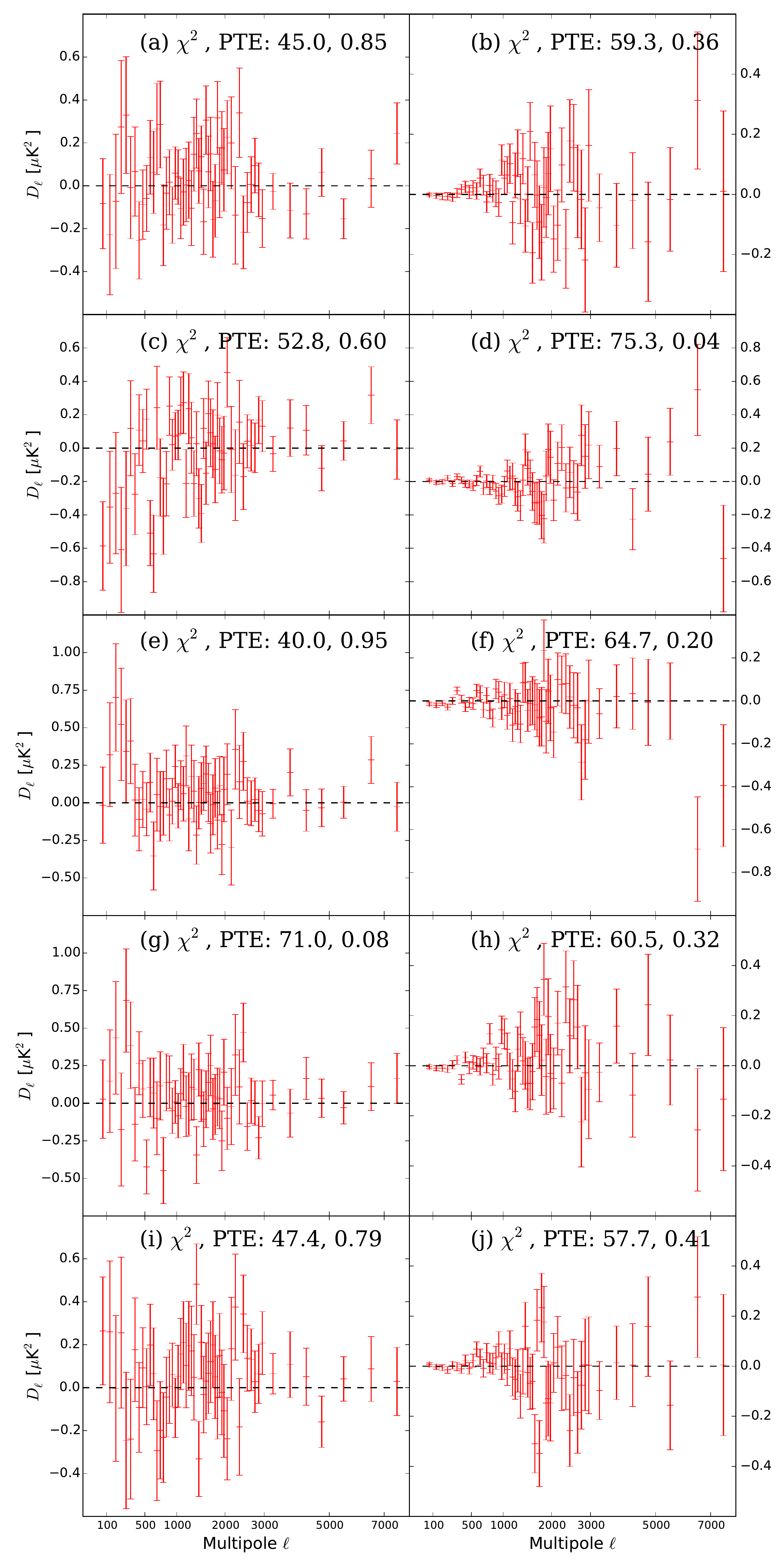}
\end{center}
\caption{Null spectra for $TE$ (left) and $EE$ (right) for the tests outlined in Section \ref{sec:consistency}.
We scale the $x$-axes to $\ell^{0.6}$.
Each row corresponds to a different null test; (a, b): Left-Right; (c, d): 1st Half-2nd Half; (e, f): Sun; (g, h): Moon; (i, j): Azimuth.}
\label{fig:null_spectra}
\vspace{0.5in}
\end{figure*}

We plot the null spectra from which the probabilities to exceed in Table \ref{tab:jacks} are calculated in Figure \ref{fig:null_spectra}.
The spectra are plotted in pseudo-$D_\ell$ space, meaning that they have not been corrected for the effects outlined in Section \ref{sec:ps} or for calibration.
When calculating $\chi^2$ for each spectrum, we use the noise-only bandpower covariance matrix (in pseudo-$D_\ell$ space) to account for correlations between bandpowers.
As discussed in Section \ref{sec:consistency}, the null spectra are free of significant systematic bias.
If a systematic were present in the data, it would be subdominant to noise, which is $\lesssim 0.5\,\mu$K$^2$ across the multipole range of interest. 
Furthermore, the amplitudes of the null spectra are much smaller than the $TE$ and $EE$ residuals plotted in Figures \ref{fig:TE} and \ref{fig:EE}.
If the null spectra were entirely attributed to unmodeled systematics and were subsequently removed, the bandpower residuals above would remain virtually unchanged, especially for $\ell \lesssim 2000$.

\end{document}